\documentclass[11pt, notitlepage]{article} 
\usepackage{a4}
\usepackage{xcolor}

\definecolor{TUMblue}{RGB}{0, 101, 189}
\definecolor{TUMlightblue}{RGB}{100,160,200}
\definecolor{TUMgreen}{RGB}{162,173,0}
\definecolor{TUMorange}{RGB}{227,114,034}
\definecolor{TUMivory}{RGB}{218,215,203}

\usepackage{natbib}

\usepackage{hyperref}
\hypersetup{
	colorlinks=true,
	linkcolor=TUMblue,
	citecolor=TUMblue,
	filecolor=TUMblue,
	urlcolor=TUMblue
}

\usepackage{etoolbox}
\makeatletter

\pretocmd{\NAT@citex}{%
	\let\NAT@hyper@\NAT@hyper@citex
	\def\NAT@postnote{#2}%
	\setcounter{NAT@total@cites}{0}%
	\setcounter{NAT@count@cites}{0}%
	\forcsvlist{\stepcounter{NAT@total@cites}\@gobble}{#3}}{}{}
\newcounter{NAT@total@cites}
\newcounter{NAT@count@cites}
\def\NAT@postnote{}

\def\NAT@hyper@citex#1{%
	\stepcounter{NAT@count@cites}%
	\hyper@natlinkstart{\@citeb\@extra@b@citeb}#1%
	\ifnumequal{\value{NAT@count@cites}}{\value{NAT@total@cites}}
	{\ifNAT@swa\else\if*\NAT@postnote*\else%
		\NAT@cmt\NAT@postnote\global\def\NAT@postnote{}\fi\fi}{}%
	\ifNAT@swa\else\if\relax\NAT@date\relax
	\else\NAT@@close\global\let\NAT@nm\@empty\fi\fi
	\hyper@natlinkend}
\renewcommand\hyper@natlinkbreak[2]{#1}

\patchcmd{\NAT@citex}
{\ifNAT@swa\else\if*#2*\else\NAT@cmt#2\fi
	\if\relax\NAT@date\relax\else\NAT@@close\fi\fi}{}{}{}
\patchcmd{\NAT@citex}
{\if\relax\NAT@date\relax\NAT@def@citea\else\NAT@def@citea@close\fi}
{\if\relax\NAT@date\relax\NAT@def@citea\else\NAT@def@citea@space\fi}{}{}

\makeatother

\usepackage{courier}
\usepackage{amssymb, graphicx}
\usepackage{subfigure}
\usepackage{amsmath}
\usepackage{amsthm}
\usepackage{verbatim}
\usepackage{dsfont}
\usepackage{geometry}
\usepackage{pdflscape}
\usepackage{multirow}
\usepackage{aliascnt}
\usepackage{verbatim}
\usepackage{graphicx}
\usepackage{float}
\usepackage{tikz}
\usetikzlibrary{calc,shapes,arrows,decorations.pathmorphing,graphs,positioning,backgrounds}
\usepackage{latexsym}
\usepackage{mathtools}
\usepackage{mathrsfs}
\usepackage{bbm}
\usepackage{shadethm}
\usepackage{enumerate}
\usepackage{colortbl}
\usepackage{framed}
\colorlet{shadecolor}{gray!25}
\usepackage{booktabs}
\usepackage{longtable}
\usepackage{multirow}
\usepackage{rotating}
\usepackage{chngpage}
\usepackage{appendix}

\newcommand{\mynewtheorem}[2]{
	\newaliascnt{#1}{dummy}
	\newtheorem{#1}[#1]{#2}
	\aliascntresetthe{#1}
	\expandafter\def\csname #1autorefname\endcsname{#2}
}

\bibpunct[\textcolor{TUMblue}{, }]{\textcolor{TUMblue}{(}}{\textcolor{TUMblue}{)}}{\textcolor{TUMblue}{;}}{\textcolor{TUMblue}{a}}{\textcolor{TUMblue}{}}{\textcolor{TUMblue}{,}}
\makeatletter
\renewcommand\eqref[1]{%
	\textup{\color{TUMblue}\tagform@{\ref{#1}}}%
}

\newcommand{\myrel}[2]{\genfrac{}{}{0pt}{1}{#1}{#2}}
\newcommand{\myrelArrowI}[2]{\genfrac{}{}{0pt}{3}{#1}{#2}}
\newcommand{\myrelArrowII}[2]{\genfrac{}{}{0pt}{0}{#1}{#2}}

\newtheorem{Theorem}{Theorem}
\newtheorem{Corollary}{Corollary}

\theoremstyle{definition}
\mynewtheorem{thm}{Theorem}
\mynewtheorem{defi}{Definition}
\mynewtheorem{lem}{Lemma}
\mynewtheorem{cor}{Corollary}
\mynewtheorem{prop}{Proposition}
\mynewtheorem{exa}{Example}
\mynewtheorem{alg}{Algorithm}
\mynewtheorem{rem}{Remark}
\mynewtheorem{bsp}{Example}

\usepackage{etoolbox}
\makeatletter
\patchcmd{\hyper@makecurrent}{%
	\ifx\Hy@param\Hy@chapterstring
	\let\Hy@param\Hy@chapapp
	\fi
}{%
	\iftoggle{inappendix}{
		\@checkappendixparam{chapter}%
		\@checkappendixparam{section}%
		\@checkappendixparam{subsection}%
		\@checkappendixparam{subsubsection}%
		\@checkappendixparam{paragraph}%
		\@checkappendixparam{subparagraph}%
	}{}%
}{}{\errmessage{failed to patch}}

\newcommand*{\@checkappendixparam}[1]{%
	\def\@checkappendixparamtmp{#1}%
	\ifx\Hy@param\@checkappendixparamtmp
	\let\Hy@param\Hy@appendixstring
	\fi
}
\makeatletter

\newtoggle{inappendix}
\togglefalse{inappendix}

\apptocmd{\appendix}{\toggletrue{inappendix}}{}{\errmessage{failed to patch}}
\apptocmd{\subappendices}{\toggletrue{inappendix}}{}{\errmessage{failed to patch}}

\usepackage{sectsty}
\allsectionsfont{\sffamily}

\usepackage{titlesec}
\titleformat{\chapter}[display]
{\normalfont\sffamily\LARGE\bfseries\centering}
{\chaptertitlename\ \thechapter}{20pt}{\LARGE}

\usepackage{caption}
\usepackage{footnote}

\newcommand{\scd}{\mathbbm{c}} 

\begin{document}
	
{	\renewcommand*{\thefootnote}{\fnsymbol{footnote}}
	\title{\textbf{\sffamily Vine copula based likelihood estimation of dependence patterns in multivariate event time data}}
	
	\date{\small \today}
\newcounter{savecntr1}
\newcounter{restorecntr1}
\newcounter{savecntr2}
\newcounter{restorecntr2}

\author{Nicole Barthel\setcounter{savecntr1}{\value{footnote}}\thanks{Department of Mathematics, Technische Universit{\"a}t M{\"u}nchen, Boltzmanstra{\ss}e 3, 85748 Garching, Germany (email: \href{mailto:nicole.barthel@tum.de}{nicole.barthel@tum.de} (corresponding author), \href{mailto:cczado@ma.tum.de}{cczado@ma.tum.de}, \href{mailto:matthias.killiches@tum.de}{matthias.killiches@tum.de})}, Candida Geerdens\setcounter{savecntr2}{\value{footnote}}\thanks{Center for Statistics, I-BioStat, Universiteit Hasselt, Agoralaan 1, B-3590 Diepenbeek, Belgium (email: \href{mailto:paul.janssen@uhasselt.be}{paul.janssen@uhasselt.be}, \href{mailto:candida.geerdens@uhasselt.be}{candida.geerdens@uhasselt.be})}, Matthias Killiches\setcounter{restorecntr1}{\value{footnote}}\setcounter{footnote}{\value{savecntr1}}\footnotemark
	\setcounter{footnote}{\value{restorecntr1}},%
	\\ Paul Janssen\setcounter{restorecntr2}{\value{footnote}}%
	\setcounter{footnote}{\value{savecntr2}}\footnotemark
	\setcounter{footnote}{\value{restorecntr2}} \ and Claudia Czado\setcounter{restorecntr1}{\value{footnote}}%
	\setcounter{footnote}{\value{savecntr1}}\footnotemark
	\setcounter{footnote}{\value{restorecntr1}}}
	
	\maketitle
}

\begin{abstract} 
	In many studies multivariate event time data are generated from clusters having a possibly complex association pattern. Flexible models are needed to capture this dependence. Vine copulas serve this purpose. Inference methods for vine copulas are available for complete data. Event time data, however, are often subject to right-censoring. As a consequence, the existing inferential tools, e.g.\ likelihood estimation, need to be adapted. A two-stage estimation approach is proposed. First, the marginal distributions are modeled. Second, the dependence structure modeled by a vine copula is estimated via likelihood maximization. Due to the right-censoring single and double integrals show up in the copula likelihood expression such that numerical integration is needed for its evaluation. For the dependence modeling a sequential estimation approach that facilitates the computational challenges of the likelihood optimization is provided. A three-dimensional simulation study provides evidence for the good finite sample performance of the proposed method. Using four-dimensional mastitis data, it is shown how an appropriate vine copula model can be selected for data at hand.
	\vspace{0.19cm}
	
\end{abstract}

\section{Introduction}

In many studies, primary interest lies in the time until a prespecified event occurs. Often, the data appear in clusters. For example, in \citet{Laevens97} time to mastitis infection in udder quarters of primiparous cows is observed. The cow is the cluster and the infection times of the four udder quarters are the clustered data. For an accurate analysis of clustered data flexible models are needed to describe the underlying dependence pattern. Copulas provide the right tools for this goal. A $d$-dimensional copula $\mathbb{C}$ is a distribution function on $[0,1]^d$ with uniformly distributed margins. According to \citet{Sklar59}, a copula is a dependence function that interconnects the marginal survival functions $S_j$, $j=1,\ldots,d$, and thereby models the joint survival function $S$, i.e. with $t_j \geq 0$

\begin{align*}
S\left(t_1,\dots,t_d\right) = \mathbb{C}\{S_1\left(t_1\right),\dots,S_d\left(t_d\right)\}.
\end{align*}

For clusters of size two, a large catalog of bivariate copula families exists. For clusters of size more than two, popular multivariate copulas such as exchangeable (EAC) and nested Archimedean copulas (NAC) \citep{embrechts2003modelling, hofert2008sampling, joe1993parametric, Nelsen2006} only induce restrictive association patterns. For instance, in EAC models all marginal copulas show exactly the same type (and even strength) of tail-dependence. In NAC models, the nesting condition limits all building blocks to stem from the same copula family leading again to the same type (but not strength) of tail-dependence. More flexible models are thus needed to capture complex association patterns present in clustered data. This is a difficult but at the same time a challenging exercise. Flexible alternatives for EAC and NAC include Joe-Hu copulas \citep{joe1996multivariate} and vine copulas \citep{aas2009pair,bedford2002vines,czado2010pair, kurowicka2010dependence,kurowicka2006uncertainty}. A Joe-Hu copula corresponds to a mixture of positive powers of max-infinitely divisible bivariate copulas. The induced association pattern is completely determined by the mixture and by the choice of bivariate copulas. The idea of a vine copula is to decompose the joint density of the clustered event times into a cascade of bivariate copula densities via conditioning. So, in both approaches bivariate copulas or bivariate copula densities are the building blocks. Given the variety of well-studied bivariate copulas, it is clear that Joe-Hu copulas and vine copulas allow a flexible modeling of the within-cluster association in clustered event time data.

For the above mentioned copula models the focus is usually on complete, i.e. non-censored, data. However, event time data are often subject to right-censoring. This means that for some observations the true event time is not observed but instead a lower (censored) time is registered. For example, in the mastitis study cows may be lost to follow-up (e.g.\ due to death) or may experience the event after the end of the study (censored at study end). The presence of right-censoring in clustered event time data complicates the statistical analysis substantially, but its incorporation is indispensable to arrive at a sound statistical analysis. Since this is not straightforward, the application of copulas to right-censored clustered data has been less explored. Recently, \citet{Geerdens2014} studied, for right-censored data, the model flexibility of Joe-Hu copulas as compared to less elaborate EAC and NAC models. Vine copulas have not yet been studied for right-censored clustered event times. Therefore, our main objective is to develop a likelihood based estimation approach using the flexible class of vine copulas. Using the theorem of \cite{Sklar59} and following the ideas in \cite{Shih95}, we proceed in two steps. In step one, the survival margins are modeled. Here, any estimation technique for univariate right-censored event time data can be used, e.g.\ maximum likelihood estimation or the nonparametric Kaplan-Meier estimator. Focus, however, lies in detecting the inherent dependence pattern using vine copula based likelihood estimation in the second step. Due to right-censoring, numerical integration is needed, making the global likelihood optimization computationally challenging. We introduce a sequential estimation approach to find a fair trade-off between the numerical demand caused by data complexity and the accuracy of the estimates.

In \autoref{Sec:VineCopulas}, we describe the construction of vine copulas; we consider trivariate and quadruple data. The mastitis data are described in detail in \autoref{Sec:Censoring}. Following the ideas in \citet{Shih95}, \autoref{Sec:MLE} contains the likelihood function for right-censored quadruple event time data. In particular, we provide the likelihood expression in terms of vine copula components and therewith extend existing vine copula concepts to the setting of right-censored clustered time-to-event data. In this section, we also discuss how to deal with numerical aspects of the presented optimization method. A simulation study is performed in \autoref{Sec:SimulationStudy} to demonstrate the good finite sample performance of our approach. In \autoref{Sec:MastitisApplication}, we revisit the mastitis data. Conclusions and remarks are collected in \autoref{Sec:Conclusion}.

\section{Vine Copulas}\label{Sec:VineCopulas}

First, we recall the definition of vine copulas  \citep{aas2009pair,bedford2002vines,czado2010pair,kurowicka2006uncertainty,kurowicka2010dependence} and explain how vine copulas are constructed following the approach taken in \citet{czado2010pair}.

The basic idea is to decompose a $d$-dimensional copula density $\mathbbm{c}$ into a product of $d\left(d-1\right)/2$ so-called pair-copulas via conditioning. The latter are copulas associated to bivariate conditional distributions. It is essential to note that the representation of $\mathbbm{c}$ in terms of pair-copulas is not unique. Depending on the conditioning strategy, there is a variety of possible decompositions. To organize the structure of a $d$-dimensional vine-copula, \citet{bedford2002vines} propose a sequence of linked trees. More precisely, a set of connected trees $\mathcal{V}\coloneqq \left(\mathcal{T}_1,\dots,\mathcal{T}_{d-1}\right)$ is called a regular vine (R-vine) on $d$ elements with the set of edges $E\left(\mathcal{V}\right) \coloneqq E_1 \cup \cdots \cup E_{d-1}$ and the set of nodes $N\left(\mathcal{V}\right) \coloneqq N_1 \cup \cdots \cup N_{d-1}$ if the following holds: 
\begin{enumerate}
	\item $\mathcal{T}_1$ is a tree with nodes $N_1=\lbrace 1,\dots,d \rbrace$ and edges $E_1$.
	\item For $j=2,\dots,d-1$, $\mathcal{T}_j$ is a tree with nodes $N_j=E_{j-1}$ and edges $E_j$.
	\item \textit{(Proximity condition)} For $j=2,\dots,d-1$, whenever two nodes of $\mathcal{T}_{j}$ are connected by an edge, the associated edges of $\mathcal{T}_{j-1}$ share a node.
\end{enumerate}	
Two four-dimensional examples of possible tree sequences, also called vine structures, are visualized in \autoref{fig:VineExamples}. We see that except for the labeling of the nodes the two examples illustrate the only possible ways to arrange the four nodes in $\mathcal{T}_1$. In the vine structure on the right, there exists a unique node in $\mathcal{T}_j$, $j=1,2,3$, that is connected to $d-j$ edges. This vine structure suggests an ordering by importance and is referred to as a C-vine. The vine structure on the left is called a D-vine. Here, no node is connected to more than two edges implying a serial ordering. In particular, in dimension three C-vines and D-vines are equivalent. In this paper, we concentrate on D-vines. The derived concepts, however, can easily be applied to C-vines \citep{Thesis_Barthel}.

\tikzstyle{ClassicalVineNode} = [ellipse, fill = white, draw = black, text = black, align = center, minimum height = .7cm, minimum width = .7cm]
\tikzstyle{TreeLabels} = [draw = none, fill = none, text = black, font = \bf]
\tikzstyle{DummyNode}  = [draw = none, fill = none, text = white]
\newcommand{\yshift}{-.5cm}
\newcommand{\yshiftlabel}{+.13cm}
\newcommand{\labelsize}{\scriptsize}
\begin{figure}
	\caption{Two examples of a four-dimensional vine structure: a D-vine on the left and a C-vine on the right}
	\label{fig:VineExamples}
	\begin{minipage}{.48\linewidth}
		\centering
		\hspace{-.55
			cm}
		\begin{tikzpicture}	[every node/.style = ClassicalVineNode, node distance =.75cm, font = \footnotesize]
		\node (1){1}
		node[DummyNode]  (Dummy12)   [right of = 1]{}
		node             (2)         [right of = Dummy12] {2}
		node[DummyNode]  (Dummy23)   [right of = 2]{}			
		node             (3)         [right of = Dummy23] {3}
		node[DummyNode]  (Dummy34)   [right of = 3]{}
		node             (4)         [right of = Dummy34] {4}
		node             (12)        [below of = Dummy12, yshift = \yshift] {12}
		node[DummyNode]  (Dummy13;2) [right of = 12]{}
		node             (23)        [below of = Dummy23, yshift = \yshift] {23}
		node[DummyNode]  (Dummy24;3) [right of = 23]{}
		node             (34)        [below of = Dummy34, yshift = \yshift] {34}
		node             (13;2)      [below of = Dummy13;2, yshift = \yshift] {13;2}
		node             (24;3)      [below of = Dummy24;3, yshift = \yshift] {24;3}
		node[TreeLabels] (T1)        [left  of = 1] {$\mathcal{T}_1$}
		node[TreeLabels] (T2)        [below of = T1, yshift = \yshift] {$\mathcal{T}_2$}
		node[TreeLabels] (T3)        [below of = T2, yshift = \yshift] {$\mathcal{T}_3$}		
		;	    	
		\draw (1) to node[draw=none, fill = none, font = \labelsize, above, yshift = \yshiftlabel] {12} (2);
		\draw (2) to node[draw=none, fill = none, font = \labelsize, above, yshift = \yshiftlabel] {23} (3);
		\draw (3) to node[draw=none, fill = none, font = \labelsize, above, yshift = \yshiftlabel] {34} (4);
		\draw (12) to node[draw=none, fill = none, font = \labelsize, above, yshift = \yshiftlabel] {13;2} (23);
		\draw (23) to node[draw=none, fill = none, font = \labelsize, above, yshift = \yshiftlabel] {24;3} (34);
		\draw (13;2) to node[draw=none, fill = none, font = \labelsize, above, yshift = \yshiftlabel] {14;23} (24;3);
		\end{tikzpicture}
	\end{minipage}
	\begin{minipage}{.48\linewidth}
		\centering
		\renewcommand{\yshift}{-.9cm}
		\renewcommand{\yshiftlabel}{-0.10cm}
		\begin{tikzpicture}	[every node/.style = ClassicalVineNode, node distance =.75cm, font = \footnotesize]
		\node (1){1}
		node             (3)         [below of = 1, yshift = 2*\yshift] {3}			
		node             (2)         [left of = 3] {2}
		node             (4)         [right of = 3] {4}
		node             (13)        [right of = 4,xshift=.25cm] {13}
		node[DummyNode]  (Dummy34)   [right of = 13]{}
		node             (14)        [right of = Dummy34] {14}
		node             (12)        [above of = Dummy34, yshift = -2*\yshift] {12}
		node  (24;1)                 [right of = 14,xshift=.5cm]{24;1}
		node  (23;1)                 [above of = 24;1, yshift = -2*\yshift]{23;1}
		node[TreeLabels] (T1)        [above  of = 1,yshift=-.1cm] {$\mathcal{T}_1$}
		node[TreeLabels] (T2)        [above of = 12,yshift=-.1cm] {$\mathcal{T}_2$}
		node[TreeLabels] (T3)        [above of = 23;1,yshift=-.1cm] {$\mathcal{T}_3$}		
		;	    	
		\draw (1) to node[draw=none, fill = none, font = \labelsize, above, rotate = 72, yshift = \yshiftlabel] {12} (2);
		\draw (1) to node[draw=none, fill = none, font = \labelsize, above, rotate = 90, yshift = \yshiftlabel] {13} (3);
		\draw (1) to node[draw=none, fill = none, font = \labelsize, above, rotate = -72, yshift = \yshiftlabel] {14} (4);
		\draw (12) to node[draw=none, fill = none, font = \labelsize, above, rotate = 72, yshift = \yshiftlabel] {23;1} (13);
		\draw (12) to node[draw=none, fill = none, font = \labelsize, above, rotate = -72, yshift = \yshiftlabel] {24;1} (14);
		\draw (23;1) to node[draw=none, fill = none, font = \labelsize, above, rotate = 90, yshift = \yshiftlabel] {34;12} (24;1);
		\end{tikzpicture}
	\end{minipage}
\end{figure}
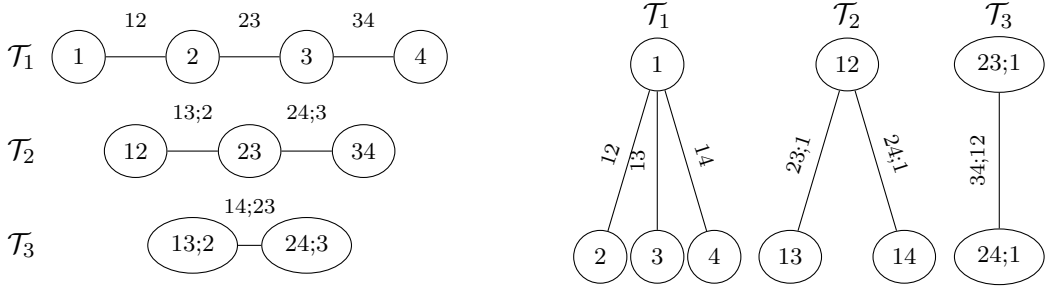

In the tree sequences, each node can be identified with a uniformly distributed random variable and each edge corresponds to a pair-copula that captures the associated dependence. To give insight into the dependencies that are captured by the examples in \autoref{fig:VineExamples} and into the labeling of the edges, we construct the vine copula density $\scd$ corresponding to the D-vine in \autoref{fig:VineExamples}. Recall that $\scd$ is a density function defined on $[0,1]^4$ and that the univariate marginals are uniform on $[0,1]$. Therefore,

\begin{align}
\scd\left(u_1,u_2,u_3,u_4\right) = \ &  \scd_{4|123}\left(u_4|u_1,u_2,u_3\right)\scd_{3|12}\left(u_3|u_1,u_2\right)\notag\\
& \times \scd_{2|1}\left(u_2|u_1\right)\label{Eq:v1}
\end{align}

with

\begin{eqnarray} \label{Eq:v2}
\scd_{2|1}\left(u_2|u_1\right)=\scd_{12}\left(u_1,u_2\right).
\end{eqnarray}

We now rewrite the conditional densities in \eqref{Eq:v1} such that at the end we have a pair-copula decomposition of $\scd$. The used conditioning strategy reflects the underlying D-vine structure. In the following derivation, $\scd_{ij;k}$ (for different $i,j,k \in \left\{1,2,3,4\right\}$) is the bivariate copula density associated with the bivariate conditional distribution of $(U_i,U_j)$ given $U_k=u_k$; and $\scd_{ij;kl}$ (for different $i,j,k,l \in \left\{1,2,3,4\right\}$) is the bivariate copula density associated with the bivariate conditional distribution of $(U_i,U_j)$ given $U_k=u_k$ and $U_l=u_l$. Further, we rely on the so-called simplifying assumption, which is commonly used for vine copulas. It states that the bivariate copulas $\scd_{ij,k}$ and $\scd_{ij;kl}$ associated with bivariate conditional distributions do not depend on the specific values of the conditioning variables, i.e.\ $\scd_{ij;k}\left(\cdot,\cdot;u_k\right) \equiv \scd_{ij;k}\left(\cdot,\cdot\right)$ and $\scd_{ij;kl}\left(\cdot,\cdot;u_k,u_l\right) \equiv \scd_{ij;kl}\left(\cdot,\cdot\right)$. See e.g.\ \cite{haff2010simplified}, \cite{stoeber2013simplified} and \cite{kurz2017testing} for discussions on the simplifying assumption.

To rewrite $\scd_{4|123}\left(u_4|u_1,u_2,u_3\right)$ in terms of pair-copulas, first note that 

\begin{align} 
\scd_{4|123}&\left(u_4|u_1,u_2,u_3\right) \nonumber\\
&= \displaystyle \frac{\scd_{14|23}\left(u_1,u_4|u_2,u_3\right) }{\scd_{1|23}\left(u_1|u_2,u_3\right) } \nonumber \\
&= \scd_{14;23}\lbrace\mathbb{C}_{1|23}(u_1|u_2,u_3), \mathbb{C}_{4|23}(u_4|u_2,u_3) \rbrace\nonumber\\
&\phantom{=\ } \times \scd_{4|23}(u_4|u_2,u_3),\label{Eq:v3}
\end{align}
\noindent
where the second equality follows from an application of Sklar's theorem \citep{Sklar59} to the bivariate conditional density $\scd_{14|23}$; $\scd_{14;23}$ is the corresponding copula density; $\mathbb{C}_{1|23}$ and $\mathbb{C}_{4|23}$ are the conditional marginals. In a similar way we obtain 

\begin{align} \label{Eq:v4}
\scd_{4|23}&(u_4|u_2,u_3)\nonumber\\  &=\frac{\scd_{24|3}(u_2,u_4|u_3)}{\scd_{2|3}(u_2|u_3)}\nonumber \\
&= \scd_{24;3}\lbrace\mathbb{C}_{2|3}(u_2|u_3), \mathbb{C}_{4|3}(u_4|u_3) \rbrace \scd_{34}(u_3,u_4)
\end{align}

and

\begin{align} \label{Eq:v5}
\scd_{3|12}&(u_3|u_1,u_2)\nonumber\\ &=\frac{\scd_{13|2}(u_1,u_3|u_2)}{\scd_{1|2}(u_1|u_2)} \nonumber \\
&= \scd_{13;2}\lbrace\mathbb{C}_{1|2}(u_1|u_2), \mathbb{C}_{3|2}(u_3|u_2) \rbrace \scd_{23}(u_2,u_3).
\end{align}

From (\ref{Eq:v1}) - (\ref{Eq:v5}) it easily follows that

\begin{align}
\scd\left(u_1, \right. & \left.\hspace{-.1cm} u_2, u_3, u_4\right) \notag\\
= \  & \scd_{12}\left(u_1, u_2\right)\scd_{23}\left(u_2, u_3\right)\scd_{34}\left(u_3, u_4\right)\notag\\
& \times \scd_{13;2}\lbrace\mathbb{C}_{1|2}\left(u_1|u_2\right), \mathbb{C}_{3|2}\left(u_3|u_2\right)\rbrace\hspace{-0.05cm}\notag\\
& \times \scd_{24;3}\lbrace\mathbb{C}_{2|3}\left(u_2|u_3\right), \mathbb{C}_{4|3}\left(u_4|u_3\right)\rbrace\notag\\
& \times \scd_{14;23}\lbrace\mathbb{C}_{1|23}\left(u_1|u_2, u_3\right), \mathbb{C}_{4|23}\left(u_4|u_2, u_3\right)\rbrace.\label{Eq:D-vineDensity4}
\end{align}

Note that only bivariate copula densities are included. Since the latter can be chosen independently from a large catalog of bivariate copula families, vine copulas proof themselves as a flexible tool to model complex association patterns. Also, it is important to note that the simplifying assumption affects the vine copula density only at the level of the bivariate pair-copulas in $\mathcal{T}_2$ and $\mathcal{T}_3$. The arguments of the pair-copulas in $\mathcal{T}_2$ and $\mathcal{T}_3$, nevertheless, do depend on the conditioning variables. They are conditional distribution functions of the form $\mathbb{C}_{i|k}$, $\mathbb{C}_{k|i}$, $\mathbb{C}_{i|jk}$ and $\mathbb{C}_{j|ik}$. Recall that these expressions are conditional marginals of bivariate conditional distributions that result from the conditioning strategy used. Thus, they are completely determined by the underlying R-vine structure. They can be computed by recursive partial differentiation of the pair-copulas of higher tree levels, i.e.\ it holds that

\begin{align*}
\textcolor{black}{\mathbb{C}}_{i|k}&\left(u_i|u_k\right)\\
& = \frac{\partial}{\partial u_k}\textcolor{black}{\mathbb{C}}_{ik}\left(u_i,u_k\right),\\
\textcolor{black}{\mathbb{C}}_{k|i}&\left(u_k|u_i\right)\\
& = \frac{\partial}{\partial u_i}\textcolor{black}{\mathbb{C}}_{ik}\left(u_i,u_k\right),\\
\textcolor{black}{\mathbb{C}}_{i|jk}&\left(u_i|u_j,u_k\right)\\
& = \frac{\partial}{\partial \textcolor{black}{\mathbb{C}}_{j|k}\left(u_j|u_k\right)}\textcolor{black}{\mathbb{C}}_{ij;k}\lbrace \textcolor{black}{\mathbb{C}}_{i|k}\left( u_i|u_k\right),\textcolor{black}{\mathbb{C}}_{j|k}\left(u_j|u_k\right)\rbrace
\end{align*}

and

\begin{align*}
\textcolor{black}{\mathbb{C}}_{j|ik}&\left(u_j|u_i,u_k\right)\\
& = \frac{\partial}{\partial \textcolor{black}{\mathbb{C}}_{i|k}\left(u_i|u_k\right)}\textcolor{black}{\mathbb{C}}_{ij;k}\lbrace \textcolor{black}{\mathbb{C}}_{i|k}\left(u_i|u_k\right),\textcolor{black}{\mathbb{C}}_{j|k}\left(u_j|u_k\right)\rbrace.
\end{align*}

Usually, these expressions are further identified with so-called h-functions that allow for a recursive notation of the associated conditional distributions \citep{Joe1997}. Recalling the simplifying assumption, we define

\begin{align*}
h_{i|k}\left(u_i| u_k\right) & \coloneqq \frac{\partial}{\partial u_k} \textcolor{black}{\mathbb{C}}_{ik}\left(u_i, u_k\right),\\
h_{k|i}\left(u_k| u_i\right) & \coloneqq  \frac{\partial}{\partial u_i} \textcolor{black}{\mathbb{C}}_{ik}\left(u_i, u_k\right),\\
h_{i|j;k}\left(u_i| u_j\right) & \coloneqq \frac{\partial}{\partial u_j} \textcolor{black}{\mathbb{C}}_{ij;k}\left(u_i, u_j\right)
\end{align*}

and

\begin{align*}
h_{j|i;k}\left(u_j| u_i\right) & \coloneqq \frac{\partial}{\partial u_i} \textcolor{black}{\mathbb{C}}_{ij;k}\left(u_i, u_j\right).
\end{align*} 

For instance, for the considered D-vine structure we have

\begin{align*}
\textcolor{black}{\mathbb{C}}_{1|23}\left(u_1|u_2,u_3\right) & = h_{1|3;2}\lbrace \textcolor{black}{\mathbb{C}}_{1|2}\left(u_1, u_2\right)\big \vert \textcolor{black}{\mathbb{C}}_{3|2}\left(u_2, u_3\right)\rbrace\\
& = h_{1|3;2}\lbrace h_{1|2}\left(u_1\vert u_2\right)\big \vert h_{3|2}\left(u_3\vert u_2\right)\rbrace.
\end{align*}

The vine copula density \eqref{Eq:D-vineDensity4} can finally be written as

\begin{align}
\textcolor{black}{\scd}\left(\right. & \left. \hspace{-0.1cm}u_1,u_2,u_3,u_4\right)\notag\\
= & \ \textcolor{black}{\scd}_{12}\left(u_{1}, u_{2}\right)\textcolor{black}{\scd}_{23}\left(u_{2}, u_{3}\right)\textcolor{black}{\scd}_{34}\left(u_{3}, u_{4}\right)\label{Eq:DensityHFunc}\\
& \times \textcolor{black}{\scd}_{13;2}\lbrace h_{1|2}\left(u_{1}|u_{2}\right),h_{3|2}\left(u_{3}|u_{2}\right)\rbrace \notag\\
& \times \textcolor{black}{\scd}_{24;3}\lbrace h_{2|3}\left(u_{2}|u_{3}\right),h_{4|3}\left(u_{4}|u_{3}\right)\rbrace \notag\\
& \times \textcolor{black}{\scd}_{14;23}\left[ h_{1|3;2}\lbrace h_{1|2}\left(u_{1}|u_{2}\right)\big\vert h_{3|2}\left(u_{3}|u_{2}\right)\rbrace,\right.\notag\\ & \phantom{\times \textcolor{black}{\scd}_{14;23}\ \ } \left. h_{4|2;3}\lbrace h_{4|3}\left(u_{4}|u_{3}\right)\big\vert h_{2|3}\left(u_{2}|u_{3}\right)\rbrace\right] \notag.
\end{align} 	

In this section, we considered a four-dimensional example to explain how a $d$-dimensional copula density can be decomposed into a product of bivariate copula densities. We further introduced h-functions to provide an expression of the density exclusively in terms of vine copula components. The presented concepts can be directly extended to $d$ dimensions. For complete data, likelihood based inference of regular (R) vine copulas is well elaborated \citep{dissmann2013selecting, brechmann2012truncated}. The algorithms to estimate the R-vine structure, the associated bivariate copula families and the copula parameters are implemented in the \texttt{R}-library \texttt{VineCopula} \citep{schepsmeier2014vinecopula}. In this paper, we assume the vine structure and the bivariate copula families of a $d$-dimensional ($d=3$ and $d=4$) R-vine model to be specified except for the parameters of the bivariate copulas. We investigate likelihood based parameter estimation given $d$-dimensional event time data. Such data are typically subject to right-censoring, which complicates the analysis in a non-trivial way.  

\section{Multivariate right-censored event time data: \newline the mastitis data}\label{Sec:Censoring}
We consider vine copula based likelihood inference for $d$-dimensional event time data subject to right-censoring. We focus on $d=3$ and $d=4$. For a sample of $n$ clusters, let $T_{ij}$ be the $j$th event time and $C_{ij}$ be the $j$th censoring time in cluster $i$ ($i=1,\ldots,n$ and $j=1,\ldots,d$). We observe $Y_{ij}=\min(T_{ij},C_{ij})$ together with the censoring indicator $\delta_{ij}=I(T_{ij}\leq C_{ij})$. Throughout, we assume that $T_{ij}$ and $C_{ij}$ are independent ($i=1,\ldots,n$ and $j=1,\ldots,d$) and that censoring is noninformative. The setting where $C_{ij}=C_i$ for all $j=1,\ldots,d$ is called common (univariate) right-censoring.
Before we discuss, in \autoref{Sec:MLE}, likelihood based estimation of the parameters of a vine copula model, we give some details on the data example mentioned in the introduction.

\renewcommand{\arraystretch}{1}
\begin{table}
	\caption{Censoring patterns of the mastitis data.}
	\label{Table:CensoringCows}	
	\centering	
	\begin{tabular}{cc}
		\hline\midrule
		\#censored observations in a cluster & \#cows\\
		\midrule
		0 & 73\\
		1 & 49\\
		2 & 36\\
		3 & 40\\
		4 & 209\\
		\midrule\midrule
		udder quarter & percentage of censoring\\
		\midrule
		front left & $64.37\%$\\
		front right & $64.37\%$\\
		rear left & $68.80\%$\\
		rear right & $67.08\%$\\
		\midrule\hline
	\end{tabular}		
\end{table}		
\renewcommand{\arraystretch}{1}

The udder infection data of \citet{Laevens97} already received considerable attention in a number of papers, e.g.\ \citet{duchateau2008frailty}, \citet{Massonnet2009} and \citet{Geerdens2014}. The study aims to quantify the impact of mastitis on the milk production and the milk quality. For this, information on the time from parturition to infection is collected for the four udder quarters of a cow. The cow is the cluster and the infection times of the four udder quarters are the clustered data.

For the $407$ primiparous cows in the study, the available data consist of the cow identification number, the minimum of the infection time and the censoring time (both in days) for each udder quarter as well as the corresponding censoring indicators, e.g.\ for the first cow the data information is given by $\{1,(67,67,119,67),(1,1,1,1)\}$, resp.\ for the last cow $\{407,(279,279,279,263),(0,0,0,1)\}$, where the ordering in a data quadruple corresponds to left front, right front, left rear and right rear. Censoring occurs at the level of the udder quarters and it is common (univariate) in the sense that the same censoring time applies to all udder quarters of an individual cow. \autoref{Table:CensoringCows} summarizes information on the censoring patterns of the mastitis data. In total, censoring is present in about $66.15\%$ of the observations. The information loss for all four udder quarters due to the high censoring rate is illustrated in \autoref{Sec:AppIllustrMastitis}.

Flexible association modeling of the mastitis data via Joe-Hu copulas has been studied by \citet{Geerdens2014}. In \autoref{Sec:MastitisApplication}, we use vine copulas to study the dependence pattern in the mastitis data; our findings confirm the need for flexible modeling.

\section{Vine based likelihood inference for four-dimensional event time data}\label{Sec:MLE}

An appropriate likelihood expression is needed to perform parametric likelihood inference for $d$-dimensional right-censored time-to-event data. We focus on quadruple data ($d=4$). We refer to \citet[\textcolor{black}{Chapter 3}]{Thesis_Barthel} for a more detailed discussion.

Based on the observed data $\boldsymbol{Y}_i$ = ($Y_{i1},\ldots,Y_{i4}$) and $\boldsymbol{\delta}_i$ = ($\delta_{i1},\ldots,\delta_{i4}$) with $Y_{ij}$ and $\delta_{ij}$ as given in \autoref{Sec:Censoring} ($i=1,\ldots,n$ and  $j=1,\ldots,4$), we define for each cluster the following joint censoring indicators: \\

\vspace{0.25cm}
\renewcommand{\arraystretch}{.85}
\begin{tabular}{l}
	no censoring:\\
	\quad $\begin{aligned}[t] \Delta_i\left(1,2,3,4\right) \coloneqq \delta_{i1}\delta_{i2}\delta_{i3}\delta_{i4}
	\end{aligned}$\\
	\\
	all components censored:\\
	\quad $\begin{aligned}[t] \Delta_i \coloneqq \prod_{j=1}^{4}\left(1-\delta_{ij}\right)
	\end{aligned}$\\
	\\
	$p$th component not censored:\\
	\quad $\begin{aligned}[t]  \Delta_i\left(p\right) \coloneqq \delta_{ip}\prod_{j=1;j\neq p }^{4}\left(1-\delta_{ij}\right)
	\end{aligned}$\\ 
	$p$th, $q$th component not censored $\text{for } p \neq q$:\\
	\quad $\begin{aligned}[t] \Delta_i\left(p,q\right) \coloneqq \delta_{ip}\delta_{iq}\prod_{j=1;j\neq p, q}^{4}\left(1-\delta_{ij}\right)
	\end{aligned}$\\ 
	\\
	$p$th, $q$th, $v$th component not censored $\text{for } w \neq p, q, v$\\
	$\text{ and } p \neq q \neq v$:\\
	\quad $\begin{aligned}[t] \Delta_i\left(p,q,v\right) \coloneqq \delta_{ip}\delta_{iq}\delta_{iv}\left(1-\delta_{iw}\right)
	\end{aligned}$\\	
\end{tabular} \\
\renewcommand{\arraystretch}{1}
\bigskip

The actual censoring in a particular cluster determines the joint censoring indicators for that cluster, e.g.\ for $\boldsymbol{\delta}_i=(1,0,0,1)$ we have $\Delta_i\left(1,4\right)=1$ and all other joint censoring indicators equal zero. Using the notation $u_{ij}=S_j(y_{ij})$ ($i=1,\ldots,n$ and $j=1,\ldots,4$) for the copula data linked to the observed data, we have for $\Delta_i\left(1,4\right)=1$ that $u_{i1}$ and $u_{i4}$ correspond to true event times. On the other hand, $u_{i2}$ and $u_{i3}$ correspond to censoring times, meaning that the copula data linked to the unknown true event times would take values smaller than $u_{i2}$ and $u_{i3}$. Therefore, the contribution to the loglikelihood is given by	

\begin{align*}
&\ell_{i,d}\left(\boldsymbol{\theta};\boldsymbol{u}_i,\boldsymbol{\delta}_i\right)\\
& = \log\left[\frac{\partial^2}{\partial u_{i1} \partial u_{i4} }\mathbb{C}\lbrace u_{i1},S_2(y_{i2}),S_3(y_{i3}),u_{i4};\boldsymbol{\theta}\rbrace \right] \bigg\vert_{\myrel{{u_{i1}=S_1\left(y_{i1}\right)}}{{u_{i4}=S_4\left(y_{i4}\right)}}},
\end{align*}
\noindent
where $\boldsymbol{u}_i=(u_{i1},u_{i2},u_{i3},u_{i4})$ and with $\boldsymbol{\theta}$ the vector collecting all parameters of the copula $\mathbb{C}$.

In general, the contribution of the $i$-th cluster to the loglikelihood is given by

\begin{align*}
& \ell_{i,d}\left(\boldsymbol{\theta};\boldsymbol{u}_i,\boldsymbol{\delta}_i\right)\\
& \coloneqq \Delta_i \log\{\mathbb{C}\left(u_{i1},u_{i2},u_{i3},u_{i4};\boldsymbol{\theta}\right)\} \phantom{\frac{\partial}{\partial u_{ip}}}\\
& \ + \sum_{p=1}^{4}\Delta_i\left(p\right)\log \{\frac{\partial}{\partial u_{ip}}\mathbb{C}\left(u_{i1},u_{i2},u_{i3},u_{i4};\boldsymbol{\theta}\right)\}\\
& \ + \sum_{p\neq q}\Delta_i\left(p,q\right)\log\{\frac{\partial^2}{\partial u_{ip} \partial u_{iq} }\mathbb{C}\left(u_{i1},u_{i2},u_{i3},u_{i4};\boldsymbol{\theta}\right)\}\\
& \ + \sum_{p\neq q \neq v}\Delta_i\left(p,q,v\right)\log\{\frac{\partial^3}{\partial u_{ip} \partial u_{iq} \partial u_{iv}}\mathbb{C}\left(u_{i1},u_{i2},u_{i3},u_{i4};\boldsymbol{\theta}\right)\}\\
& \ + \Delta_i\left(1,2,3,4\right) \log\{\scd\left(u_{i1},u_{i2},u_{i3}\,u_{i4};\boldsymbol{\theta}\right)\}.
\end{align*}

The loglikelihood for four-dimensional time-to-event data subject to right-censoring is therefore given by

\begin{align}
\log L_{n,d}\left(\boldsymbol{\theta};\boldsymbol{u}_1,\dots,\boldsymbol{u}_n,\boldsymbol{\delta}_1,\dots,\boldsymbol{\delta}_n\right) & \coloneqq \sum_{i=1}^{n} \ell_{i,d}\left(\boldsymbol{\theta};\boldsymbol{u}_i,\boldsymbol{\delta}_i\right). 
\label{Eq:logLik_uLevel}
\end{align}

\citet{Massonnet2009} and \citet{Geerdens2014} use this likelihood expression to model dependencies within the mastitis data. \citet{Shih95} and \citet{andersen2005two} consider similar versions for bivariate event time data.

Once we have decided on the vine structure to be used, we need the vine version of the partial derivatives in \eqref{Eq:logLik_uLevel}. For instance, for the D-vine given in \autoref{fig:VineExamples} we have

\begin{align*}
\frac{\partial^2 \mathbb{C}\left(u_{i1},u_{i2},u_{i3},u_{i4}\right)}{\partial u_{i1} \partial u_{i4}}\hspace{-3cm}\\
= & \  \int_{0}^{u_{i2}} \int_{0}^{u_{i3}} \scd_{12}\left(u_{i1}, v_{i2}\right)\scd_{23}\left(v_{i2}, v_{i3}\right)\scd_{34}\left(v_{i3}, u_{i4}\right)\\
& \times \scd_{13;2}\{\mathbb{C}_{1|2}\left(u_{i1}|v_{i2}\right), \mathbb{C}_{3|2}\left(v_{i3}|v_{i2}\right)\}\\
& \times \scd_{24;3}\{\mathbb{C}_{2|3}\left(v_{i2}|v_{i3}\right), \mathbb{C}_{4|3}\left(u_{i4}|v_{i3}\right)\}\\
& \times \scd_{14;23}\{\mathbb{C}_{1|23}\left(u_{i1}|v_{i2}, v_{i3}\right),\mathbb{C}_{4|23}\left(u_{i4}|v_{i2}, v_{i3}\right)\}dv_{i3} dv_{i2}\\
= & \int_{0}^{u_{i2}} \int_{0}^{u_{i3}} \scd_{12}\left(u_{i1}, v_{i2}\right)\scd_{23}\left(v_{i2}, v_{i3}\right)\scd_{34}\left(v_{i3}, u_{i4}\right)\\
& \times \scd_{13;2}\{h_{1|2}\left(u_{i1}|v_{i2}\right),h_{3|2}\left(v_{i3}|v_{i2}\right)\}\\
& \times \scd_{24;3}\lbrace h_{2|3}\left(v_{i2}|v_{i3}\right),h_{4|3}\left(u_{i4}|v_{i3}\right)\}\\
& \times \scd_{14;23}\left[h_{1|3;2}\{h_{1|2}\left(u_{i1}|v_{i2}\right)\big\vert h_{3|2}\left(v_{i3}|v_{i2}\right)\},\right.\\
& \phantom{\times \scd_{14;23} \ \ } \left. h_{4|2;3}\{h_{4|3}\left(u_{i4}|v_{i3}\right)\big\vert h_{2|3}\left(v_{i2}|v_{i3}\right)\}\right] dv_{i3} dv_{i2}.
\end{align*}

The complete collection of vine equivalents of the partial derivatives is derived in \citet[\textcolor{black}{Chapter 3}]{Thesis_Barthel} and is given in \autoref{Sec:PartDeriv} (\autoref{theo:D-vineDerivatives} and \autoref{Corollary:D-vineDerivatives}).

\newpage
\subsection*{Practical implementation}
We end this section by two remarks concerning the practical implementation of the presented optimization problem.

First, note that in practice, the marginal survival functions, which are assumed to be known in the above discussion, are typically unknown. We therefore use the two-stage estimation procedure described in \citet{Shih95}. A parametric approach can be applied. In stage one, we assume $S_j(\cdot)$ to be known up to some parameter vector $\boldsymbol{\alpha}_j$, i.e.\ $S_j(\cdot)=S_j(\cdot,\boldsymbol{\alpha}_j)$ ($j=1,\ldots,4$). We obtain the maximum likelihood estimate (MLE) $\hat{\boldsymbol{\alpha}}_j$ of $\boldsymbol{\alpha}_j$ and calculate $\hat{u}_{ij}=S_j(y_{ij},\hat{\boldsymbol{\alpha}}_j)$ ($i=1,\ldots,n$ and $j=1,\ldots,4$). In stage two, we replace $u_{ij}$ by the pseudo-observation $\hat{u}_{ij}$ ($i=1,\ldots,n$ and $j=1,\ldots,4$) and maximize the loglikelihood with respect to $\boldsymbol{\theta}$. Alternatively, a semiparametric approach can be applied. In stage one, we estimate the marginals nonparametrically. We obtain the Kaplan-Meier estimate (KME) $\hat{S}_j(\cdot)$ of $S_j(\cdot)$ ($j=1,\dots,4$) and calculate the pseudo-observations $\hat{u}_{ij}=\hat{S}_j(y_{ij})$. In stage two, we use the latter as substitutes for $u_{ij}$ and maximize the loglikelihood with respect to $\boldsymbol{\theta}$.

Second, due to right-censoring the use of single and double integrals and hence numerical integration cannot be avoided when evaluating the loglikelihood. Thus, appropriate starting values are indispensable for a reasonable trade-off between numerical demand and accuracy of the estimates. Due to the rapidly increasing number of parameters for vine copulas in higher dimensions this issue also arises for complete data. Herein, the so-called sequential estimation approach of  \citet{dissmann2013selecting} is usually applied. It splits up a $d$-dimensional estimation problem into $d(d-1)/2$ bivariate ones. First, the parameters of the $d-1$ bivariate copulas in $\mathcal{T}_1$ are estimated. Next, the parameter estimates are used to obtain estimates of the h-functions. These estimates are needed as arguments in the pair-copulas in $\mathcal{T}_2$ when estimating the $d - 2$ copula parameters in $\mathcal{T}_2$, etc.
\citet{haff2013parameter} provide asymptotic properties for this approach. Since it makes the estimation of high-dimensional vine copula models tractable and computationally easy while showing excellent estimation performance, analysis for complete data often exclusively rely on the sequential estimation approach.

In the setting with right-censored quadruple data, we can mimic this idea and estimate the parameters of the three bivariate copulas in $\mathcal{T}_1$ separately by using the bivariate version of the loglikelihood given in \eqref{Eq:logLik_uLevel}. However, by construction the arguments in $\mathcal{T}_2$ and $\mathcal{T}_3$ are not associated with observed (event or censored) times. As a consequence, estimation via the two-dimensional version of \eqref{Eq:logLik_uLevel} is no longer feasible. Instead, after having obtained the parameter estimates for $\mathcal{T}_1$, we substitute them in the loglikelihood \eqref{Eq:logLik_uLevel}, which we then maximize with respect to the remaining copula parameters in $\mathcal{T}_2$ and $\mathcal{T}_3$. By doing so, we achieve dimension reduction by at least $3$ for $d=4$. We refer to this approach as $\mathcal{T}_1$-sequential estimation. Finally, we use the estimates of the $\mathcal{T}_1$-sequential approach as starting values to solve the computationally heavy optimization problem with respect to all 6 parameters ($d=4$) of the vine copula model simultaneously (step 2 in the two-stage estimation procedure of \citet{Shih95}). Bootstrap standard errors of the estimates for both the global estimation approach and the $\mathcal{T}_1$-sequential estimation approach are obtained by using for a fitted model the resampling scheme given in \autoref{Sec:VineCopBootAlg}. 

For our calculations, we rely on standard optimization methods and the \texttt{VineCopula} package in \texttt{R} \citep{schepsmeier2014vinecopula}, in which the evaluation of h-functions, of the cumulative distribution function and of the density function is implemented for many parametric bivariate copulas.

\section{Simulation study}\label{Sec:SimulationStudy}

We investigate the finite sample performance of the loglikelihood approach presented in \autoref{Sec:MLE} through a simulation study. To cover a broad range of simulation settings while keeping the numerical effort for a large number of replications reasonable, we restrict ourselves to three dimensions. The goal is to assess the impact of right-censoring on vine copula based estimation of the within-cluster association. For this purpose, various degrees of right-censoring, different types of tail-dependence and different strengths of dependence are considered. A more elaborate study can be found in \citet[\textcolor{black}{Chapter 4}]{Thesis_Barthel}.

\subsection{Considered scenarios}
To generate multivariate right-censored time-to-event data with a dependence structure specified by a vine copula, we simulate in a first step complete copula data using the \texttt{R}-package \texttt{VineCopula} \citep{schepsmeier2014vinecopula}. We assume the copula $\mathbb{C}$ to be a vine copula with density

\begin{equation*}
\begin{split}
\scd&\left(u_1, u_2, u_3\right)\\
& =\scd_{12}\left(u_1, u_2\right)\scd_{23}\left(u_2, u_3\right)\scd_{13;2}\{\mathbb{C}_{1|2}\left(u_1|u_2\right), \mathbb{C}_{3|2}\left(u_3|u_2\right)\}.
\end{split}
\label{eq:3D-vineDensity}
\end{equation*}

Recall that in dimension three, all vine structures are equivalent up to the labeling of the nodes. Here, the copulas $\mathbb{C}_{12}$ and $\mathbb{C}_{23}$ are assumed to arise from the same copula family. We investigate both the scenario of lower tail-dependent copulas using the Clayton family and the scenario of upper tail-dependent copulas using the Gumbel family. For ease of comparison, we take Kendall's $\tau$ to be the same in both tail-dependence scenarios; we set $\tau_{12} = 0.6$ and $\tau_{23} = 0.6$ assuming strong dependencies. We assume $\mathbb{C}_{13;2}$ to be a Frank copula, which has no tail dependence, with moderate dependence $\tau_{13;2}=0.3$. Two extra simulation settings considering $\tau_{12} = \tau_{23} = \tau_{13;2} = 0.1$ (weak dependencies) and $\tau_{12} = \tau_{23} = \tau_{13;2} = 0.3$ (moderate dependencies) are included in \autoref{Sec:ExtSimStudy}. The three copula families are common choices covering the three standard tail-dependence scenarios for bivariate data. Recall that in a vine copula model, these families can be arbitrarily combined allowing for complex dependence structures such as asymmetric tail-dependence behavior.

In a second step, the inverse probability integral transform is applied to the marginal copula data to obtain the true event times. Note that the proposed modeling strategy handles marginal and dependence modeling separately with no restrictions with regard to the marginal estimation. Thus, the settings for the marginal survival functions mainly serve the purpose to define the transformation from copula data to data on the actual time scale without distorting the dependence structure, which is our focus. Given that the Weibull is a commonly used parametric survival function, we assume this form for the margins of the event times as well as for the censoring mechanism, i.e.\ $S\left(t\right) = \exp\left(-\left(\frac{t}{\lambda} \right)^{\alpha} \right)$ with shape parameter $\alpha$ and scale parameter $\lambda$ (in accordance with the parametrization used in \texttt{R}). The parameter choices are given in \autoref{table:1SimStudySetup} and are inspired by the marginal estimates of the trivariate tumorigenesis data in \citet{Mantel77}. The latter motivated the extensive simulation study in  \citet[\textcolor{black}{Section 4.1.1}]{Thesis_Barthel}, on which we build our investigations. To assess the effect of censoring, we investigate the performance of the estimation procedure for complete data as well as for a moderate overall censoring rate of 25\% and for a heavy censoring rate of 65\%. Note that the margins are affected to a different extent by the censoring mechanism as caused by distinct survival functions.

Finally, the observed data are obtained by taking the minima of the true event times and the associated censoring times. To this data we apply a two-stage approach for known margins as well as for parametrically (MLE) and nonparametrically (KME) estimated margins as described in \autoref{Sec:MLE}. In case of complete event time data, we use the empirical distribution functions (ECDF) as nonparametric estimates for the marginals. All scenarios are investigated for samples of size 200 and 500. Each sample is replicated 200 times.

\begin{table}
	\captionof{table}{Specification of the Weibull parameters of the survival function for each of the event times $T_1, T_2, T_3$ and of the two common censoring distributions leading to 25\%, resp. 65\% overall common right-censoring. Further, the individual censoring rates for each of the three margins are shown.}
	\label{table:1SimStudySetup}
	\footnotesize
	\centering
	\begin{tabular}{l c c c c c c} \midrule
		& & \multicolumn{3}{c}{Event times} & \multicolumn{2}{c}{\hspace{-.3cm}Censoring times}\\
		& & $T_1$ & $T_2$ & $T_3$ & 25\% & 65\% \\
		\hline
		\midrule
		\multirow{2}{*}{Weibull parameters} & $\alpha$ & 3.39 & 4.20 & 3.53 & 6.72 & 6.72 \\ 		
		& $\lambda$ & 3.32 & 2.21 & 2.68 & 3.11 & 2.17\\
		\midrule
		\multirow{2}{*}{Marginal censoring} & & 52\% & 12\% & 29\% & x & \\
		& & 82\% & 49\% & 67\% &  & x \\
		\hline\hline	
	\end{tabular}
\end{table}
\normalsize

\subsection{Results}
We visualize the results of the simulations in \autoref{fig:200+500_65} and \autoref{fig:500_0+25+65}, where the true Kendall's $\tau$ values are indicated by a horizontal line. \autoref{fig:200+500_65} shows satisfactory performance of the estimators when common right-censoring is present, even in case of heavy censoring (65\%). The two-stage approaches with (non)parametrically estimated margins benefit the most from an increasing sample size. In particular, due to the comparable performance of the parametric and the semiparametric estimation approach, the latter qualifies as an appropriate tool when working with real data. It allows a flexible estimation of the marginals and excludes the risk to misspecify the underlying parametric models. \autoref{fig:500_0+25+65} shows the censoring effect. Comparing the upper and lower parts illustrates the impact of the marginal censoring rates. Given \autoref{table:1SimStudySetup} we indeed expect that $\tau_{23}$ can be estimated in a more accurate way than $\tau_{12}$. Also, the method is more sensitive to a higher common right-censoring rate, especially when estimating the parameters of a lower tail-dependent copula, as can be seen by comparing the left-hand side and right-hand side of \autoref{fig:500_0+25+65}. This is due to the lack of information in the data for small copula values, i.e.\ high event times (see also \autoref{fig:ScatterplotCows12}). Overall, we can conclude that the presented method is on target for all investigated parameters in the underlying R-vine models.

A detailed summary of the simulation results can be found in \autoref{Table:CCF_65_200+500} and \autoref{Table:CCF_0+25_500} (Clayton for $\mathcal{T}_1$ and Frank for $\mathcal{T}_2$) and \autoref{Table:GGF_65_200+500} and \autoref{Table:GGF_0+25_500} (Gumbel for $\mathcal{T}_1$ and Frank for $\mathcal{T}_2$). Here, $\theta$ is the true parameter value, $\bar{\theta}$ is the  mean estimate, $\hat{b}(\bar{\theta})$ is the estimated bias, $s^2(\bar{\theta})$ is the estimated squared standard error and mse($\bar{\theta}$) is the estimated mean squared error of $\bar{\theta}$. The same performance measures are given for the corresponding Kendall's $\tau$ values. \autoref{Table:CCF_65_200+500_3} to \autoref{Table:GGF_0+25_500_1} in \autoref{Sec:ExtSimStudy} show similar results for the two extra simulation settings considering weak and moderate dependencies for all three bivariate copulas.

\begin{landscape}
	\centering
	\begin{figure}[ht]
		\caption{Boxplots of the estimated Kendall's $\tau$ values for 65\% common right-censored event time data with Clayton copulas (left) and Gumbel copulas (right) in $\mathcal{T}_1$, true $\tau_{12}=0.6$, $\tau_{23}=0.6$,  $\tau_{13;2}=0.3$ and sample sizes 200 and 500. Known margins, parametrically estimated (MLE) and nonparametrically (KME) estimated margins are considered.}
		\label{fig:200+500_65}
		\begin{minipage}{0.5\linewidth}
			\centering
			Clayton copulas in $\mathcal{T}_1$, Frank copula in $\mathcal{T}_2$
			\includegraphics[width=0.88\linewidth]{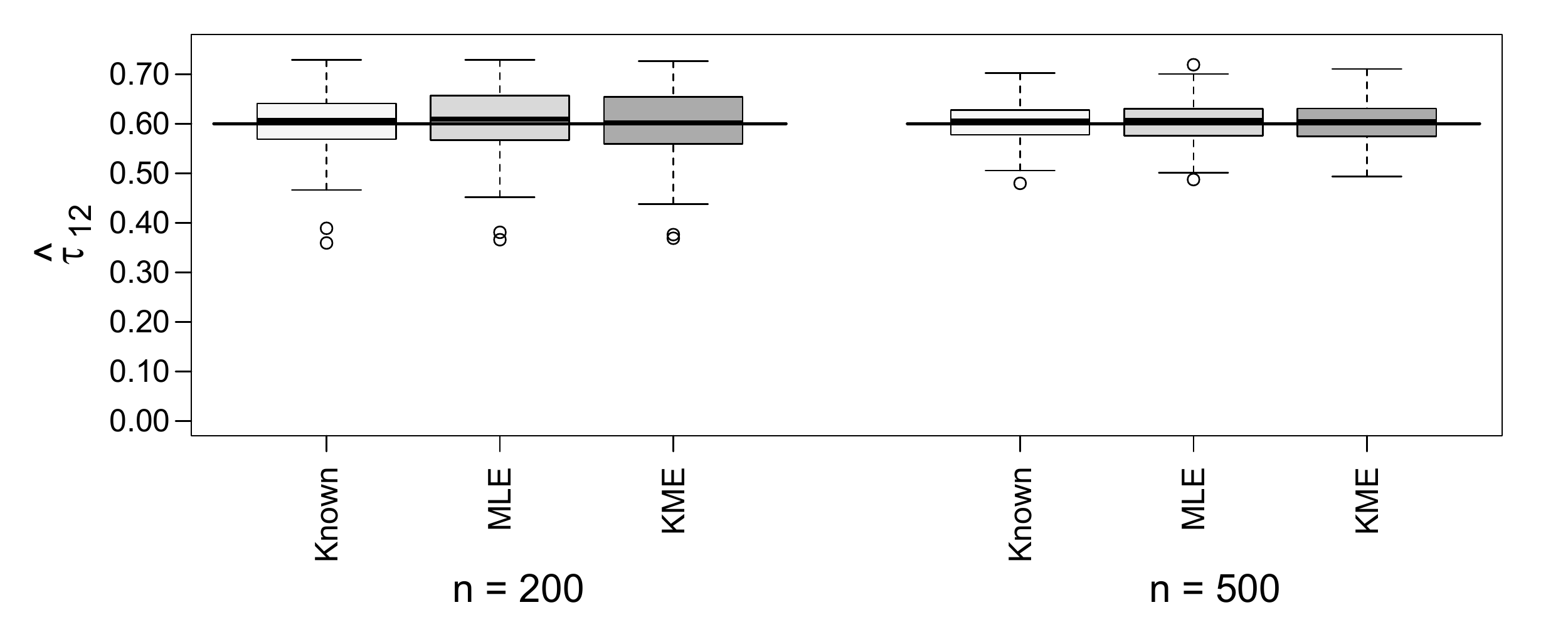}	
			\includegraphics[width=0.88\linewidth]{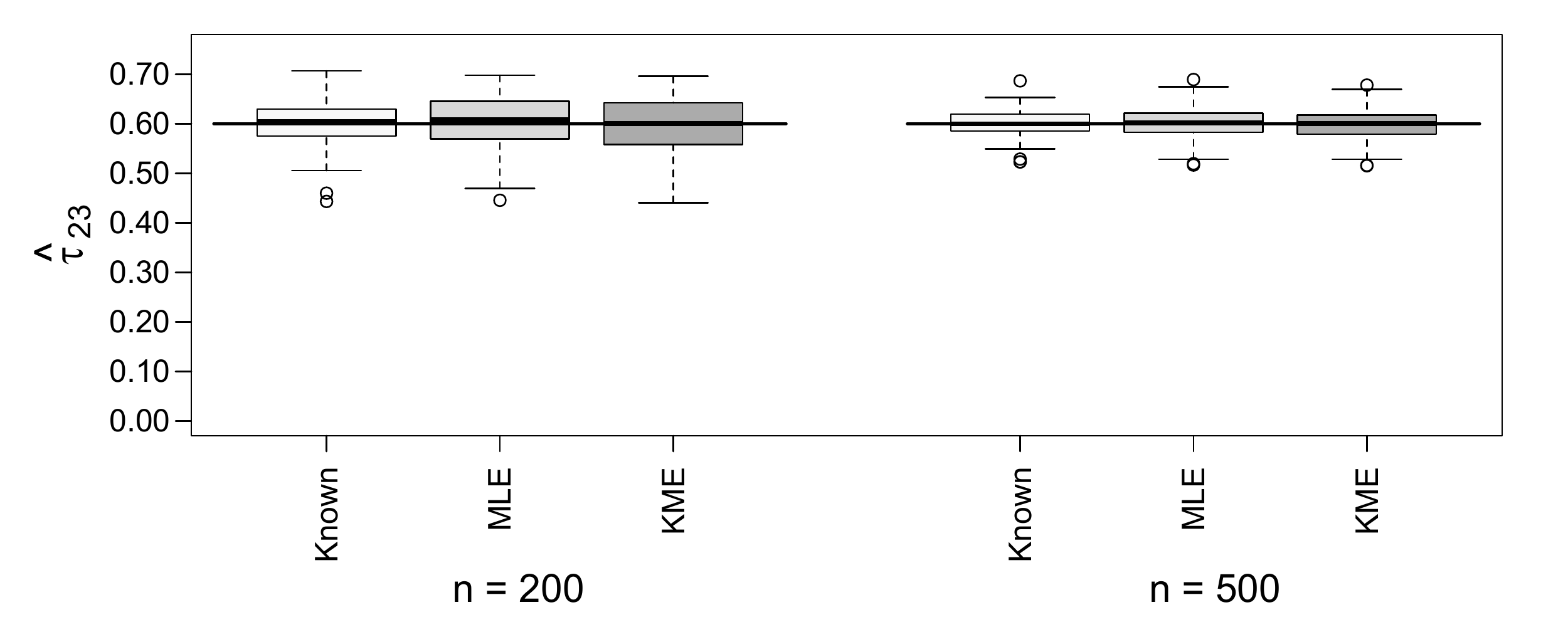}
			\includegraphics[width=0.88\linewidth]{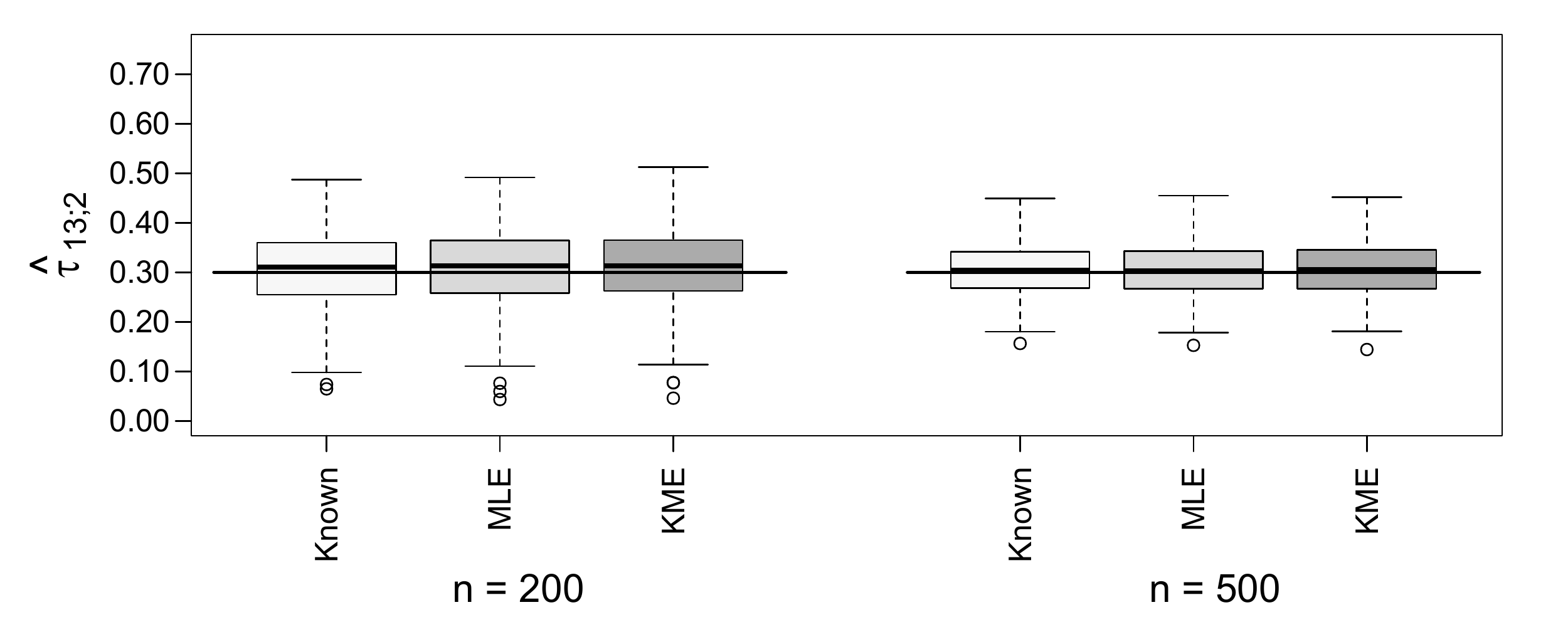}
		\end{minipage}
		\begin{minipage}{0.5\linewidth}
			\centering
			Gumbel copulas in $\mathcal{T}_1$, Frank copula in $\mathcal{T}_2$
			\includegraphics[width=0.88\linewidth]{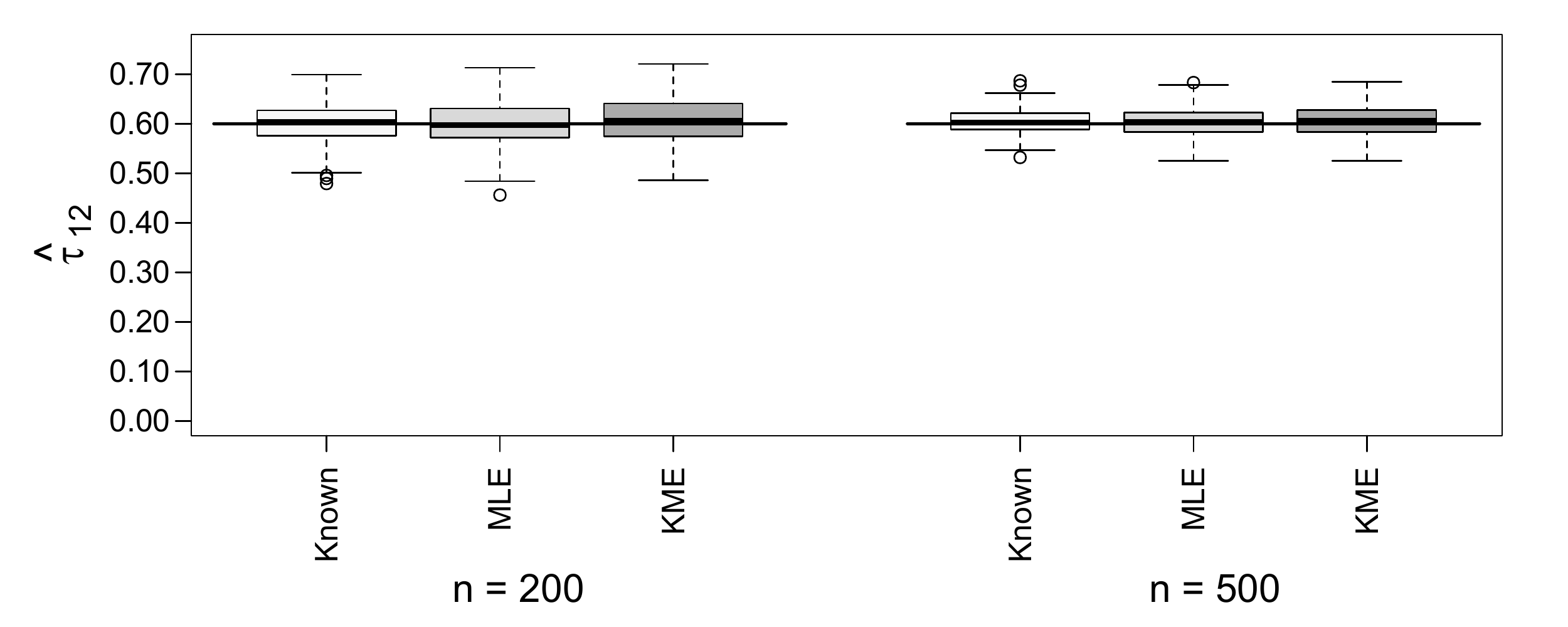}	
			\includegraphics[width=0.88\linewidth]{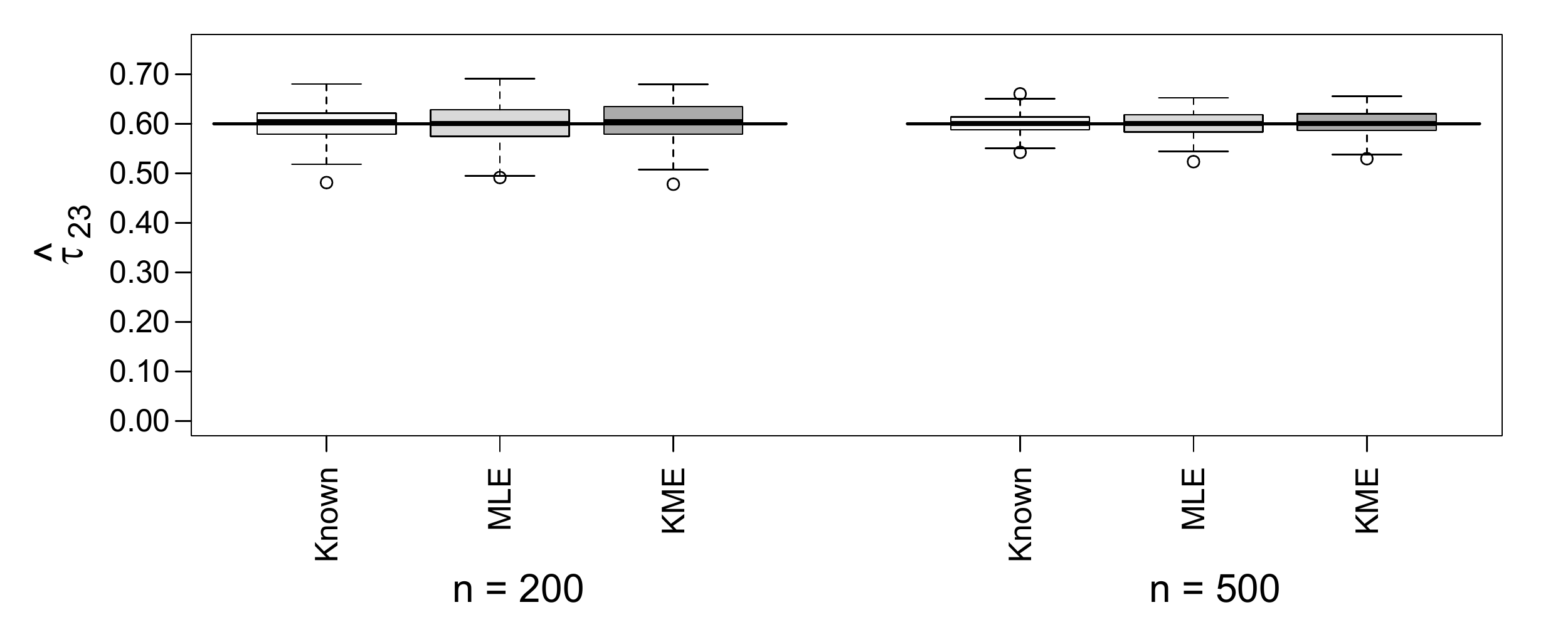}
			\includegraphics[width=0.88\linewidth]{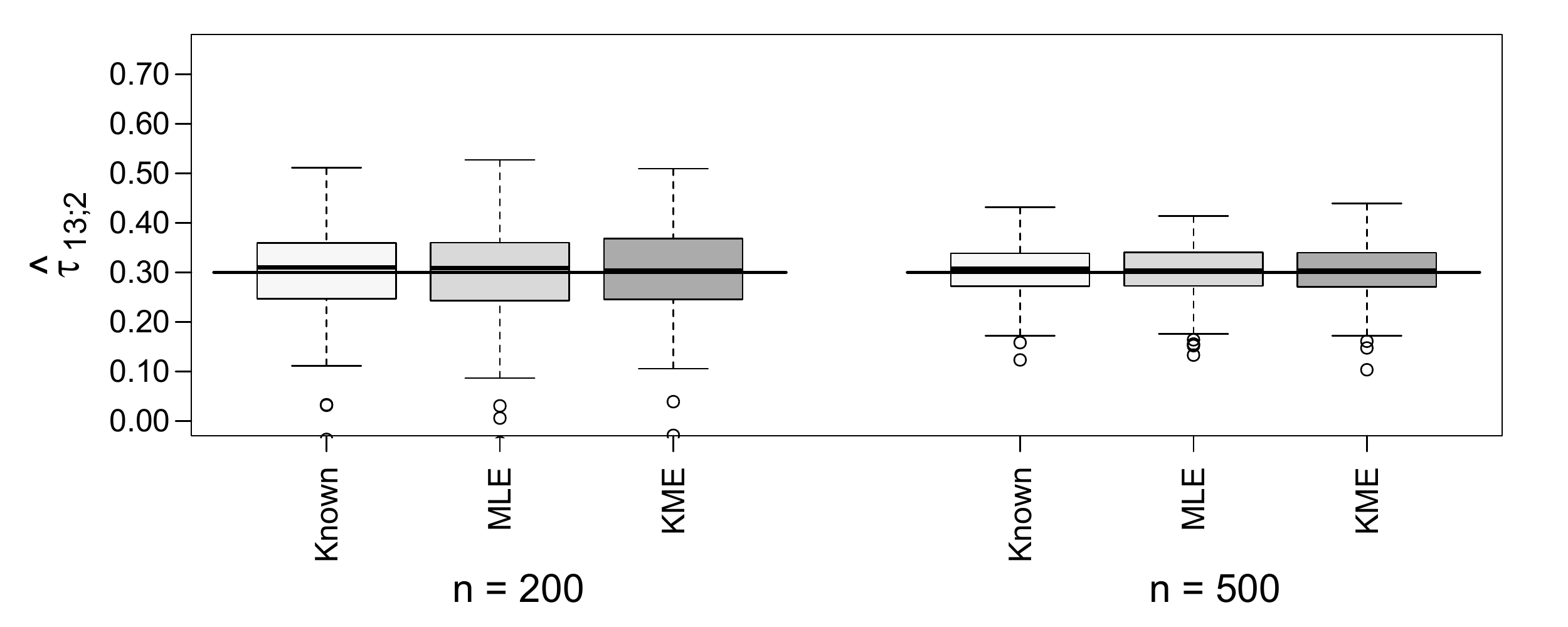}	
		\end{minipage}
	\end{figure}	
\end{landscape}

\begin{landscape}
	\centering
	\begin{figure}[ht]
		\caption{Boxplots of the estimated Kendall's $\tau$ values in $\mathcal{T}_1$ for an increasing percentage of common right-censoring with Clayton copulas (left) and Gumbel copulas (right) in $\mathcal{T}_1$, true $\tau_{12}=0.6$,  $\tau_{23}=0.6$ and sample size 500. Known margins, parametrically estimated margins (MLE) and nonparametrically estimated margins (ECDF/KME) are considered.}
		\label{fig:500_0+25+65}
		\begin{minipage}{0.5\linewidth}
			\centering
			Clayton copulas in $\mathcal{T}_1$			
			\includegraphics[width=1\linewidth]{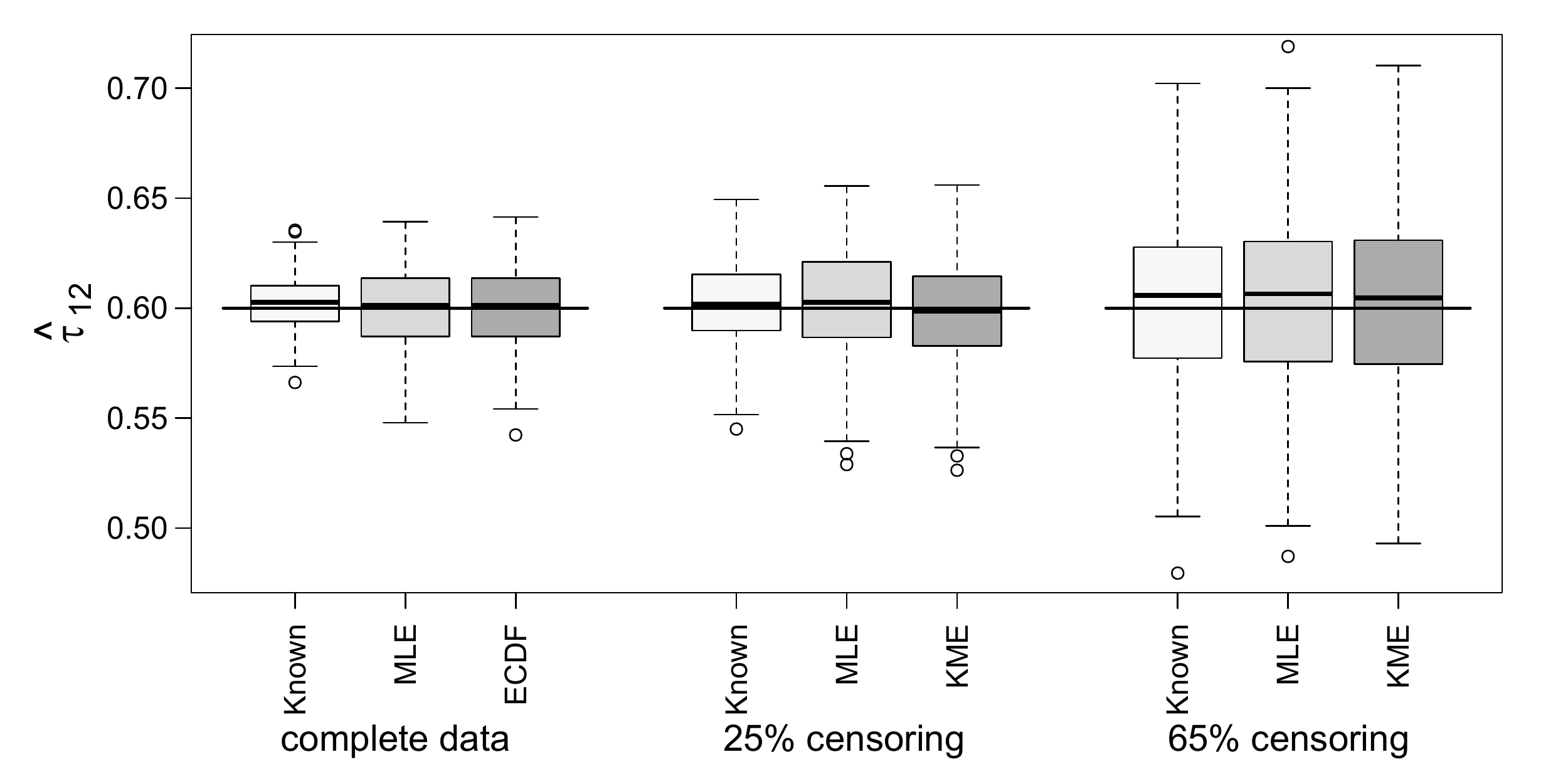}
			\includegraphics[width=1\linewidth]{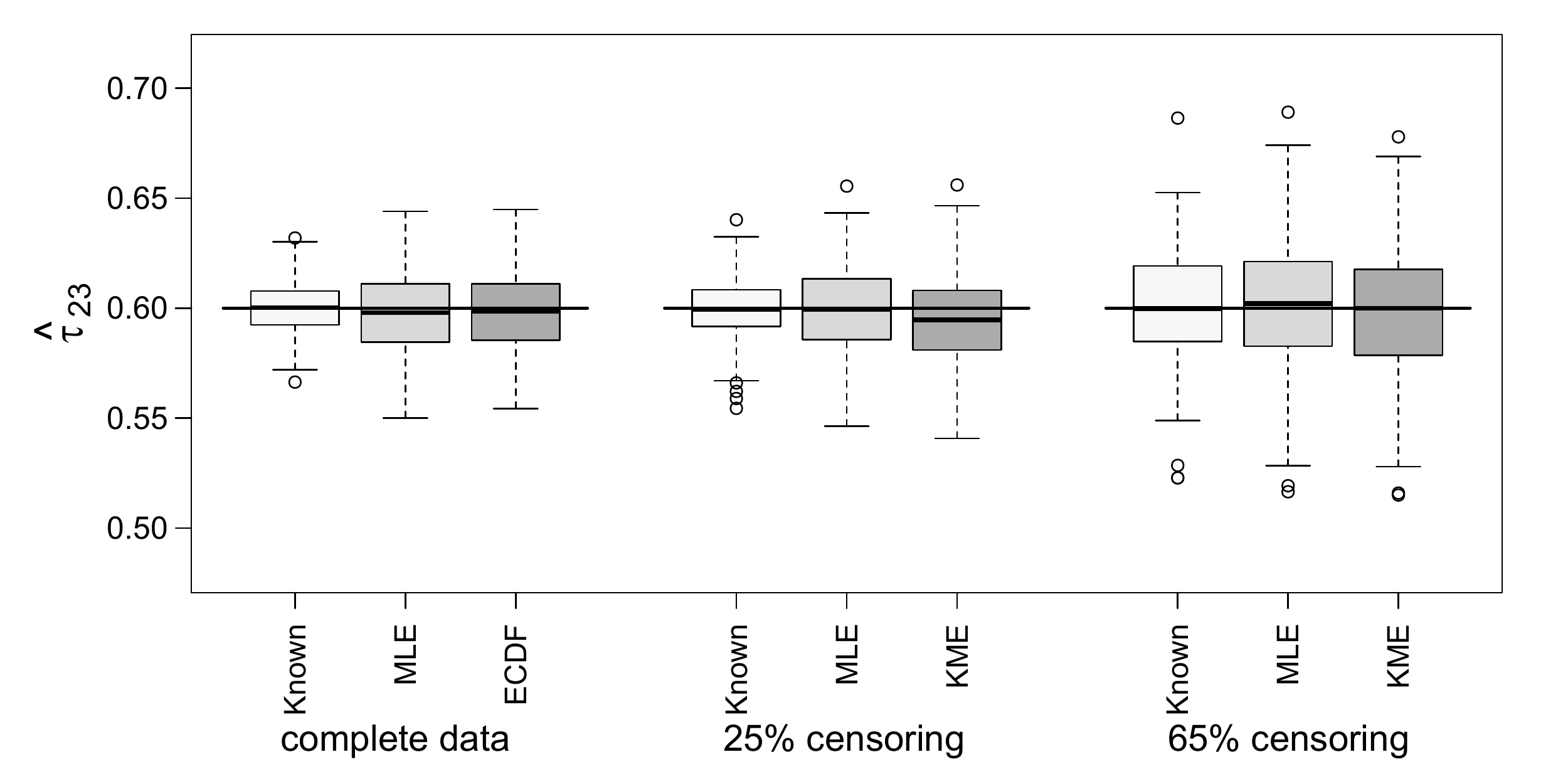}
		\end{minipage}
		\begin{minipage}{0.5\linewidth}
			\centering
			Gumbel copulas in $\mathcal{T}_1$
			\includegraphics[width=1\linewidth]{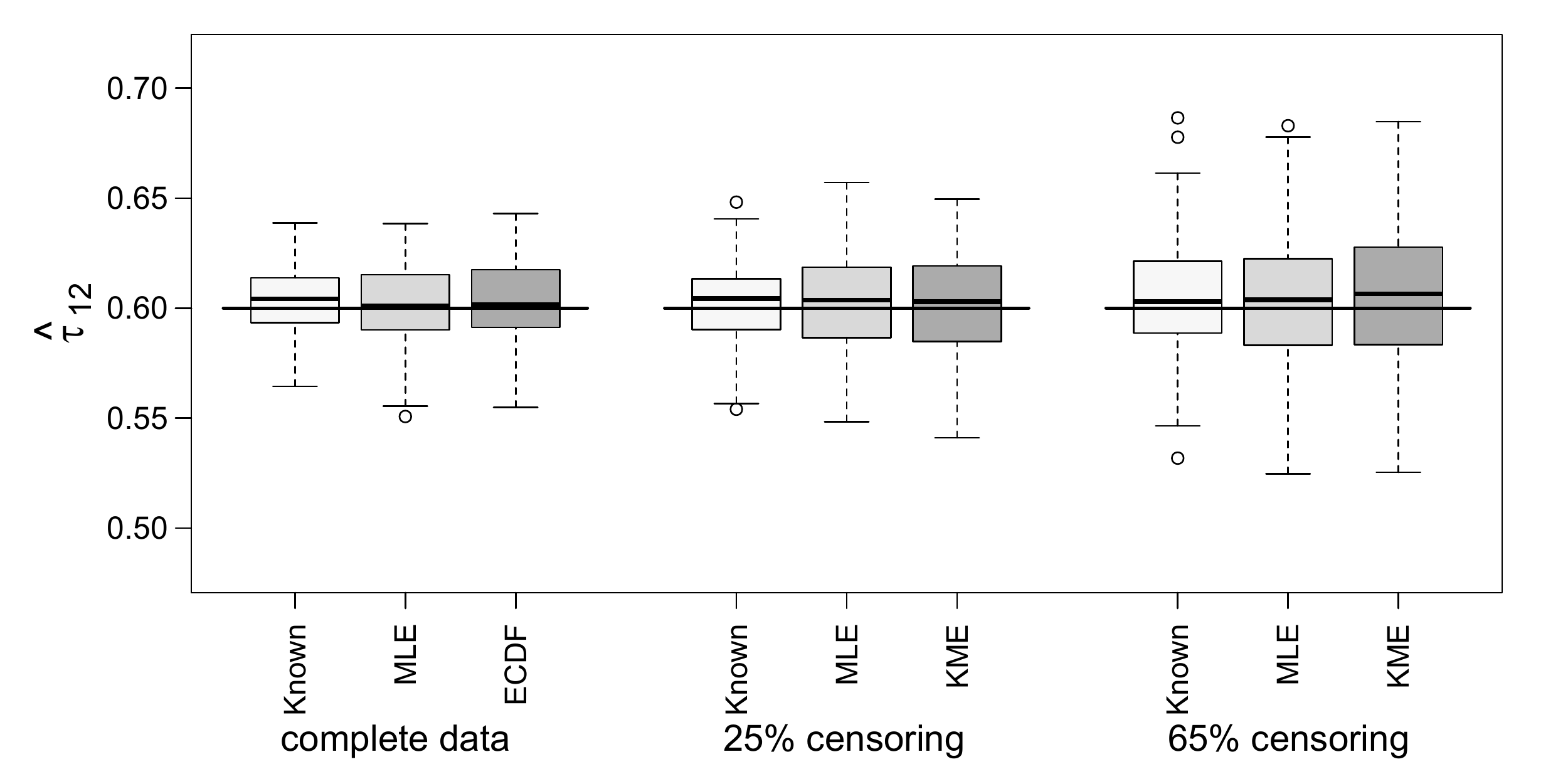}
			\includegraphics[width=1\linewidth]{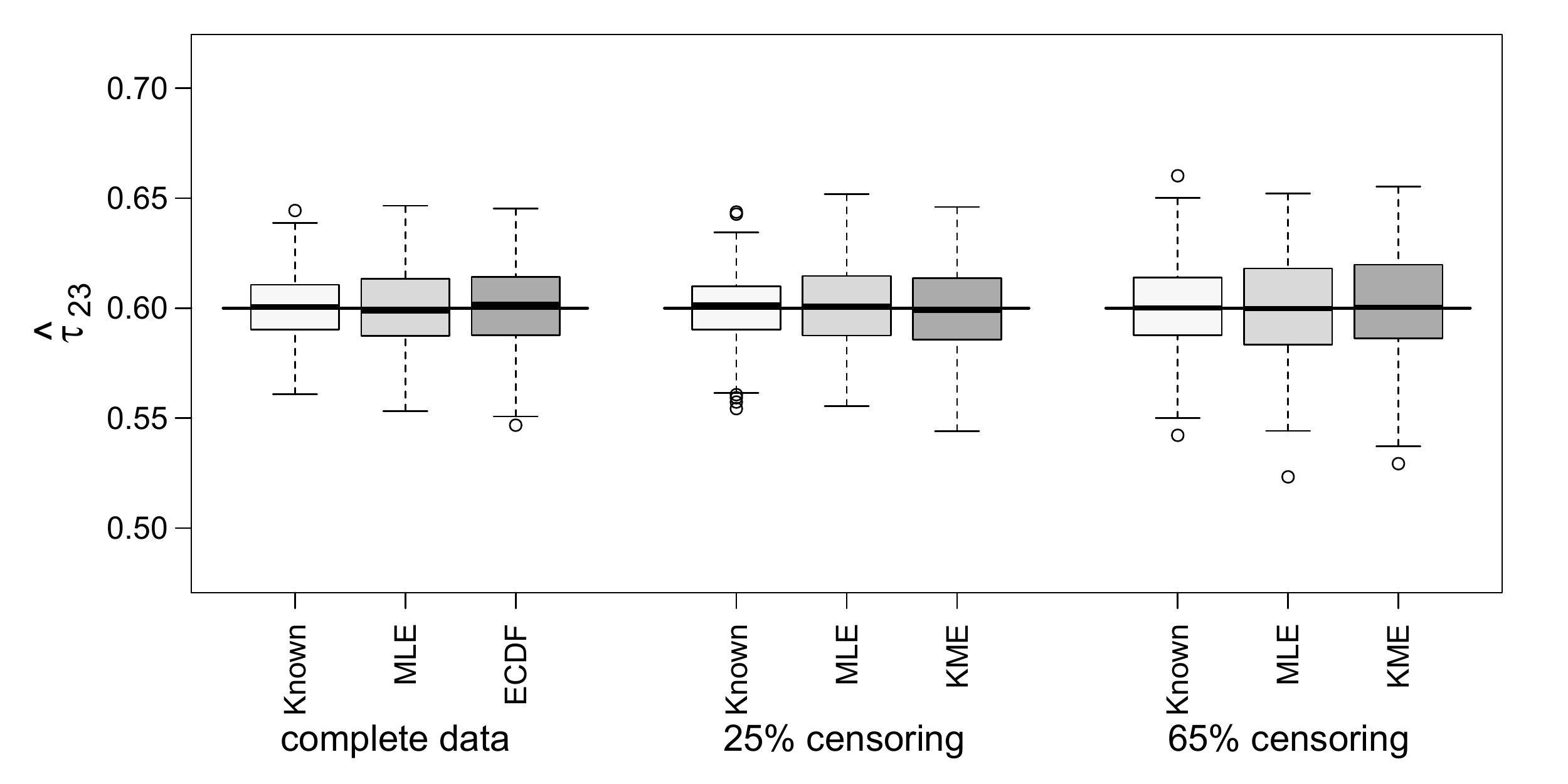}
		\end{minipage}
	\end{figure}	
\end{landscape}

\begin{landscape}
	\begin{table}[ht]
		\centering
		\captionof{table}{Performance measures for the estimation of the copula parameters and Kendall's $\tau$ values in case of 65\% common right-censored event time data with sample sizes 200 and 500. The copula combination Clayton (C), Clayton (C), Frank (F) with true $\tau_{12}=0.6$,  $\tau_{23}=0.6$ and $\tau_{13;2}=0.3$ is investigated. Known margins, parametrically estimated margins (MLE) and nonparametrically estimated margins (KME) are considered.}
		\label{Table:CCF_65_200+500}
		\begin{tabular}{cccccccccccccc}
			\midrule
			&  &  &  & \multicolumn{5}{c}{Copula parameter} &  \multicolumn{5}{c}{Kendall's $\tau$} \\
			&  &  &  & $\theta$ & $\bar{\theta}$ &  $\hat{b}\left(\bar{\theta}\right)$ & $s^2\left(\bar{\theta}\right)$ & $\widehat{mse}\left(\bar{\theta}\right)$ & $\tau$ & $\bar{\tau}$ & $\hat{b}\left(\bar{\tau}\right)$ & $s^2\left(\bar{\tau}\right)$ & $\widehat{mse}\left(\bar{\tau}\right)$ \\
			\hline
			\cmidrule{1-14}
			\multirow{9}{*} {\begin{sideways} $n = 200,\ 65\% \text{ censoring}$ \end{sideways}} & \multirow{3}{*} {\begin{sideways} Known \end{sideways}} & C & $\theta_{12}$ & 3.00 & 3.14 & 0.1430 & 0.5762 & 0.5966 & 0.60 &  0.60 & 0.0025 & 0.0036 & 0.0036 \\
			&  & C & $\theta_{23}$ & 3.00 & 3.06 &  0.0609 & 0.2499 & 0.2536 & 0.60 & 0.60 &  0.0009 & 0.0016 & 0.0016 \\
			&  & F & $\theta_{13;2}$ & 2.92 & 3.03 & 0.1155 & 0.8131 & 0.8264 & 0.30 & 0.31 &  0.0052 & 0.0062 & 0.0062 \\
			\cmidrule{2-14}
			& \multirow{3}{*} {\begin{sideways} MLE \end{sideways}} & C & $\theta_{12}$ & 3.00 & 3.22 &  0.2247 & 0.6998 & 0.7504 & 0.60 & 0.61 &  0.0073 & 0.0040 & 0.0041 \\
			&  & C & $\theta_{23}$ & 3.00 & 3.12 &  0.1240 & 0.3782 & 0.3936 & 0.60 & 0.60 &  0.0039 & 0.0024 & 0.0024 \\
			&  & F & $\theta_{13;2}$ & 2.92 & 3.04 &  0.1190 & 0.8641 & 0.8783 & 0.30 & 0.31 &  0.0052 & 0.0066 & 0.0066 \\
			\cmidrule{2-14}
			& \multirow{3}{*} {\begin{sideways} KME \end{sideways}} & C & $\theta_{12}$ & 3.00 & 3.15 &  0.1542 & 0.6681 & 0.6919 &  0.60 & 0.60 & 0.0021 & 0.0041 & 0.0041 \\
			&  & C & $\theta_{23}$ & 3.00 & 3.05 &  0.0455 & 0.3857 & 0.3877 & 0.60 & 0.60 &  -0.0025 & 0.0025 & 0.0026 \\
			&  & F & $\theta_{13;2}$ & 2.92 & 3.07 &  0.1548 & 0.8911 & 0.9151 & 0.30 & 0.31 &  0.0081 & 0.0066 & 0.0067 \\
			\hline
			\cmidrule{1-14}
			\multirow{9}{*} {\begin{sideways} $n = 500,\ 65\% \text{ censoring}$ \end{sideways}} & \multirow{3}{*} {\begin{sideways} Known \end{sideways}} & C & $\theta_{12}$ & 3.00 & 3.10 &  0.0983 & 0.2246 & 0.2343 & 0.60 & 0.60 & 0.0044 & 0.0013 & 0.0013 \\
			&  & C & $\theta_{23}$ & 3.00 & 3.03 &  0.0318 & 0.1083 & 0.1093 & 0.60 & 0.60 &  0.0008 & 0.0007 & 0.0007 \\
			&  & F & $\theta_{13;2}$ & 2.92 & 3.00 &  0.0855 & 0.3681 & 0.3754 & 0.30 & 0.31 & 0.0052 & 0.0027 & 0.0028 \\
			\cmidrule{2-14}
			& \multirow{3}{*} {\begin{sideways} MLE \end{sideways}} & C & $\theta_{12}$ & 3.00 & 3.11 &  0.1143 & 0.2703 & 0.2833 & 0.60 & 0.61 &  0.0051 & 0.0015 & 0.0015 \\
			&  & C & $\theta_{23}$ & 3.00 & 3.05 &  0.0460 & 0.1364 & 0.1386 & 0.60 & 0.60 &  0.0016 & 0.0008 & 0.0008 \\
			&  & F & $\theta_{13;2}$ & 2.92 & 3.00 &  0.0855 & 0.3771 & 0.3844 & 0.30 & 0.31 &  0.0052 & 0.0028 & 0.0028 \\
			\cmidrule{2-14}
			& \multirow{3}{*} {\begin{sideways} KME \end{sideways}} & C & $\theta_{12}$ & 3.00 & 3.09 &  0.0887 & 0.2713 & 0.2791 & 0.60 &  0.60 & 0.0030 & 0.0016 & 0.0016 \\
			&  & C & $\theta_{23}$ & 3.00 & 3.00 &  0.0042 & 0.1323 & 0.1323 & 0.60 & 0.60 &  -0.0017 & 0.0008 & 0.0008 \\
			&  & F & $\theta_{13;2}$ & 2.92 & 3.02 &  0.0988 & 0.3867 & 0.3964 & 0.30 & 0.31 &  0.0063 & 0.0029 & 0.0029 \\
			\hline
			\hline
		\end{tabular}
	\end{table}
\end{landscape}

\begin{landscape}
	\begin{table}[ht]
		\centering
		\captionof{table}{Performance measures for the estimation of the copula parameters and Kendall's $\tau$ values in case of complete and 25\% common right-censored event time data with sample size 500. The copula combination Clayton (C), Clayton (C), Frank (F) with true $\tau_{12}=0.6$,  $\tau_{23}=0.6$ and $\tau_{13;2}=0.3$ is investigated. Known margins, parametrically estimated margins (MLE) and nonparametrically estimated margins (KME) are considered.}
		\label{Table:CCF_0+25_500}
		\begin{tabular}{cccccccccccccc}
			\midrule
			&  &  &  & \multicolumn{5}{c}{Copula parameter} &  \multicolumn{5}{c}{Kendall's $\tau$} \\
			&  &  &  & $\theta$ & $\bar{\theta}$ & $\hat{b}\left(\bar{\theta}\right)$ & $s^2\left(\bar{\theta}\right)$ & $\widehat{mse}\left(\bar{\theta}\right)$ & $\tau$ & $\bar{\tau}$ &  $\hat{b}\left(\bar{\tau}\right)$ & $s^2\left(\bar{\tau}\right)$ & $\widehat{mse}\left(\bar{\tau}\right)$ \\
			\hline
			\cmidrule{1-14}
			\multirow{9}{*} {\begin{sideways} $n = 500,\ 25\% \text{ censoring}$ \end{sideways}} & \multirow{3}{*} {\begin{sideways} Known \end{sideways}} & C & $\theta_{12}$ & 3.00 & 3.03 &  0.0282 & 0.0537 & 0.0545 & 0.60 & 0.60 & 0.0014 & 0.0003 & 0.0003 \\
			&  & C & $\theta_{23}$ & 3.00 &  3.00 & -0.0021 & 0.0322 & 0.0322 & 0.60 & 0.60 & -0.0007 & 0.0002 & 0.0002 \\
			&  & F & $\theta_{13;2}$ & 2.92 & 2.94 &  0.0275 & 0.1431 & 0.1439 & 0.30 & 0.30 & 0.0015 & 0.0011 & 0.0011 \\
			\cmidrule{2-14}
			& \multirow{3}{*} {\begin{sideways} MLE \end{sideways}} & C & $\theta_{12}$ & 3.00 & 3.03 &  0.0345 & 0.0872 & 0.0884 & 0.60 & 0.60 &  0.0014 & 0.0006 & 0.0006 \\
			&  & C & $\theta_{23}$ & 3.00 & 3.00 &  -0.0028 & 0.0599 & 0.0599 &  0.60 & 0.60 & -0.0012 & 0.0004 & 0.0004 \\
			&  & F & $\theta_{13;2}$ & 2.92 & 2.95 &  0.0292 & 0.1478 & 0.1487 &  0.30 & 0.30 & 0.0017 & 0.0011 & 0.0011 \\
			\cmidrule{2-14}
			& \multirow{3}{*} {\begin{sideways} KME \end{sideways}} & C & $\theta_{12}$ & 3.00 & 2.99 &  -0.0113 & 0.0904 & 0.0905 & 0.60 & 0.60 & -0.0024 & 0.0006 & 0.0006 \\
			&  & C & $\theta_{23}$ & 3.00 & 2.94 &  -0.0629 & 0.0621 & 0.0661 & 0.60 & 0.59 &  -0.0061 & 0.0004 & 0.0005 \\
			&  & F & $\theta_{13;2}$ & 2.92 & 2.96 &  0.0391 & 0.1497 & 0.1512 & 0.30 & 0.30 & 0.0025 & 0.0011 & 0.0012 \\
			\hline
			\cmidrule{1-14}
			\multirow{9}{*} {\begin{sideways} $n = 500, \text{ complete data}$ \end{sideways}} & \multirow{3}{*} {\begin{sideways} Known \end{sideways}} & C & $\theta_{12}$ & 3.00 & 3.04 &  0.0364 & 0.0235 & 0.0248 & 0.60 &  0.60 & 0.0025 & 0.0001 & 0.0002 \\
			&  & C & $\theta_{23}$ & 3.00 &  3.00 & 0.0036 & 0.0239 & 0.0239 & 0.60 & 0.60 & -0.0001 & 0.0002 & 0.0002 \\
			&  & F & $\theta_{13;2}$ & 2.92 & 2.96 &  0.0457 & 0.0916 & 0.0937 & 0.30 & 0.30 & 0.0035 & 0.0007 & 0.0007 \\
			\cmidrule{2-14}
			& \multirow{3}{*} {\begin{sideways} MLE \end{sideways}} & C & $\theta_{12}$ & 3.00 & 3.01 &  0.0110 & 0.0514 & 0.0515 & 0.60 & 0.60 &  0.0001 & 0.0003 & 0.0003 \\
			&  & C & $\theta_{23}$ & 3.00 & 2.98 &  -0.0247 & 0.0517 & 0.0523 &  0.60 & 0.60 & -0.0028 & 0.0003 & 0.0003 \\
			&  & F & $\theta_{13;2}$ & 2.92 & 2.96 &  0.0408 & 0.0912 & 0.0929 &  0.30 & 0.30 & 0.0030 & 0.0007 & 0.0007 \\
			\cmidrule{2-14}
			& \multirow{3}{*} {\begin{sideways} ECDF \end{sideways}} & C & $\theta_{12}$ & 3.00 & 3.02 &  0.0153 & 0.0543 & 0.0545 & 0.60 & 0.60 &  0.0004 & 0.0003 & 0.0003 \\
			&  & C & $\theta_{23}$ & 3.00 & 2.98 & -0.0176 & 0.0551 & 0.0555 &  0.60 & 0.60 & -0.0023 & 0.0004 & 0.0004 \\
			&  & F & $\theta_{13;2}$ & 2.92 & 2.97 &  0.0481 & 0.0977 & 0.1000 &  0.30 & 0.30 & 0.0036 & 0.0007 & 0.0008 \\
			\hline
			\hline
		\end{tabular}
	\end{table}
\end{landscape}

\begin{landscape}
	\begin{table}[ht]
		\centering
		\captionof{table}{Performance measures for the estimation of the copula parameters and Kendall's $\tau$ values in case of 65\% common right-censored event time data with sample sizes 200 and 500. The copula combination Gumbel (G), Gumbel (G), Frank (F) with true $\tau_{12}=0.6$,  $\tau_{23}=0.6$ and $\tau_{13;2}=0.3$ is investigated. Known margins, parametrically estimated margins (MLE) and nonparametrically estimated margins (KME) are considered.}
		\label{Table:GGF_65_200+500}
		\begin{tabular}{cccccccccccccc}
			\midrule
			&  &  &  & \multicolumn{5}{c}{Copula parameter} &  \multicolumn{5}{c}{Kendall's $\tau$} \\
			&  &  &  & $\theta$ & $\bar{\theta}$ &  $\hat{b}\left(\bar{\theta}\right)$ & $s^2\left(\bar{\theta}\right)$ & $\widehat{mse}\left(\bar{\theta}\right)$ & $\tau$ & $\bar{\tau}$ &  $\hat{b}\left(\bar{\tau}\right)$ & $s^2\left(\bar{\tau}\right)$ & $\widehat{mse}\left(\bar{\tau}\right)$ \\
			\hline
			\cmidrule{1-14}
			\multirow{9}{*} {\begin{sideways} $n = 200,\ 65\% \text{ censoring}$ \end{sideways}} & \multirow{3}{*} {\begin{sideways} Known \end{sideways}} & G & $\theta_{12}$ & 2.50 & 2.53 &  0.0265 & 0.0641 & 0.0648 & 0.60 &  0.60 & 0.0003 & 0.0016 & 0.0016 \\
			&  & G & $\theta_{23}$ & 2.50 & 2.52 &  0.0201 & 0.0396 & 0.0400 &  0.60 & 0.60 & 0.0007 & 0.0010 & 0.0010 \\
			&  & F & $\theta_{13;2}$ & 2.92 & 2.99 &  0.0705 & 0.8978 & 0.9028 &  0.30 & 0.30 & 0.0008 & 0.0069 & 0.0069 \\
			\cmidrule{2-14}
			& \multirow{3}{*} {\begin{sideways} MLE \end{sideways}} & G & $\theta_{12}$ & 2.50 & 2.52 & 0.0158 & 0.0827 & 0.0830 & 0.60 & 0.60 &  -0.0026 & 0.0021 & 0.0021 \\
			&  & G & $\theta_{23}$ & 2.50 & 2.53 &  0.0250 & 0.0570 & 0.0577 &  0.60 & 0.60 & 0.0004 & 0.0014 & 0.0014 \\
			&  & F & $\theta_{13;2}$ & 2.92 & 3.00 &  0.0783 & 0.9986 & 1.0048 &  0.30 & 0.30 & 0.0009 & 0.0076 & 0.0076 \\
			\cmidrule{2-14}
			& \multirow{3}{*} {\begin{sideways} KME \end{sideways}} & G & $\theta_{12}$ & 2.50 & 2.58 &  0.0820 & 0.1069 & 0.1136 & 0.60 & 0.61 &  0.0067 & 0.0023 & 0.0024 \\
			&  & G & $\theta_{23}$ & 2.50 & 2.56 &  0.0558 & 0.0634 & 0.0665 & 0.60 & 0.60 &  0.0049 & 0.0015 & 0.0015 \\
			&  & F & $\theta_{13;2}$ & 2.92 & 3.00 &  0.0805 & 0.9979 & 1.0044 &  0.30 & 0.30 & 0.0011 & 0.0075 & 0.0075 \\
			\hline
			\cmidrule{1-14}
			\multirow{9}{*} {\begin{sideways} $n = 500,\ 65\% \text{ censoring}$ \end{sideways}} & \multirow{3}{*} {\begin{sideways} Known \end{sideways}} & G & $\theta_{12}$ & 2.50 & 2.54 &  0.0376 & 0.0291 & 0.0305 & 0.60 & 0.60 &  0.0042 & 0.0007 & 0.0007 \\
			&  & G & $\theta_{23}$ & 2.50 & 2.51 &  0.0106 & 0.0170 & 0.0171 &  0.60 & 0.60 & 0.0006 & 0.0004 & 0.0004 \\
			&  & F & $\theta_{13;2}$ & 2.92 & 2.97 &  0.0494 & 0.3800 & 0.3825 &  0.30 & 0.30 & 0.0020 & 0.0030 & 0.0030 \\
			\cmidrule{2-14}
			& \multirow{3}{*} {\begin{sideways} MLE \end{sideways}} & G & $\theta_{12}$ & 2.50 & 2.53 &  0.0324 & 0.0376 & 0.0386 & 0.60 & 0.60 &  0.0029 & 0.0009 & 0.0009 \\
			&  & G & $\theta_{23}$ & 2.50 & 2.51 &  0.0107 & 0.0243 & 0.0244 &  0.60 & 0.60 & 0.0002 & 0.0006 & 0.0006 \\
			&  & F & $\theta_{13;2}$ & 2.92 & 2.97 &  0.0507 & 0.3892 & 0.3918 &  0.30 & 0.30 & 0.0021 & 0.0030 & 0.0030 \\
			\cmidrule{2-14}
			& \multirow{3}{*} {\begin{sideways} KME \end{sideways}} & G & $\theta_{12}$ & 2.50 & 2.55 &  0.0524 & 0.0445 & 0.0472 & 0.60 & 0.61 &  0.0056 & 0.0010 & 0.0011 \\
			&  & G & $\theta_{23}$ & 2.50 &  2.52 & 0.0162 & 0.0254 & 0.0257 & 0.60 & 0.60 &  0.0010 & 0.0006 & 0.0006 \\
			&  & F & $\theta_{13;2}$ & 2.92 & 2.98 &  0.0578 & 0.4216 & 0.4249 & 0.30 & 0.30 &  0.0025 & 0.0033 & 0.0033 \\
			\hline\hline
		\end{tabular}
	\end{table}
\end{landscape}

\begin{landscape}
	\begin{table}[ht]
		\centering
		\captionof{table}{Performance measures for the estimation of the copula parameters and Kendall's $\tau$ values in
			case of complete and 25\% common right-censored event time data with sample size 500. The copula combination Gumbel (G), Gumbel (G), Frank (F) with true $\tau_{12}=0.6$,  $\tau_{23}=0.6$ and $\tau_{13;2}=0.3$ is investigated. Known margins, parametrically estimated margins (MLE) and nonparametrically estimated margins (ECDF/KME) are considered.}
		\label{Table:GGF_0+25_500}
		\begin{tabular}{cccccccccccccc}
			\midrule
			&  &  &  & \multicolumn{5}{c}{Copula parameter} &  \multicolumn{5}{c}{Kendall's $\tau$} \\
			&  &  &  & $\theta$ & $\bar{\theta}$ & $\hat{b}\left(\bar{\theta}\right)$ & $s^2\left(\bar{\theta}\right)$ & $\widehat{mse}\left(\bar{\theta}\right)$ & $\tau$ & $\bar{\tau}$ & $\hat{b}\left(\bar{\tau}\right)$ & $s^2\left(\bar{\tau}\right)$ & $\widehat{mse}\left(\bar{\tau}\right)$ \\
			\hline
			\cmidrule{1-14}
			\multirow{9}{*} {\begin{sideways} $n = 500,\  25\% \text{ censoring}$ \end{sideways}} & \multirow{3}{*} {\begin{sideways} Known \end{sideways}} & G & $\theta_{12}$ & 2.50 & 2.52 & 0.0181 & 0.0129 & 0.0132 & 0.60 &  0.60 & 0.0021 & 0.0003 & 0.0003 \\
			&  & G & $\theta_{23}$ & 2.50 & 2.51 & 0.0052 & 0.0100 & 0.0101 &  0.60 & 0.60 & 0.0002 & 0.0003 & 0.0003 \\
			&  & F & $\theta_{13;2}$ & 2.92 & 2.93 &  0.0107 & 0.1517 & 0.1518 &  0.30 & 0.30 & 0.0000 & 0.0012 & 0.0012 \\
			\cmidrule{2-14}
			& \multirow{3}{*} {\begin{sideways} MLE \end{sideways}} & G & $\theta_{12}$ & 2.50 & 2.52 &  0.0207 & 0.0198 & 0.0203 & 0.60 & 0.60 &  0.0021 & 0.0005 & 0.0005 \\
			&  & G & $\theta_{23}$ & 2.50 & 2.51 & 0.0084 & 0.0148 & 0.0148 &  0.60 & 0.60 & 0.0004 & 0.0004 & 0.0004 \\
			&  & F & $\theta_{13;2}$ & 2.92 & 2.92 & 0.0027 & 0.1529 & 0.1529 &  0.30 & 0.30 & -0.0007 & 0.0012 & 0.0012 \\
			\cmidrule{2-14}
			& \multirow{3}{*} {\begin{sideways} KME \end{sideways}} & G & $\theta_{12}$ & 2.50 & 2.52 & 0.0193 & 0.0207 & 0.0210 & 0.60 & 0.60 &  0.0018 & 0.0005 & 0.0005 \\
			&  & G & $\theta_{23}$ & 2.50 & 2.50 & 0.0018 & 0.0155 & 0.0155 &  0.60 & 0.60 & -0.0007 & 0.0004 & 0.0004 \\
			&  & F & $\theta_{13;2}$ & 2.92 & 2.93 & 0.0106 & 0.1602 & 0.1603 &  0.30 & 0.30 & -0.0001 & 0.0012 & 0.0012 \\
			\hline
			\cmidrule{1-14}
			\multirow{9}{*} {\begin{sideways} $n = 500, \text{ complete data}$ \end{sideways}} & \multirow{3}{*} {\begin{sideways} Known \end{sideways}} & G & $\theta_{12}$ & 2.50 & 2.52 & 0.0212 & 0.0078 & 0.0083 & 0.60 &  0.60 & 0.0029 & 0.0002 & 0.0002 \\
			&  & G & $\theta_{23}$ & 2.50 & 2.51 & 0.0064 & 0.0086 & 0.0086 &  0.60 & 0.60 & 0.0005 & 0.0002 & 0.0002 \\
			&  & F & $\theta_{13;2}$ & 2.92 & 2.96 &  0.0388 & 0.0997 & 0.1012 &  0.30 & 0.30 & 0.0028 & 0.0008 & 0.0008 \\
			\cmidrule{2-14}
			& \multirow{3}{*} {\begin{sideways} MLE \end{sideways}} & G & $\theta_{12}$ & 2.50 & 2.51 &  0.0136 & 0.0123 & 0.0125 & 0.60 & 0.60 &  0.0014 & 0.0003 & 0.0003 \\
			&  & G & $\theta_{23}$ & 2.50 & 2.50 &  -0.0000 & 0.0131 & 0.0131 &  0.60 & 0.60 & -0.0008 & 0.0003 & 0.0003 \\
			&  & F & $\theta_{13;2}$ & 2.92 & 2.95 & 0.0276 & 0.0983 & 0.0991 &  0.30 & 0.30 & 0.0018 & 0.0008 & 0.0008 \\
			\cmidrule{2-14}
			& \multirow{3}{*} {\begin{sideways} ECDF \end{sideways}} & G & $\theta_{12}$ & 2.50 & 2.53 & 0.0255 & 0.0135 & 0.0141 & 0.60 & 0.60 &  0.0032 & 0.0003 & 0.0003 \\
			&  & G & $\theta_{23}$ & 2.50 & 2.51 & 0.0094 & 0.0144 & 0.0145 &  0.60 & 0.60 & 0.0006 & 0.0004 & 0.0004 \\
			&  & F & $\theta_{13;2}$ & 2.92 & 2.95 & 0.0279 & 0.1040 & 0.1048 & 0.30 & 0.30 & 0.0018 & 0.0008 & 0.0008 \\
			\hline
			\hline
		\end{tabular}
	\end{table}
\end{landscape}

\section{Real data application: The mastitis data}\label{Sec:MastitisApplication}
\normalsize
In this section, we investigate the dependence structure present in the mastitis data by fitting several vine copula models. According to Laevens (personal communication and \cite{Laevens97}), there is no biological rule that could provide guidance for the dependence modeling. The primary goal is therefore to illustrate how the introduced methodology can be applied to real data. In particular, we give insights about the effect of right-censoring in the context of copula estimation. Also, earlier investigations of the mastitis data using e.g.\ EAC models \citep{Massonnet2009,Geerdens2014} assumed equal correlations between all pairs of udder quarters inducing rather restrictive association patterns. We will see that these less elaborate models do not sufficiently fit the data. 

\begin{figure}[H]
	\centering
	\caption{Pairs plot of the two front udder quarters of the mastitis data based on pseudo-observations generated via Kaplan-Meier estimates of the marginals. Observations shown as $\bullet$ are event times for both udder quarters; $\leftarrow$ is an event time only for FR; $\downarrow$ is an event time only for FL; censored in both components is shown as \hspace{-.3cm}  $\myrelArrowI{}{\myrelArrowI{\myrelArrowI{\myrelArrowII{\leftarrow}{}}{}}{}}\hspace{-.42cm}\downarrow$.}
	\label{fig:ScatterplotCows12}
	\vspace{-.5cm}
	\includegraphics[width=.4\linewidth]{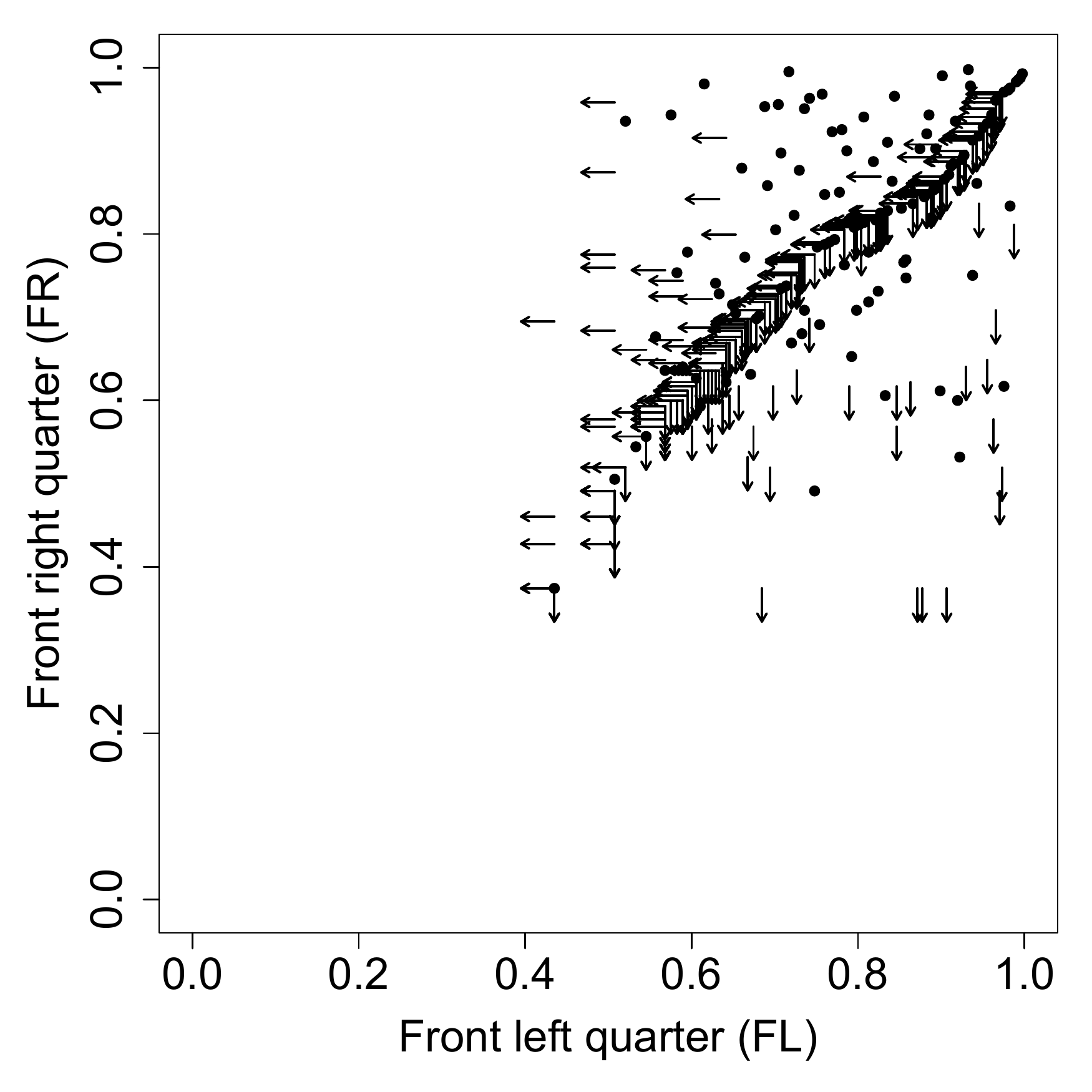}
\end{figure}

Before starting the discussion on model selection, recall that we observe 407 clusters (cows) with 66.15\% censoring (see \autoref{Table:CensoringCows}). Further, it is important to note that the information loss due to right-censoring complicates accurate model selection and implies -- as will be seen from the further discussion -- the need for careful comparison of possible models.
Pairs plots can be used to demonstrate the information loss in a graphical way. In \autoref{fig:ScatterplotCows12}, we give the pairs plot for the data points $(\hat{u}_i^{\textup{FL}},\hat{u}_i^{\textup{FR}})$ with $\hat{u}_{i}^{\textup{FL}}=\hat{S}_{\textup{FL}}(y_{i}^{\textup{FL}})$ and $\hat{u}_{i}^{\textup{FR}}=\hat{S}_{\textup{FR}}(y_{i}^{\textup{FR}})$, where $y_{i}^{\textup{FL}}$ and $y_{i}^{\textup{FR}}$ are the observed infection times for the front left udder quarter and the front right udder quarter of cow $i$, $i = 1,\ldots,407$, and $\hat{S}_{\textup{FL}}$ and $\hat{S}_{\textup{FR}}$ are the corresponding Kaplan-Meier estimates. Given the heavy censoring, the scatter plot for the data points $(y_i^{\textup{FL}}, y_i^{\textup{FR}})$ would contain only a few points in the upper right corner (of the first quadrant), which in turn leads to an almost empty lower left corner in \autoref{fig:ScatterplotCows12} (see the annotation in the caption of \autoref{fig:ScatterplotCows12} for the graphical representation of censored observations). The Kaplan-Meier survival functions estimated for the four udder quarters and pairs plots for all udder quarter pairs are given in \autoref{fig:KM} and \autoref{fig:ScatterplotsMastitis} of \autoref{Sec:AppIllustrMastitis}.

Given the good performance of the two-stage semiparametric estimation in the simulation study, we flexibly model the marginal survival functions using the Kaplan-Meier estimator and thus do not imply any parametric assumptions for the underyling data. We maximize the loglikelihood \eqref{Eq:logLik_uLevel} over all copula parameters using the parameter estimates obtained from the $\mathcal{T}_1$-sequential approach as starting values. We consider vine copula models based on one parameter bivariate copulas such that all considered models have the same number of parameters (six). In this case, the AIC and BIC both select the model that gives the highest loglikelihood. We therefore use the loglikelihood for model selection. The use of the loglikelihood value as well as AIC and BIC for model selection in the context of semiparametric copula estimation for right-censored data has been studied in \citet{Chen2010129} and in \citet{Geerdens2014}.

In the following, we assume a D-vine tree structure for the mastitis data. All possible 12 D-vines are represented in \autoref{Table:LogLikli_MastitisModels} by their first tree level, since the latter uniquely determines the whole D-vine structure. For all D-vines the same type of copula is assumed in $\mathcal{T}_1$, however allowing for different parameters. We consider the Clayton, Gumbel or Frank copula, respectively. With this choice we account for possible lower and upper tail-dependence as well as for no tail-dependence inherent in the underlying data. In particular, asymmetric tail-dependence behavior is modeled through combination of the copula families in the considered D-vine copula models. Further, Frank copulas are taken in the two lower tree levels. By doing so, 36 models are investigated in total. \autoref{Table:LogLikli_MastitisModels} shows the loglikelihood values for the considered models obtained via simultaneous estimation of all six parameters. The loglikelihood values obtained through the $\mathcal{T}_1$-sequential estimation approach are shown in brackets. In general, D-vine structures that capture the dependence along the two flanks perform best, whereas D-vines with two diagonals would generally not be selected. Further, the choice of Frank and Clayton copulas in $\mathcal{T}_1$ is superior to the one of Gumbel copulas. Models with Frank copulas perform slightly better than those with Clayton copulas. Recall that for heavily censored copula data the lower left corner of a pairs plot is empty. However, there might be a considerable amount of observed event times in the upper right corner, where, therefore, most of the information is located (see \autoref{fig:ScatterplotCows12}). Since Clayton and Frank copulas behave similar in the upper right corner, i.e.\ for early event times, it is clear that the information loss in case of heavy right-censoring makes it difficult to distinguish between Clayton and Frank copulas.

To further explore this finding we consider for the D-vine with structure (c) (the best performing structure in \autoref{Table:LogLikli_MastitisModels}) the 24 additional vine models (besides C-C-C, G-G-G and F-F-F in $\mathcal{T}_1$) having structure (c), where we allow combinations of Clayton, Gumbel and Frank copulas in $\mathcal{T}_1$. The loglikelihood values are listed in \autoref{Table:LogLikli_MastitisCombinationsModels1342}. The model with all dependencies captured by Frank copulas remains the best (see \autoref{Table:LogLikli_MastitisModels}), but models which combine Clayton and Frank copulas in $\mathcal{T}_1$ perform equally well. The estimated copula parameters of the four best models are given in \autoref{Table:MastitisBestModels1342} together with their corresponding estimated Kendall's $\tau$ values and tail-dependence coefficients. A strong lower tail-dependence and thus a strong association between late event times is detected for Clayton copulas in $\mathcal{T}_1$. Further, the strength of overall dependence detected for the three udder pairs in $\mathcal{T}_1$ is higher for Clayton copulas as compared to Frank copulas. The fact that it is difficult to distinguish between Clayton and Frank is a typical consequence of the information loss caused by the heavy censoring in the data. To obtain standard errors 100 bootstrap replications are used, both for global likelihood estimation and for $\mathcal{T}_1$-sequential likelihood estimation. Using 100 bootstrap samples, the estimates for the standard error of the various parameters are already quite accurate. Details on the bootstrap algorithm for right-censored event time data modeled via a vine copula are given in \autoref{Sec:VineCopBootAlg}. Detailed bootstrapping results for the mastitis data are available in \autoref{Sec:VineCopBootRes}.

The results are in line with the findings in \citet{Geerdens2014}, where a Joe-Hu copula that combines a Clayton Laplace transform with bivariate Frank copulas is in the top three of the best models. Both analyses, using a vine copula or a Joe-Hu copula, stress the need for flexible copula models for the mastitis data.

We conclude by an important remark on the practical implementation of the optimization procedure. A comparison of the estimation results for both considered estimation methods qualifies the $\mathcal{T}_1$-sequential estimation approach as an important simplification and a valid alternative for the computationally extensive full loglikelihood optimization. Given that heavy censoring goes along with numerical challenges in the full optimization approach, the $\mathcal{T}_1$-sequential approach is the estimation method to apply in practice.

\tikzstyle{ClassicalVineNode} = [rectangle, fill = white, draw = black, text = black, font = \footnotesize, align = center, minimum height = .4cm, minimum width = .4cm]
\tikzstyle{TreeLabels} = [draw = none, fill = none, text = black, font = \scriptsize]
\tikzstyle{DummyNode}  = [draw = none, fill = none, text = white]
\tikzstyle{DummyEdge}  = [draw = white]
\renewcommand{\labelsize}{\scriptsize}
\newcommand{\shift}{.7cm}

\begin{table}[H]
	\centering
	\small
	\renewcommand{\arraystretch}{1.15}
	\caption{D-vine structures considered for the mastitis data and corresponding loglikelihood values obtained via simultaneous estimation of all six parameters ($\mathcal{T}_1$-sequential estimation). Frank copulas are taken in $\mathcal{T}_2$ and $\mathcal{T}_3$.}
	\label{Table:LogLikli_MastitisModels}
	\vspace*{-1cm}
	\begin{tabular}{p{.01cm}p{1.2cm}p{1.2cm}p{1.2cm}p{1.2cm}p{1.2cm}p{1.2cm}p{.01cm}}	\centering
		& & & & & & &\\	
		\multicolumn{8}{c}{FRONT}\\	
		\multirow{8}{*}{\rotatebox{90}{LEFT}} & \hspace{.5cm}(a) & \hspace{.5cm}(b) & \hspace{.5cm}(c) & \hspace{.5cm}(d) & \hspace{.5cm}(e) & \hspace{.5cm}(f) & \multirow{8}{*}{\rotatebox{-90}{\hspace{-.5cm} RIGHT}} \\
		& \multirow{3}{*}{
			\centering		
			\begin{tikzpicture}	[every node/.style = ClassicalVineNode, node distance = 1]
			\node (1){1}
			node             (2)         [right of = 1, xshift=\shift]  {2}
			node             (3)         [below of = 1, yshift=-\shift] {3}
			node             (4)         [right of = 3, xshift=\shift]  {4}
			;
			\begin{scope}[on background layer]
			\draw (1) to (2);
			\draw (2) to (4);
			\draw (1) to (3);
			\end{scope}		
			\end{tikzpicture}	
		} & 	\multirow{3}{*}{
			\centering
			\begin{tikzpicture}	[every node/.style = ClassicalVineNode, node distance = 1]
			\node (1){1}
			node             (2)         [right of = 1, xshift=\shift]  {2}
			node             (3)         [below of = 1, yshift=-\shift] {3}
			node             (4)         [right of = 3, xshift=\shift]  {4}
			;
			\begin{scope}[on background layer]
			\draw (1) to (2);
			\draw (3) to (4);
			\draw (1) to (3);
			\end{scope}		
			\end{tikzpicture}			
		} & 	\multirow{3}{*}{
			\centering
			\begin{tikzpicture}	[every node/.style = ClassicalVineNode, node distance = 1]
			\node (1){1}
			node             (2)         [right of = 1, xshift=\shift]  {2}
			node             (3)         [below of = 1, yshift=-\shift] {3}
			node             (4)         [right of = 3, xshift=\shift]  {4}
			;
			\begin{scope}[on background layer]
			\draw (2) to (4);
			\draw (3) to (4);
			\draw (1) to (3);
			\end{scope}		
			\end{tikzpicture}	
		} & \multirow{3}{*}{
			\centering
			\begin{tikzpicture}	[every node/.style = ClassicalVineNode, node distance = 1]
			\node (1){1}
			node             (2)         [right of = 1, xshift=\shift]  {2}
			node             (3)         [below of = 1, yshift=-\shift] {3}
			node             (4)         [right of = 3, xshift=\shift]  {4}
			;
			\begin{scope}[on background layer]
			\draw (1) to (2);
			\draw (2) to (4);
			\draw (3) to (4);
			\end{scope}		
			\end{tikzpicture}
		} & \multirow{3}{*}{
			\centering
			\begin{tikzpicture}	[every node/.style = ClassicalVineNode, node distance = 1]
			\node (1){1}
			node             (2)         [right of = 1, xshift=\shift]  {2}
			node             (3)         [below of = 1, yshift=-\shift] {3}
			node             (4)         [right of = 3, xshift=\shift]  {4}
			;
			\begin{scope}[on background layer]
			\draw (1) to (2);
			\draw (3) to (4);
			\draw (2) to (3);
			\end{scope}		
			\end{tikzpicture}
		} & \multirow{3}{*}{
			\centering
			\begin{tikzpicture}	[every node/.style = ClassicalVineNode, node distance = 1]
			\node (1){1}
			node             (2)         [right of = 1, xshift=\shift]  {2}
			node             (3)         [below of = 1, yshift=-\shift] {3}
			node             (4)         [right of = 3, xshift=\shift]  {4}
			;
			\begin{scope}[on background layer]
			\draw (2) to (4);
			\draw (1) to (3);
			\draw (1) to (4);
			\end{scope}		
			\end{tikzpicture}
		} & \\
		& & & & & & &\\
		& & & & & & &\\
		& \hspace{.5cm}(g) & \hspace{.5cm}(h) & \hspace{.5cm}(i) & \hspace{.5cm}(j) & \hspace{.5cm}(k) & \hspace{.5cm}(l) &\\
		& \multirow{3}{*}{
			\centering
			\begin{tikzpicture}	[every node/.style = ClassicalVineNode, node distance = 1]
			\node (1){1}
			node             (2)         [right of = 1, xshift=\shift]  {2}
			node             (3)         [below of = 1, yshift=-\shift] {3}
			node             (4)         [right of = 3, xshift=\shift]  {4}
			;
			\begin{scope}[on background layer]
			\draw (1) to  (2);
			\draw (3) to  (4);
			\draw (1) to (4);
			\end{scope}		
			\end{tikzpicture}
		} & \multirow{3}{*}{
			\centering
			\begin{tikzpicture}	[every node/.style = ClassicalVineNode, node distance = 1]
			\node (1){1}
			node             (2)         [right of = 1, xshift=\shift]  {2}
			node             (3)         [below of = 1, yshift=-\shift] {3}
			node             (4)         [right of = 3, xshift=\shift]  {4}
			;
			\begin{scope}[on background layer]
			\draw (2) to (4);
			\draw (1) to (3);
			\draw (2) to (3);
			\end{scope}		
			\end{tikzpicture}
		} & \multirow{3}{*}{
			\centering
			\begin{tikzpicture}	[every node/.style = ClassicalVineNode, node distance = 1]
			\node (1){1}
			node             (2)         [right of = 1, xshift=\shift]  {2}
			node             (3)         [below of = 1, yshift=-\shift] {3}
			node             (4)         [right of = 3, xshift=\shift]  {4}
			;
			\begin{scope}[on background layer]
			\draw (1) to (2);
			\draw (2) to (3);
			\draw (1) to (4);
			\end{scope}		
			\end{tikzpicture}
		} & \multirow{3}{*}{
			\centering
			\begin{tikzpicture}	[every node/.style = ClassicalVineNode, node distance = 1]
			\node (1){1}
			node             (2)         [right of = 1, xshift=\shift]  {2}
			node             (3)         [below of = 1, yshift=-\shift] {3}
			node             (4)         [right of = 3, xshift=\shift]  {4}
			;
			\begin{scope}[on background layer]
			\draw (1) to (3);
			\draw (2) to (3);
			\draw (1) to (4);
			\end{scope}		
			\end{tikzpicture}
		} & \multirow{3}{*}{
			\centering
			\begin{tikzpicture}	[every node/.style = ClassicalVineNode, node distance = 1]
			\node (1){1}
			node             (2)         [right of = 1, xshift=\shift]  {2}
			node             (3)         [below of = 1, yshift=-\shift] {3}
			node             (4)         [right of = 3, xshift=\shift]  {4}
			;
			\begin{scope}[on background layer]
			\draw (3) to (4);
			\draw (2) to (3);
			\draw (1) to (4);
			\end{scope}		
			\end{tikzpicture}
		} &  \multirow{3}{*}{
			\centering
			\begin{tikzpicture}	[every node/.style = ClassicalVineNode, node distance = 1]
			\node (1){1}
			node             (2)         [right of = 1, xshift=\shift]  {2}
			node             (3)         [below of = 1, yshift=-\shift] {3}
			node             (4)         [right of = 3, xshift=\shift]  {4}
			;
			\begin{scope}[on background layer]
			\draw (2) to (4);
			\draw (2) to (3);
			\draw (1) to (4);
			\end{scope}		
			\end{tikzpicture}
		} & \\
		& & & & & & &\\
		& & & & & & &\\
		\multicolumn{8}{c}{REAR}\\
		& & & & & & &\\ 
	\end{tabular}
	
	\centering	
	\renewcommand{\arraystretch}{1}
	\vspace*{-.25cm}
	\begin{tabular}{p{.5cm}p{.1cm}p{2.0cm}p{2.0cm}p{2.0cm}}		
		\midrule
		\multicolumn{2}{c}{\multirow{2}{*}{D-vine}} &  \multicolumn{3}{c}{Common family in $\mathcal{T}_1$}\\
		& & \centering Clayton & \centering Gumbel & \quad \quad \  Frank\\	
		\hline
		\cmidrule{1-5}	
		\multirow{24}{*}{\rotatebox{90}{loglikelihood}}
		& (a) & -138.73 (-138.82) & -153.45 (-153.69) & -137.24 (-137.30) \\
		& (b) & -139.19 (-139.25) & -148.91 (-148.99) & -136.05 (-136.10) \\
		& (c) &\cellcolor[gray]{0.9}-127.93 (-127.96) &\cellcolor[gray]{0.9} -142.98 (-143.09) &\cellcolor[gray]{0.9}-124.70 (-124.82) \\
		& (d) & -138.47 (-138.56) & -147.99 (-148.25) & -134.91 (-134.95) \\
		& (e) & -141.45 (-141.56) & -142.29 (-142.44) & -137.62 (-137.65) \\
		& (f) &\cellcolor[gray]{0.9}-129.78 (-129.88) &\cellcolor[gray]{0.9}-145.89 (-145.99) & \cellcolor[gray]{0.9}-130.10 (-130.16) \\
		& (g) & -138.05 (-138.42) & -143.71 (-144.80) & -135.95 (-136.41) \\
		& (h) &\cellcolor[gray]{0.9}-132.53 (-132.60) & \cellcolor[gray]{0.9}-145.33 (-145.40) &\cellcolor[gray]{0.9}-131.17 (-131.28) \\
		& (i) & -140.63 (-140.89) & -145.50 (-145.98) & -139.28 (-139.46) \\
		& (j) & -133.55 (-132.57) & -143.81 (-143.88) & -134.71 (-134.82) \\
		& (k) & -137.42 (-137.63) & -141.61 (-141.81) & -136.92 (-137.17) \\
		& (l) & -134.37 (-134.50) & -145.29 (-145.73) & -134.23 (-134.29) \\
		\cmidrule{1-5}
		\hline	
	\end{tabular}			
	\renewcommand{\arraystretch}{1}
\end{table}

\renewcommand{\arraystretch}{1.4}
\begin{table}[H]
	\centering	
	\caption{Loglikelihood values obtained via simultaneous estimation of all six parameters (the $\mathcal{T}_1$-sequential estimation approach). The considered models all have D-vine structure (c) and combinations of Clayton (C), Gumbel (G) and Frank (F) copulas in $\mathcal{T}_1$. Frank copulas are taken in $\mathcal{T}_2$ and $\mathcal{T}_3$.}
	\label{Table:LogLikli_MastitisCombinationsModels1342}
	\begin{tabular}{p{0.5cm}ccc}
		\cmidrule{1-4}
		& \multicolumn{3}{c}{Families in $\mathcal{T}_1$: fam$_{13}$--fam$_{34}$--fam$_{24}$} \\
		\hline\hline	
		\multirow{16}{*}{\rotatebox{90}{loglikelihood}}  & C--C--F & C--F--C & F--C--C\\
		& -127.31 (-127.58)  & -125.73 (-125.83) & -128.67 (-128.78)\\
		\cmidrule{2-4}
		& C--F--F& F--C--F & F--F--C \\
		& -125.39 (-125.49) & -127.81 (-127.90) & -125.66 (-125.77)\\
		\cmidrule{2-4}	
		& C--C--G & C--G--C & G--C--C\\
		& -141.21 (-141.28) & -137.64 (-137.66) & -145.48 (-145.56) \\
		\cmidrule{2-4}
		& C--G--G & G--C--G & G--G--C \\	
		& -138.50 (-138.53) & -155.25 (-155.28) & -143.59 (-143.61)\\
		\cmidrule{2-4}
		& C--G--F & C--F--G & G--C--F\\
		& -135.04 (-135.08) & -137.86 (-138.14) & -144.30 (-144.37)\\
		\cmidrule{2-4}
		& F--C--G & G--F--C & F--G--C\\
		&  -141.29 (-141.36) & -141.28 (-141.47) & -136.04 (-136.08)\\
		\cmidrule{2-4}	
		& G--G--F & G--F--G & F--G--G \\ 
		&  -136.40 (-136.42) & -139.75 (-139.77) & -149.23 (-149.27)\\
		\cmidrule{2-4}
		& G--F--F & F--G--F &  F--F--G \\ &  
		-139.38 (-139.56) & -132.76 (-132.77) & -137.16 (-137.43) \\
		\hline\hline
	\end{tabular}			
\end{table}		
\renewcommand{\arraystretch}{1}

\begin{landscape}
	\begin{table}[ht]
		\centering	
		\footnotesize
		\caption{Estimated copula parameters, Kendall's $\tau$ values and tail-dependence coefficients for the four best models fitted to the mastitis data with underlying D-vine structure (c). Results for both the $\mathcal{T}_1$-sequential estimation approach and for joint estimation of all six parameters are shown. Standard errors are obtained using the bootstrap algorithm described in \autoref{Sec:VineCopBootAlg} and are given in parenthesis.}
		\label{Table:MastitisBestModels1342}
		\begin{tabular}{lcccccccc}
			\cmidrule{1-9}
			& \multicolumn{4}{c}{$\mathcal{T}_1$-sequential estimation} & \multicolumn{4}{c}{Global estimation}  \\
			& \multirow{2}{*}{logll} & \multirow{2}{*}{Parameter} & \multirow{2}{*}{Kendall's $\tau$} & Lower tail & \multirow{2}{*}{logll} & \multirow{2}{*}{Parameter} & \multirow{2}{*}{Kendall's $\tau$} & Lower tail \\
			& &  & & dependence & & &  & dependence\\
			\hline
			\cmidrule{1-9}	
			F; $\hat{\theta}_{13}$ & & 6.38 (0.81) & 0.53 (0.04) & -- & & 6.56 (0.80) & 0.54 (0.04) & -- \\
			F; $\hat{\theta}_{34}$ & & 6.34 (0.79)  & 0.53 (0.04) &  -- & & 6.34 (0.75) & 0.53 (0.04) &  --\\
			F; $\hat{\theta}_{24}$ & \multirow{2}{*}{-124.82} & 6.77 (0.80) & 0.55 (0.04) &   -- & \multirow{2}{*}{-124.70} & 6.99 (0.77) & 0.56 (0.03) &\\
			F; $\hat{\theta}_{14;3}$ &  & 1.67 (0.57) & 0.18 (0.06) &   -- &  & 1.68 (0.55) & 0.18 (0.06) &\\
			F; $\hat{\theta}_{23;4}$ & & 2.81 (0.57) & 0.29 (0.05) &  -- & & 2.79 (0.55) & 0.29 (0.05) &  --\\
			F; $\hat{\theta}_{12;34}$ & & 3.72 (0.63) & 0.37 (0.05) &  -- & & 3.71 (0.65) & 0.37 (0.05) &  --\\
			\cmidrule{1-9}
			C; $\hat{\theta}_{13}$ & & 3.60 (0.58) & 0.64 (0.04) &  0.82 (0.03) & & 3.78 (0.58) & 0.65 (0.04) & 0.83 (0.02) \\
			F; $\hat{\theta}_{34}$ & & 6.34 (0.79) & 0.53 (0.04) & -- & & 6.39 (0.75) & 0.53 (0.04) & --\\
			F; $\hat{\theta}_{24}$ & \multirow{2}{*}{-125.49} & 6.77 (0.79) & 0.55 (0.04) & -- & \multirow{2}{*}{-125.39} & 6.93 (0.74) & 0.56 (0.03) & --\\
			F; $\hat{\theta}_{14;3}$ &  & 1.49 (0.58) & 0.16 (0.06) & -- &  & 1.51 (0.53) & 0.16 (0.05) & --\\
			F; $\hat{\theta}_{23;4}$ & & 2.81 (0.53) & 0.29 (0.05) & -- & & 2.78 (0.51) & 0.29 (0.05) & --\\
			F; $\hat{\theta}_{12;34}$ & & 3.48 (0.63) & 0.35 (0.05) & -- & & 3.48 (0.61) & 0.35 (0.05) & --\\
			\cmidrule{1-9}
			F; $\hat{\theta}_{13}$ &  & 6.38 (0.81) &  0.53 (0.04) & -- &  & 6.51 (0.79) & 0.54 (0.04) & --\\
			F; $\hat{\theta}_{34}$ &  & 6.34 (0.79) &  0.53 (0.04) & --  &  & 6.36 (0.72) & 0.53 (0.04) & --\\
			C; $\hat{\theta}_{24}$ & \multirow{2}{*}{-125.77} & 3.90 (0.60) & 0.66 (0.03) & 0.84 (0.02) & \multirow{2}{*}{-125.66} & 4.10 (0.61) & 0.67 (0.03) & 0.84 (0.02) \\
			F; $\hat{\theta}_{14;3}$ & & 1.54 (0.55) & 0.17 (0.06) & -- & & 1.57 (0.55) & 0.17 (0.06) & --\\
			F; $\hat{\theta}_{23;4}$ &  & 2.76 (0.55) & 0.29 (0.05) & --  &  & 2.79 (0.55) & 0.29 (0.05) & --\\
			F; $\hat{\theta}_{12;34}$ & & 3.86 (0.64) & 0.38 (0.05) & -- & & 3.86 (0.65) & 0.38 (0.05) & --\\
			\cmidrule{1-9}
			C; $\hat{\theta}_{13}$ &  & 3.60 (0.58) & 0.64 (0.04) & 0.82 (0.03) &  & 3.75 (0.61) & 0.65 (0.04) & 0.83 (0.03) \\
			F; $\hat{\theta}_{34}$ &  & 6.34 (0.79) & 0.53 (0.04) & -- &  & 6.40 (0.74) & 0.53 (0.04) & --\\
			C; $\hat{\theta}_{24}$ & \multirow{2}{*}{-125.83} & 3.90 (0.59) & 0.66 (0.03) & 0.84 (0.02) & \multirow{2}{*}{-125.73} & 4.04 (0.59) & 0.67 (0.03) & 0.84 (0.02) \\
			F; $\hat{\theta}_{14;3}$ & & 1.36 (0.57) & 0.15 (0.06) & -- & & 1.39 (0.55) & 0.15 (0.06) & --\\
			F; $\hat{\theta}_{23;4}$ &  & 2.71 (0.53) & 0.28 (0.05) & -- &  & 2.72 (0.51) & 0.28 (0.05) & --\\
			F; $\hat{\theta}_{12;34}$ &  & 3.70 (0.64) & 0.37 (0.05) & -- &  & 3.71 (0.63) & 0.37 (0.05) & --\\	
			\cmidrule{1-9}
			\hline	
		\end{tabular}	
	\end{table}	
	\renewcommand{\arraystretch}{1}
\end{landscape}

\normalsize
\section{Discussion}\label{Sec:Conclusion}
In this paper, we investigate likelihood based inference for clustered right-censored event times using vine copulas. Prior to this work, vine theory has only been developed for complete data. The estimation procedure is conducted in two subsequent steps (two-stage approach). First, the marginal distributions are estimated considering standard parametric and nonparametric estimation techniques for univariate right-censored data. Second, the dependence structure is modeled. \autoref{theo:D-vineDerivatives} and \autoref{Corollary:D-vineDerivatives} (see \ref{Sec:PartDeriv}) provide the likelihood contributions for right-censored quadruple data in terms of vine copula components. For right-censored trivariate data a simulation study gives evidence that the presented estimators are on target. Several R-vine models are fitted to the four-dimensional mastitis data using both a full and a sequential  estimation approach. The results qualify the latter as the preferable estimation technique in practice. It provides comparable estimation results while significantly simplifying the numerically challenging optimization problem. Our findings for the mastitis data are in line with \citet{Geerdens2014}, where the Joe-Hu family is used for flexible dependence modeling in right-censored event time data. Both methods, the one based on vine copulas as well as the one based on Joe-Hu copulas, stress the need for more flexible copula models as compared to less elaborated ones such as exchangeable (EAC) and nested Archimedean copulas (NAC). For all models, the data complexity due to right-censoring makes the statistical analysis of multivariate event time data highly challenging with regard to numerical demand and computational manageability.

Having the basic methodology at hand provides room for further research on the use of vine copulas in the presence of censoring. Often a data set includes one or more covariates. In the context of copula models a covariate can affect the survival margins and/or the dependence structure. If the covariate is at the level of the cluster and only takes a few values, the data set can be split into several subsets and the proposed copula modeling can be used for each subset separately. If the covariate at the level of the cluster is continuous, then one can model the copula parameters as a function of the covariate (e.g.\ linear) and further proceed as in this paper. If a covariate is not at the level of the cluster it is not possible to discuss its impact on the association and the covariate can only be included in the margins by using e.g.\ a Cox model in the first estimation step. Nonparametric marginal estimation is more involved when covariates are present. An option is to apply the Beran estimator or an extended version of it \citep{beran1981nonparametric}. Using ideas from this paper, we currently look at applications for recurrent data. The first results of this ongoing project look promising and will be reported in an upcoming manuscript.

\section*{Computational aspects}
All computations were conducted on a customary Windows 7 Lenovo laptop with Intel(R) Core(TM) i5-3320M CPU @ 2.60 GHz and 8 GB RAM.

For scenarios in the simulation study with a sample size of 500, the computation time for one loglikelihood optimization, i.e.\ for one replication, ranges from on average 0.15 seconds for complete data to 3.2 minutes in case of 65\% censoring. This observation is due to an increasing amount of integrals which need to be evaluated with an increasing percentage of censored observations. For censored data, the three estimators using known margins, parametrically and nonparametrically estimated margins show comparable performance in terms of computation time.

For the mastitis data, the loglikelihood function for a given vine copula model and therewith its optimization highly depends on the underlying R-vine structure. The latter determines, based on the censoring pattern within the clusters, the number of double integrals that need to be evaluated in each iteration step. The complexity of the double integrals on the other hand depends on the bivariate building blocks of the vine copula model. The complete loglikelihood optimization, i.e.\ finding starting values via the $\mathcal{T}_1$-sequential estimation approach and subsequent full optimization with respect to all parameters, takes on average 3 hours for (in terms of computational complexity) moderate loglikelihood expressions. Calculations for numerically highly complex loglikelihood functions took up to 2 days.

\section*{Acknowledgements}

The authors wish to thank Dr.\ H.\ Laevens (Catholic University College Sint-Lieven, Sint-
Niklaas, Belgium) for the permission to use the mastitis data as a real data example for this research and Dr.\ Ulf Schepsmeier for discussions on early drafts of the paper. Also, we thank the Co-Editor, the Associate Editor and, last but not least, the two reviewers. Based on their constructive comments and good suggestions we could further improve the content of our manuscript as well as the presentation.

Parts of the numerical calculations were performed on a Linux cluster supported by
DFG grant INST 95/919-1 FUGG.

Funding: This work was supported by the  Deutsche Forschungsgemeinschaft [DFG CZ 86/4-1]; the Research Foundation Flanders (FWO), Scientific Research Community [W000817N];
	and the Interuniversity Attraction Poles Programme [IAP-network P7/06], Belgian Science Policy Office.

\bibliographystyle{elsarticle-harv} 
\bibliography{References}

\appendix
\section{Partial derivatives of a four-dimensional D-vine}\label{Sec:PartDeriv}	

\begin{Theorem}
	For the copula density \eqref{Eq:D-vineDensity4} the following holds:
	\normalsize
	\begin{enumerate}
		\item[1.\phantom{(a)}] $\begin{aligned}[t] \mathbb{C}\left(u_1,u_2,u_3,u_4\right)\hspace{-1cm}\\
		= & \  \int_{0}^{u_2} \int_{0}^{u_3} \scd_{23}\left(v_2, v_3\right) \mathbb{C}_{14;23}\{\mathbb{C}_{1|23}\left(u_1|v_2, v_3\right), \mathbb{C}_{4|23}\left(u_4|v_2, v_3\right) \}dv_3 dv_2\end{aligned}$
		\item[2.(a)]$\begin{aligned}[t] \frac{\partial \mathbb{C}\left(u_1,u_2,u_3,u_4\right)}{\partial u_1}\hspace{-1.3cm}\\
		= & \  \int_{0}^{u_2} \int_{0}^{u_3} \scd_{12}\left(u_1, v_2\right)\scd_{23}\left(v_2, v_3\right)\scd_{13;2}\{\mathbb{C}_{1|2}\left(u_1|v_2\right), \mathbb{C}_{3|2}\left(v_3|v_2\right)\}\\
		& \times \frac{\partial}{\partial \tilde{u}_1} \mathbb{C}_{14;23}\{\tilde{u}_1, \mathbb{C}_{4|23}\left(u_4|v_2, v_3\right)\}\bigg \vert_{\tilde{u}_1 = \mathbb{C}_{1|23}\left(u_1|v_2, v_3\right)} dv_3 dv_2\end{aligned}$
		\item[\phantom{4.}(b)] $\begin{aligned}[t] \frac{\partial \mathbb{C}\left(u_1,u_2,u_3,u_4\right)}{\partial u_2}\hspace{-1.3cm}\\
		= & \   \int_{0}^{u_3} \scd_{23}\left(u_2, v_3\right) \mathbb{C}_{14;23}\{\mathbb{C}_{1|23}\left(u_1|u_2, v_3\right), \mathbb{C}_{4|23}\left(u_4|u_2, v_3\right) \}dv_3 \end{aligned}$
		\item[\phantom{4.}(c)] $\begin{aligned}[t] \frac{\partial \mathbb{C}\left(u_1,u_2,u_3,u_4\right)}{\partial u_3}\hspace{-1.3cm}\\
		= & \  \int_{0}^{u_2} \scd_{23}\left(v_2, u_3\right) \mathbb{C}_{14;23}\{\mathbb{C}_{1|23}\left(u_1|v_2, u_3\right), \mathbb{C}_{4|23}\left(u_4|v_2, u_3\right) \} dv_2\end{aligned}$
		\item[\phantom{4.}(d)] $\begin{aligned}[t] \frac{\partial \mathbb{C}\left(u_1,u_2,u_3,u_4\right)}{\partial u_4}\hspace{-1.3cm}\\
		= & \  \int_{0}^{u_2} \int_{0}^{u_3} \scd_{23}\left(v_2, v_3\right)\scd_{34}\left(v_3, u_4\right)\scd_{24;3}\{\mathbb{C}_{2|3}\left(v_2|v_3\right), \mathbb{C}_{4|3}\left(u_4|v_3\right)\}\\
		& \times \frac{\partial}{\partial \tilde{u}_4} \mathbb{C}_{14;23}\{\mathbb{C}_{1|23}\left(u_1|v_2, v_3\right), \tilde{u}_4\}\bigg \vert_{\tilde{u}_4 = \mathbb{C}_{4|23}\left(u_4|v_2, v_3\right)} dv_3 dv_2\end{aligned}$
		\item[3.(a)]$\begin{aligned}[t] \frac{\partial^2 \mathbb{C}\left(u_1,u_2,u_3,u_4\right)}{\partial u_1 \partial u_2}\hspace{-1.4cm}\\
		= & \  \int_{0}^{u_3} \scd_{12}\left(u_1, u_2\right)\scd_{23}\left(u_2, v_3\right)\scd_{13;2}\{\mathbb{C}_{1|2}\left(u_1|u_2\right), \mathbb{C}_{3|2}\left(v_3|u_2\right)\}\\
		& \times \frac{\partial}{\partial \tilde{u}_1} \mathbb{C}_{14;23}\{\tilde{u}_1, \mathbb{C}_{4|23}\left(u_4|u_2, v_3\right)\}\bigg \vert_{\tilde{u}_1 = \mathbb{C}_{1|23}\left(u_1|u_2, v_3\right)} dv_3\end{aligned}$
		\item[\phantom{4.}(b)] $\begin{aligned}[t] \frac{\partial^2 \mathbb{C}\left(u_1,u_2,u_3,u_4\right)}{\partial u_1 \partial u_3}\hspace{-1.4cm}\\
		= & \  \int_{0}^{u_2} \scd_{12}\left(u_1, v_2\right)\scd_{23}\left(v_2, u_3\right)\scd_{13;2}\{\mathbb{C}_{1|2}\left(u_1|v_2\right), \mathbb{C}_{3|2}\left(u_3|v_2\right)\}\\
		& \times \frac{\partial}{\partial \tilde{u}_1} \mathbb{C}_{14;23}\{\tilde{u}_1, \mathbb{C}_{4|23}\left(u_4|v_2, u_3\right)\}\bigg \vert_{\tilde{u}_1 = \mathbb{C}_{1|23}\left(u_1|v_2, u_3\right)} dv_2\end{aligned}$
		\item[\phantom{4.}(c)] $\begin{aligned}[t] \frac{\partial^2 \mathbb{C}\left(u_1,u_2,u_3,u_4\right)}{\partial u_1 \partial u_4}\hspace{-1.4cm}\\
		= & \  \int_{0}^{u_2} \int_{0}^{u_3} \scd_{12}\left(u_1, v_2\right)\scd_{23}\left(v_2, v_3\right)\scd_{34}\left(v_3, u_4\right)\\
		& \times \scd_{13;2}\{\mathbb{C}_{1|2}\left(u_1|v_2\right), \mathbb{C}_{3|2}\left(v_3|v_2\right)\}\scd_{24;3}\{\mathbb{C}_{2|3}\left(v_2|v_3\right), \mathbb{C}_{4|3}\left(u_4|v_3\right)\}\\
		& \times \scd_{14;23}\{\mathbb{C}_{1|23}\left(u_1|v_2, v_3\right),\mathbb{C}_{4|23}\left(u_4|v_2, v_3\right)\}dv_3 dv_2\end{aligned}$
		\item[\phantom{4.}(d)] $\begin{aligned}[t] \frac{\partial^2 \mathbb{C}\left(u_1,u_2,u_3,u_4\right)}{\partial u_2 \partial u_3}\hspace{-1.4cm}\\
		= & \   \scd_{23}\left(u_2, u_3\right)\mathbb{C}_{14;23}\{\mathbb{C}_{1|23}\left(u_1|u_2, u_3\right), \mathbb{C}_{4|23}\left(u_4|u_2, u_3\right)\}\end{aligned}$
		\item[\phantom{4.}(e)] $\begin{aligned}[t] \frac{\partial^2 \mathbb{C}\left(u_1,u_2,u_3,u_4\right)}{\partial u_2 \partial u_4}\hspace{-1.4cm}\\
		= & \  \int_{0}^{u_3} \scd_{23}\left(u_2, v_3\right)\scd_{34}\left(v_3, u_4\right)\scd_{24;3}\{\mathbb{C}_{2|3}\left(u_2|v_3\right), \mathbb{C}_{4|3}\left(u_4|v_3\right)\}\\
		& \times \frac{\partial}{\partial \tilde{u}_4} \mathbb{C}_{14;23}\{\mathbb{C}_{1|23}\left(u_1|u_2, v_3\right), \tilde{u}_4\}\bigg \vert_{\tilde{u}_4 = \mathbb{C}_{4|23}\left(u_4|u_2, v_3\right)} dv_3\end{aligned}$
		\item[\phantom{4.}(f)] $\begin{aligned}[t] \frac{\partial^2 \mathbb{C}\left(u_1,u_2,u_3,u_4\right)}{\partial u_3 \partial u_4}\hspace{-1.4cm}\\
		= & \   \int_{0}^{u_2} \scd_{23}\left(v_2, u_3\right)\scd_{34}\left(u_3, u_4\right)\scd_{24;3}\{\mathbb{C}_{2|3}\left(v_2|u_3\right), \mathbb{C}_{4|3}\left(u_4|u_3\right)\}\\
		& \times \frac{\partial}{\partial \tilde{u}_4} \mathbb{C}_{14;23}\{\mathbb{C}_{1|23}\left(u_1|v_2, u_3\right), \tilde{u}_4\}\bigg \vert_{\tilde{u}_4 = \mathbb{C}_{4|23}\left(u_4|v_2, u_3\right)} dv_2\end{aligned}$
		\item[4.(a)]$\begin{aligned}[t] \frac{\partial^3 \mathbb{C}\left(u_1,u_2,u_3,u_4\right)}{\partial u_1 \partial u_2 \partial u_3} \hspace{-1.4cm}\\
		= & \   \scd_{12}\left(u_1, u_2\right)\scd_{23}\left(u_2, u_3\right)\scd_{13;2}\{\mathbb{C}_{1|2}\left(u_1|u_2\right), \mathbb{C}_{3|2}\left(u_3|u_2\right)\}\\
		& \times \frac{\partial}{\partial \tilde{u}_1} \mathbb{C}_{14;23}\{\tilde{u}_1, \mathbb{C}_{4|23}\left(u_4|u_2, u_3\right)\}\bigg \vert_{\tilde{u}_1 = \mathbb{C}_{1|23}\left(u_1|u_2, u_3\right)}\end{aligned}$
		\item[\phantom{4.}(b)] $\begin{aligned}[t] \frac{\partial^3 \mathbb{C}\left(u_1,u_2,u_3,u_4\right)}{\partial u_1 \partial u_2 \partial u_4}\hspace{-1.4cm}\\
		= & \   \int_{0}^{u_3} \scd_{12}\left(u_1, u_2\right)\scd_{23}\left(u_2, v_3\right)\scd_{34}\left(v_3, u_4\right)\\
		& \times \scd_{13;2}\{\mathbb{C}_{1|2}\left(u_1|u_2\right), \mathbb{C}_{3|2}\left(v_3|u_2\right)\}\scd_{24;3}\{\mathbb{C}_{2|3}\left(u_2|v_3\right), \mathbb{C}_{4|3}\left(u_4|v_3\right)\}\\
		& \times \scd_{14;23}\{\mathbb{C}_{1|23}\left(u_1|u_2, v_3\right),\mathbb{C}_{4|23}\left(u_4|u_2, v_3\right)\} dv_3\end{aligned}$
		\item[\phantom{4.}(c)] $\begin{aligned}[t] \frac{\partial^3 \mathbb{C}\left(u_1,u_2,u_3,u_4\right)}{\partial u_1 \partial u_3 \partial u_4}\hspace{-1.4cm}\\
		= & \   \int_{0}^{u_2} \scd_{12}\left(u_1, v_2\right)\scd_{23}\left(v_2, u_3\right)\scd_{34}\left(u_3, u_4\right)\\
		& \times \scd_{13;2}\{\mathbb{C}_{1|2}\left(u_1|v_2\right), \mathbb{C}_{3|2}\left(u_3|v_2\right)\}\scd_{24;3}\{\mathbb{C}_{2|3}\left(v_2|u_3\right), \mathbb{C}_{4|3}\left(u_4|u_3\right)\}\\
		& \times \scd_{14;23}\{\mathbb{C}_{1|23}\left(u_1|v_2, u_3\right), \mathbb{C}_{4|23}\left(u_4|v_2, u_3\right)\} dv_2\end{aligned}$
		\item[\phantom{4.}(d)] $\begin{aligned}[t] \frac{\partial^3 \mathbb{C}\left(u_1,u_2,u_3,u_4\right)}{\partial u_2 \partial u_3 \partial u_4} \hspace{-1.4cm}\\
		= & \    \scd_{23}\left(u_2, u_3\right)\scd_{34}\left(u_3, u_4\right)\scd_{24;3}\{\mathbb{C}_{2|3}\left(u_2|u_3\right), \mathbb{C}_{4|3}\left(u_4|u_3\right)\}\\
		& \times \frac{\partial}{\partial \tilde{u}_4} \mathbb{C}_{14;23}\{\mathbb{C}_{1|23}\left(u_1|u_2, u_3\right), \tilde{u}_4\}\bigg \vert_{\tilde{u}_4 = \mathbb{C}_{4|23}\left(u_4|u_2, u_3\right)}\end{aligned}$
		\item[5.\phantom{(a)}] $\begin{aligned}[t] \scd\left(u_1, u_2, u_3, u_4\right)\hspace{-0.9cm}\\
		= & \   \scd_{12}\left(u_1, u_2\right)\scd_{23}\left(u_2, u_3\right)\scd_{34}\left(u_3, u_4\right)\scd_{13;2}\{\mathbb{C}_{1|2}\left(u_1|u_2\right), \mathbb{C}_{3|2}\left(u_3|u_2\right)\}\\
		& \times \scd_{24;3}\{\mathbb{C}_{2|3}\left(u_2|u_3\right), \mathbb{C}_{4|3}\left(u_4|u_3\right)\}\scd_{14;23}\{\mathbb{C}_{1|23}\left(u_1|u_2, u_3\right), \mathbb{C}_{4|23}\left(u_4|u_2, u_3\right)\}\end{aligned}$
	\end{enumerate}
	\label{theo:D-vineDerivatives}
\end{Theorem}
\normalsize

\newpage
\begin{Corollary} \leavevmode 
	In terms of h-functions, for the copula density \eqref{Eq:D-vineDensity4} the following holds:
	\normalsize
	\begin{enumerate}
		\item[1.\phantom{(a)}] $\begin{aligned}[t] \mathbb{C}\left(u_1,u_2,u_3,u_4\right)\hspace{-1.5cm}\\
		= & \  \int_{0}^{u_2} \int_{0}^{u_3} \mathbb{C}_{14;23}\left[h_{1|3;2}\{h_{1|2}\left(u_1|v_2\right)\big\vert h_{3|2}\left(v_3|v_2\right)\}, h_{4|2;3}\{h_{4|3}\left(u_4|v_3\right)\big\vert h_{2|3}\left(v_2|v_3\right)\}\right]\\
		& \times \scd_{23}\left(v_2,v_3\right)dv_3 dv_2 \end{aligned}$
		\item[2.(a)]  $\begin{aligned}[t] \frac{\partial \mathbb{C}\left(u_1,u_2,u_3,u_4\right)}{\partial u_1}\hspace{-1.8cm}\\
		= & \  \int_{0}^{u_2} \int_{0}^{u_3} \scd_{12}\left(u_1, v_2\right)\scd_{23}\left(v_2, v_3\right)\scd_{13;2}\{h_{1|2}\left(u_1|v_2\right), h_{3|2}\left(v_3|v_2\right)\}\\
		& \times h_{4|1;23}\left[h_{4|2;3}\{h_{4|3}\left(u_4|v_3\right)\big\vert h_{2|3}\left(v_2|v_3\right)\}\bigg\vert h_{1|3;2}\{h_{1|2}\left(u_1|v_2\right)\big\vert h_{3|2}\left(v_3|v_2\right)\}\right]dv_3 dv_2\end{aligned}$
		\item[\phantom{2.}(b)] $\begin{aligned}[t] \frac{\partial \mathbb{C}\left(u_1,u_2,u_3,u_4\right)}{\partial u_2}\hspace{-1.8cm}\\
		= & \  \int_{0}^{u_3 }\mathbb{C}_{14;23}\left[h_{1|3;2}\{h_{1|2}\left(u_1|u_2\right)\big\vert h_{3|2}\left(v_3|u_2\right)\}, h_{4|2;3}\{h_{4|3}\left(u_4|v_3\right)\big\vert h_{2|3}\left(u_2|v_3\right)\}\right]\\
		& \times \scd_{23}\left(u_2,v_3\right) dv_3 \end{aligned}$
		\item[\phantom{2.}(c)] $\begin{aligned}[t] \frac{\partial \mathbb{C}\left(u_1,u_2,u_3,u_4\right)}{\partial u_3}\hspace{-1.8cm}\\
		= & \  \int_{0}^{u_2} \mathbb{C}_{14;23}\left[h_{1|3;2}\{h_{1|2}\left(u_1|v_2\right)\big\vert h_{3|2}\left(u_3|v_2\right)\}, h_{4|2;3}\{h_{4|3}\left(u_4|u_3\right)\big\vert h_{2|3}\left(v_2|u_3\right)\}\right]\\
		& \times \scd_{23}\left(v_2,u_3\right) dv_2 \end{aligned}$
		\item[\phantom{2.}(d)] $\begin{aligned}[t] \frac{\partial \mathbb{C}\left(u_1,u_2,u_3,u_4\right)}{\partial u_4}\hspace{-1.8cm}\\
		= & \  \int_{0}^{u_2} \int_{0}^{u_3} \scd_{23}\left(v_2, v_3\right)\scd_{34}\left(v_3, u_4\right)\scd_{24;3}\{h_{2|3}\left(v_2|v_3\right),h_{4|3}\left(u_4|v_3\right)\}\\
		& \times h_{1|4;23}\left[h_{1|3;2}\{h_{1|2}\left(u_1|v_2\right)\big\vert h_{3|2}\left(v_3|v_2\right)\}\bigg\vert  h_{4|2;3}\{h_{4|3}\left(u_4|v_3\right)\big\vert h_{2|3}\left(v_2|v_3\right)\}\right]dv_3 dv_2\end{aligned}$
		\item[3.(a)] $\begin{aligned}[t] \frac{\partial^2 \mathbb{C}\left(u_1,u_2,u_3,u_4\right)}{\partial u_1 \partial u_2}\hspace{-1.9cm}\\
		= & \  \int_{0}^{u_3} \scd_{12}\left(u_1, u_2\right)\scd_{23}\left(u_2, v_3\right)\scd_{13;2}\{h_{1|2}\left(u_1|u_2\right),h_{3|2}\left(v_3|u_2\right)\}\\
		& \times  h_{4|1;23}\left[h_{4|2;3}\{h_{4|3}\left(u_4|v_3\right)\big\vert h_{2|3}\left(u_2|v_3\right)\}\bigg\vert h_{1|3;2}\{h_{1|2}\left(u_1|u_2\right)\big\vert h_{3|2}\left(v_3|u_2\right)\}\right]dv_3\end{aligned}$
		\item[\phantom{3.}(b)] $\begin{aligned}[t] \frac{\partial^2 \mathbb{C}\left(u_1,u_2,u_3,u_4\right)}{\partial u_1 \partial u_3}\hspace{-1.9cm}\\
		= & \  \int_{0}^{u_2} \scd_{12}\left(u_1, v_2\right)\scd_{23}\left(v_2, u_3\right)\scd_{13;2}\{h_{1|2}\left(u_1|v_2\right),h_{3|2}\left(u_3|v_2\right)\}\\
		& \times  h_{4|1;23}\left[h_{4|2;3}\{h_{4|3}\left(u_4|u_3\right)\big\vert h_{2|3}\left(v_2|u_3\right)\}\bigg\vert h_{1|3;2}\{h_{1|2}\left(u_1|v_2\right)\big\vert h_{3|2}\left(u_3|v_2\right)\}\right]dv_2\end{aligned}$
		\item[\phantom{3.}(c)] $\begin{aligned}[t] \frac{\partial^2 \mathbb{C}\left(u_1,u_2,u_3,u_4\right)}{\partial u_1 \partial u_4}\hspace{-1.9cm}\\
		= & \  \int_{0}^{u_2} \int_{0}^{u_3} \scd_{12}\left(u_1, v_2\right)\scd_{23}\left(v_2, v_3\right)\scd_{34}\left(v_3, u_4\right)\\
		& \times \scd_{13;2}\{h_{1|2}\left(u_1|v_2\right),h_{3|2}\left(v_3|v_2\right)\}\scd_{24;3}\{h_{2|3}\left(v_2|v_3\right),h_{4|3}\left(u_4|v_3\right)\}\\
		& \times \scd_{14;23}\left[h_{1|3;2}\{h_{1|2}\left(u_1|v_2\right)\big\vert h_{3|2}\left(v_3|v_2\right)\}, h_{4|2;3}\{h_{4|3}\left(u_4|v_3\right)\big\vert h_{2|3}\left(v_2|v_3\right)\}\right]dv_3 dv_2\end{aligned}$
		\item[\phantom{3.}(d)] $\begin{aligned}[t] \frac{\partial^2 \mathbb{C}\left(u_1,u_2,u_3,u_4\right)}{\partial u_2 \partial u_3}\hspace{-1.9cm}\\
		= & \  \scd_{23}\left(u_2,u_3\right)\mathbb{C}_{14;23}\left[h_{1|3;2}\{h_{1|2}\left(u_1|u_2\right)\big\vert h_{3|2}\left(u_3|u_2\right)\}, h_{4|2;3}\{h_{4|3}\left(u_4|u_3\right)\big\vert h_{2|3}\left(u_2|u_3\right)\}\right]\end{aligned}$
		\item[\phantom{3.}(e)] $\begin{aligned}[t] \frac{\partial^2 \mathbb{C}\left(u_1,u_2,u_3,u_4\right)}{\partial u_2 \partial u_4}\hspace{-1.9cm}\\
		= & \  \int_{0}^{u_3} \scd_{23}\left(u_2, v_3\right)\scd_{34}\left(v_3, u_4\right)\scd_{24;3}\{h_{2|3}\left(u_2|v_3\right),h_{4|3}\left(u_4|v_3\right)\}\\
		& \times h_{1|4;23}\left[h_{1|3;2}\{h_{1|2}\left(u_1|u_2\right)\big\vert h_{3|2}\left(v_3|u_2\right)\}\bigg\vert  h_{4|2;3}\{h_{4|3}\left(u_4|v_3\right)\big\vert h_{2|3}\left(u_2|v_3\right)\}\right]dv_3\end{aligned}$
		\item[\phantom{3.}(f)] $\begin{aligned}[t] \frac{\partial^2 \mathbb{C}\left(u_1,u_2,u_3,u_4\right)}{\partial u_3 \partial u_4}\hspace{-1.9cm}\\
		= & \  \int_{0}^{u_2} \scd_{23}\left(v_2, u_3\right)\scd_{34}\left(u_3, u_4\right)\scd_{24;3}\{h_{2|3}\left(v_2|u_3\right),h_{4|3}\left(u_4|u_3\right)\}\\
		& \times h_{1|4;23}\left[h_{1|3;2}\{h_{1|2}\left(u_1|v_2\right)\big\vert h_{3|2}\left(u_3|v_2\right)\}\bigg\vert  h_{4|2;3}\{h_{4|3}\left(u_4|u_3\right)\big\vert h_{2|3}\left(v_2|u_3\right)\}\right]dv_2\end{aligned}$
		\item[4.(a)] $\begin{aligned}[t] \frac{\partial^3 \mathbb{C}\left(u_1,u_2,u_3,u_4\right)}{\partial u_1 \partial u_2 \partial u_3}\hspace{-1.9cm}\\
		= & \  \scd_{12}\left(u_1, u_2\right)\scd_{23}\left(u_2, u_3\right)\scd_{13;2}\{h_{1|2}\left(u_1|u_2\right),h_{3|2}\left(u_3|u_2\right)\}\\
		& \times h_{4|1;23}\left[h_{4|2;3}\{h_{4|3}\left(u_4|u_3\right)\big\vert h_{2|3}\left(u_2|u_3\right)\}\bigg\vert h_{1|3;2}\{h_{1|2}\left(u_1|u_2\right)\big\vert h_{3|2}\left(u_3|u_2\right)\}\right]\end{aligned}$
		\item[\phantom{4.}(b)] $\begin{aligned}[t] \frac{\partial^3 \mathbb{C}\left(u_1,u_2,u_3,u_4\right)}{\partial u_1 \partial u_2 \partial u_4}\hspace{-1.9cm}\\
		= & \  \int_{0}^{u_3} \scd_{14;23}\left[h_{1|3;2}\{h_{1|2}\left(u_1|u_2\right)\big\vert h_{3|2}\left(v_3|u_2\right)\}, h_{4|2;3}\{h_{4|3}\left(u_4|v_3\right)\big\vert h_{2|3}\left(u_2|v_3\right)\}\right]\\
		& \times \scd_{12}\left(u_1, u_2\right)\scd_{23}\left(u_2, v_3\right)\scd_{34}\left(v_3, u_4\right)\scd_{13;2}\{h_{1|2}\left(u_1|u_2\right),h_{3|2}\left(v_3|u_2\right)\}\\
		& \times \scd_{24;3}\{h_{2|3}\left(u_2|v_3\right),h_{4|3}\left(u_4|v_3\right)\}dv_3\end{aligned}$
		\item[\phantom{4.}(c)] $\begin{aligned}[t] \frac{\partial^3 \mathbb{C}\left(u_1,u_2,u_3,u_4\right)}{\partial u_1 \partial u_3 \partial u_4}\hspace{-1.9cm}\\
		= & \  \int_{0}^{u_2} \scd_{14;23}\left[h_{1|3;2}\{h_{1|2}\left(u_1|v_2\right)\big\vert h_{3|2}\left(u_3|v_2\right)\}, h_{4|2;3}\{h_{4|3}\left(u_4|u_3\right)\big\vert h_{2|3}\left(v_2|u_3\right)\}\right]\\
		& \times \scd_{12}\left(u_1, v_2\right)\scd_{23}\left(v_2, u_3\right)\scd_{34}\left(u_3, u_4\right)\scd_{13;2}\{h_{1|2}\left(u_1|v_2\right),h_{3|2}\left(u_3|v_2\right)\}\\
		& \times \scd_{24;3}\{h_{2|3}\left(v_2|u_3\right),h_{4|3}\left(u_4|u_3\right)\} dv_2\end{aligned}$
		\item[\phantom{4.}(d)] $\begin{aligned}[t] \frac{\partial^3 \mathbb{C}\left(u_1,u_2,u_3,u_4\right)}{\partial u_2 \partial u_3 \partial u_4}\hspace{-1.9cm}\\
		= & \  \scd_{23}\left(u_2, u_3\right)\scd_{34}\left(u_3, u_4\right)\scd_{24;3}\{h_{2|3}\left(u_2|u_3\right),h_{4|3}\left(u_4|u_3\right)\}\\
		& \times h_{1|4;23}\left[h_{1|3;2}\{h_{1|2}\left(u_1|u_2\right)\big\vert h_{3|2}\left(u_3|u_2\right)\}\bigg\vert  h_{4|2;3}\{h_{4|3}\left(u_4|u_3\right)\big\vert h_{2|3}\left(u_2|u_3\right)\}\right]\end{aligned}$
		\item[5.\phantom{(a)}] $\begin{aligned}[t] \scd\left(u_1, u_2, u_3, u_4\right)\hspace{-1.5cm}\\
		= & \ \scd_{12}\left(u_1, u_2\right)\scd_{23}\left(u_2, u_3\right)\scd_{34}\left(u_3, u_4\right)\\
		& \times \scd_{13;2}\{h_{1|2}\left(u_1|u_2\right),h_{3|2}\left(u_3|u_2\right)\}\scd_{24;3}\{h_{2|3}\left(u_2|u_3\right),h_{4|3}\left(u_4|u_3\right)\}\\
		& \times \scd_{14;23}\left[h_{1|3;2}\{h_{1|2}\left(u_1|u_2\right)\big\vert h_{3|2}\left(u_3|u_2\right)\}, h_{4|2;3}\{h_{4|3}\left(u_4|u_3\right)\big\vert h_{2|3}\left(u_2|u_3\right)\}\right]\\\end{aligned}$
	\end{enumerate}
	\label{Corollary:D-vineDerivatives}
\end{Corollary}
\normalsize	

\section{Vine copula bootstrap algorithm}\label{Sec:VineCopBootAlg}
Under common (univariate) right-censoring, as it is present e.g.\ in the mastitis data, standard errors for the estimated parameters of a vine copula can be obtained using a parametric bootstrap algorithm \citep{davison1997bootstrap, Massonnet2009}. The following steps are based on the procedure in \cite{Geerdens2014}:

	\begin{itemize}
		\item[] \textit{Step 1:}\\ Fit the vine copula model of interest to the copula data $\left(\widehat{u}_{ij},\delta_{ij}\right)$, $i=1,\ldots,n$ and $j=1,\ldots,4$, where $y_{ij} = \min\left(t_{ij},c_i\right)$, $\delta_{ij} = I(t_{ij}\leq c_{i})$ and $\widehat{u}_{ij} = \widehat{S}_j\left(y_{ij}\right)$ with $\widehat{S}_j$ the Kaplan-Meier estimate based on $\left(y_{ij},\delta_{ij}\right)$. Obtain the vector of copula parameter estimates $\widehat{\boldsymbol{\theta}}$, which maximizes the loglikelihood \eqref{Eq:logLik_uLevel}. 
		\bigskip
		\item[] \textit{Step 2:}\\ Obtain the Kaplan-Meier estimate $\widehat{G}$ of the censoring distribution $G$ based on the observations $\left(\max\left(y_{i1},y_{i2},y_{i3},y_{i4}\right),1-\delta_{i1}\delta_{i2}\delta_{i3}\delta_{i4}\right)$, $i=1,\ldots,n$. 
		\bigskip
		\item[] \textit{Step 3:}\\ Generate $B$ bootstrap samples in the following way: For $b=1,\ldots,B$, $i=1,\ldots,n$ and $j=1,\ldots,4$, 
		\begin{itemize}
			\item[] \textit{Step 3.1:}\\ sample vine copula data $\left(u^{(b)}_{i1},u^{(b)}_{i2},u^{(b)}_{i3},u^{(b)}_{i4}\right)$ from the fitted vine copula model with parameter vector $\widehat{\boldsymbol{\theta}}$.
			\bigskip
			\item[] \textit{Step 3.2:}\\ generate event times $\left(t^{(b)}_{i1},t^{(b)}_{i2},t^{(b)}_{i3},t_{i4}^{(b)}\right)$ via $t^{(b)}_{ij} = \widehat{S}^{-1}_j\left(u^{(b)}_{ij}\right)$. 
			\bigskip
			\item[] \textit{Step 3.3:}\\ generate $c^{(b)}_i$ from $\widehat{G}$.
			\bigskip
			\item[] \textit{Step 3.4:}\\ obtain observed data by setting $y^{(b)}_{ij} = \min\left(t^{(b)}_{ij}, c^{(b)}_i\right)$ and $\delta^{(b)}_{ij} = I\left(t^{(b)}_{ij} \leq c^{(b)}_i\right)$.
			\bigskip
			\item[] \textit{Step 3.5:}\\ set $\widehat{u}^{(b)}_{ij} = \widehat{S}^{(b)}_j\left(y^{(b)}_{ij}\right)$ with $\widehat{S}^{(b)}_j$ the Kaplan-Meier estimate based on $\left(y^{(b)}_{ij}, \delta^{(b)}_{ij}\right)$.
			\bigskip
			\item[] \textit{Step 3.6:}\\ given the bootstrap data $\left(\widehat{u}^{(b)}_{ij}, \delta^{(b)}_{ij}\right)$, fit the vine copula model of interest by maximizing the loglikelihood \eqref{Eq:logLik_uLevel} to obtain $\widehat{\boldsymbol{\theta}}^{(b)}$ for bootstrap sample $b$.
		\end{itemize}
		\item[]\textit{Step 4:}\\ Obtain the bootstrap standard errors using $\widehat{\boldsymbol{\theta}}^{(1)},\ldots,\widehat{\boldsymbol{\theta}}^{(B)}$.
	\end{itemize}

	\section{Supplementary material}		
	\subsection{Exploration of the mastitis data}\label{Sec:AppIllustrMastitis}
	\begin{figure}[H]
		\centering
		\caption{Kaplan-Meier estimates of the four udder quarters of the mastitis data illustrating the high censoring rate for all four marginals.}
		\label{fig:KM}
		\includegraphics[width=.4\linewidth]{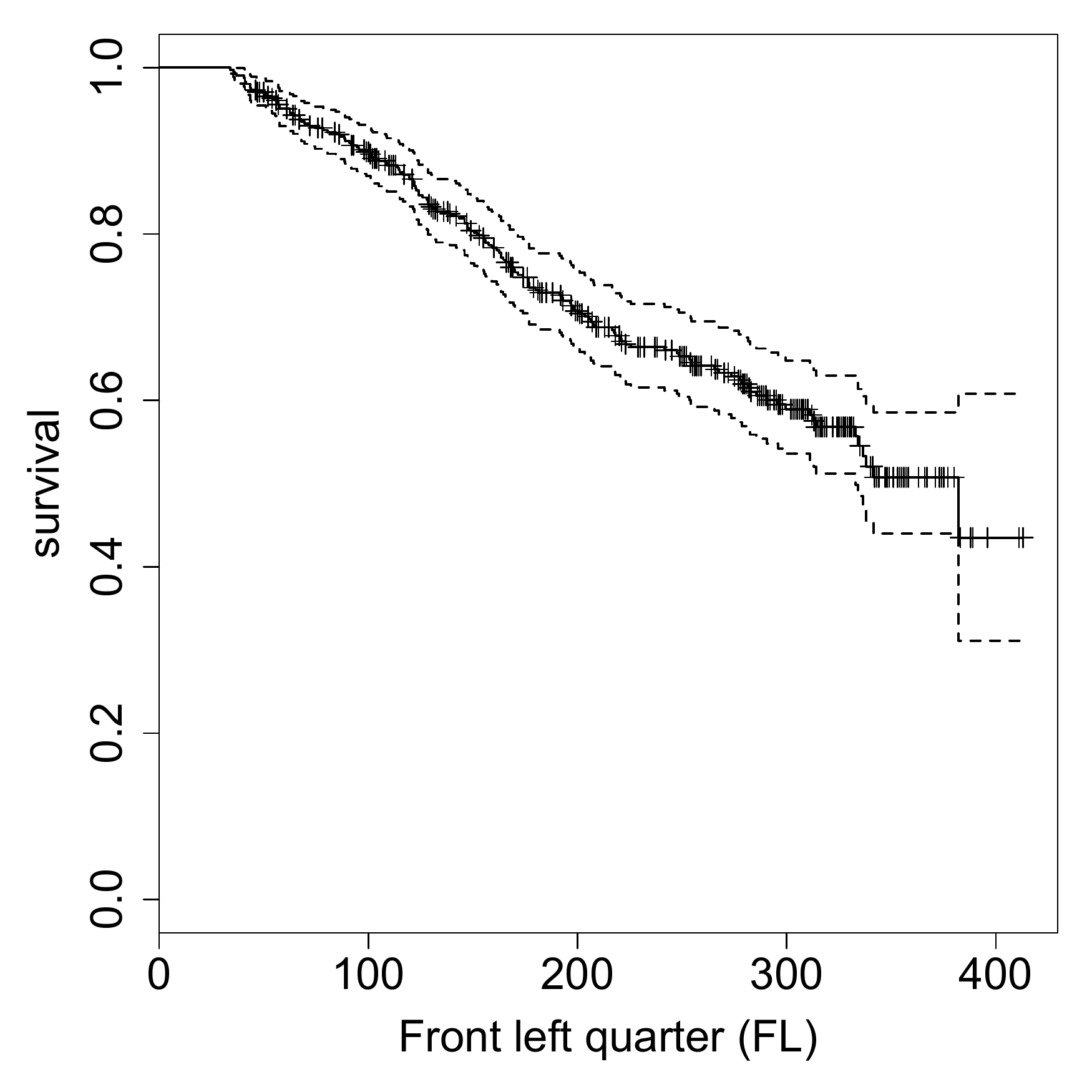}\includegraphics[width=.4\linewidth]{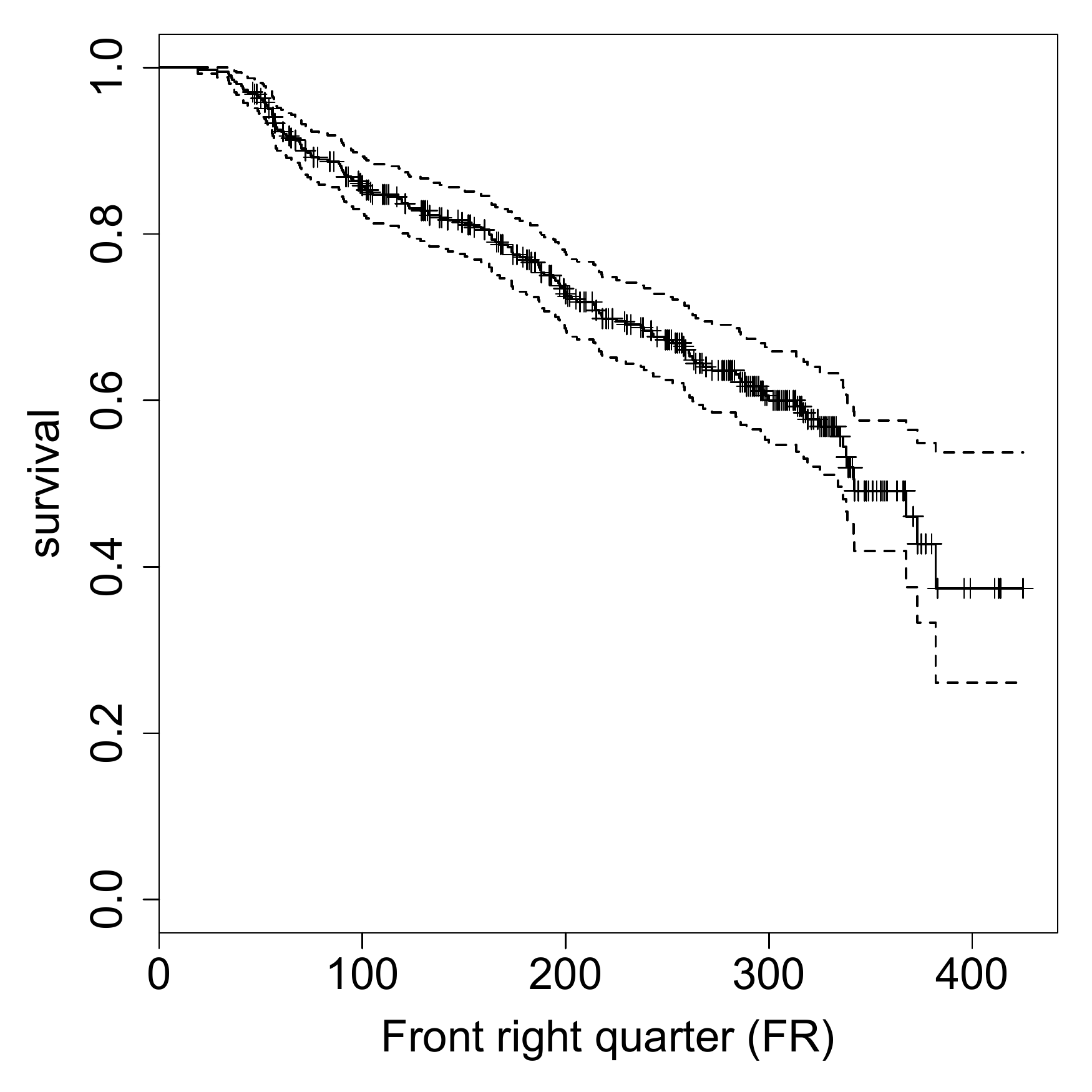}\\
		\includegraphics[width=.4\linewidth]{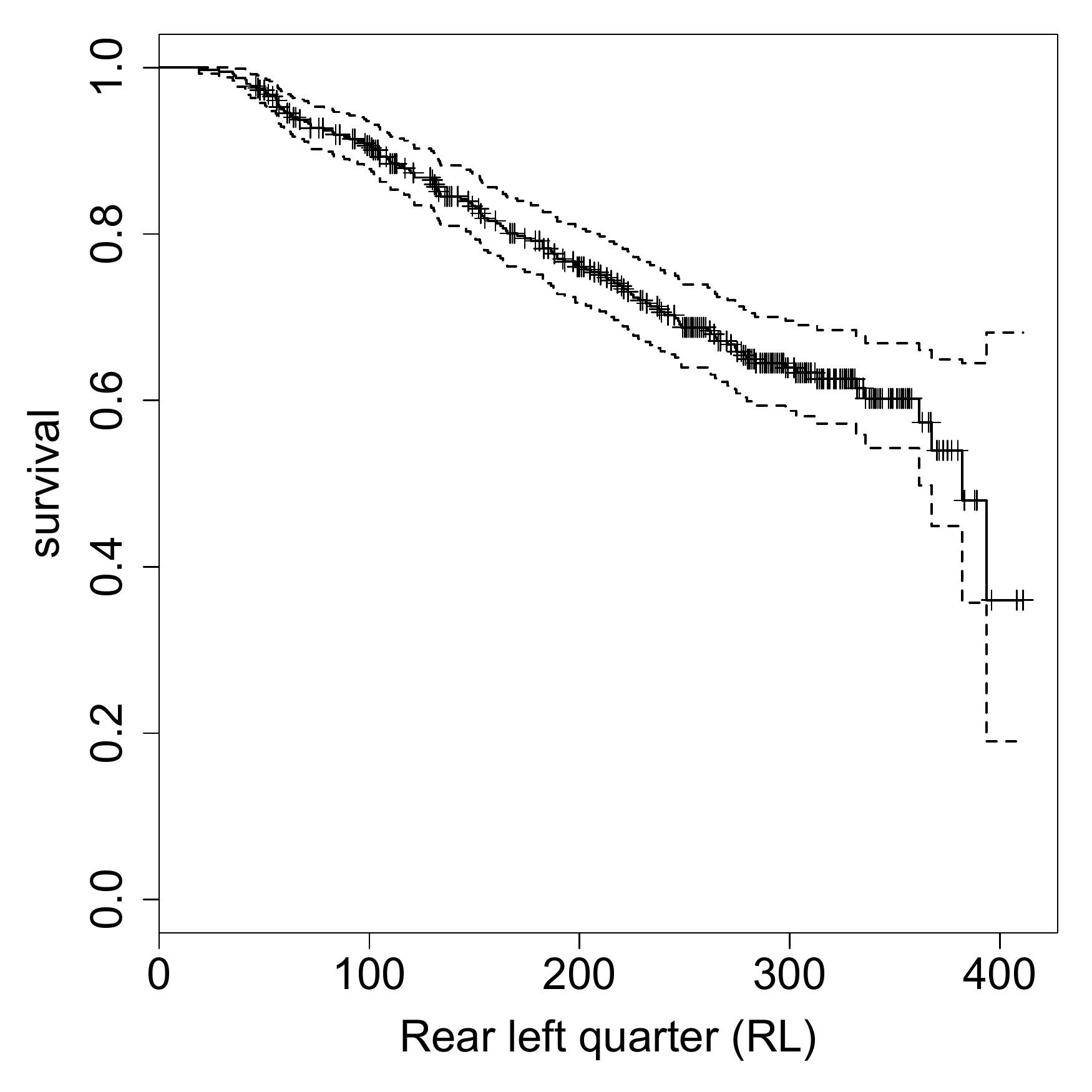}\includegraphics[width=.4\linewidth]{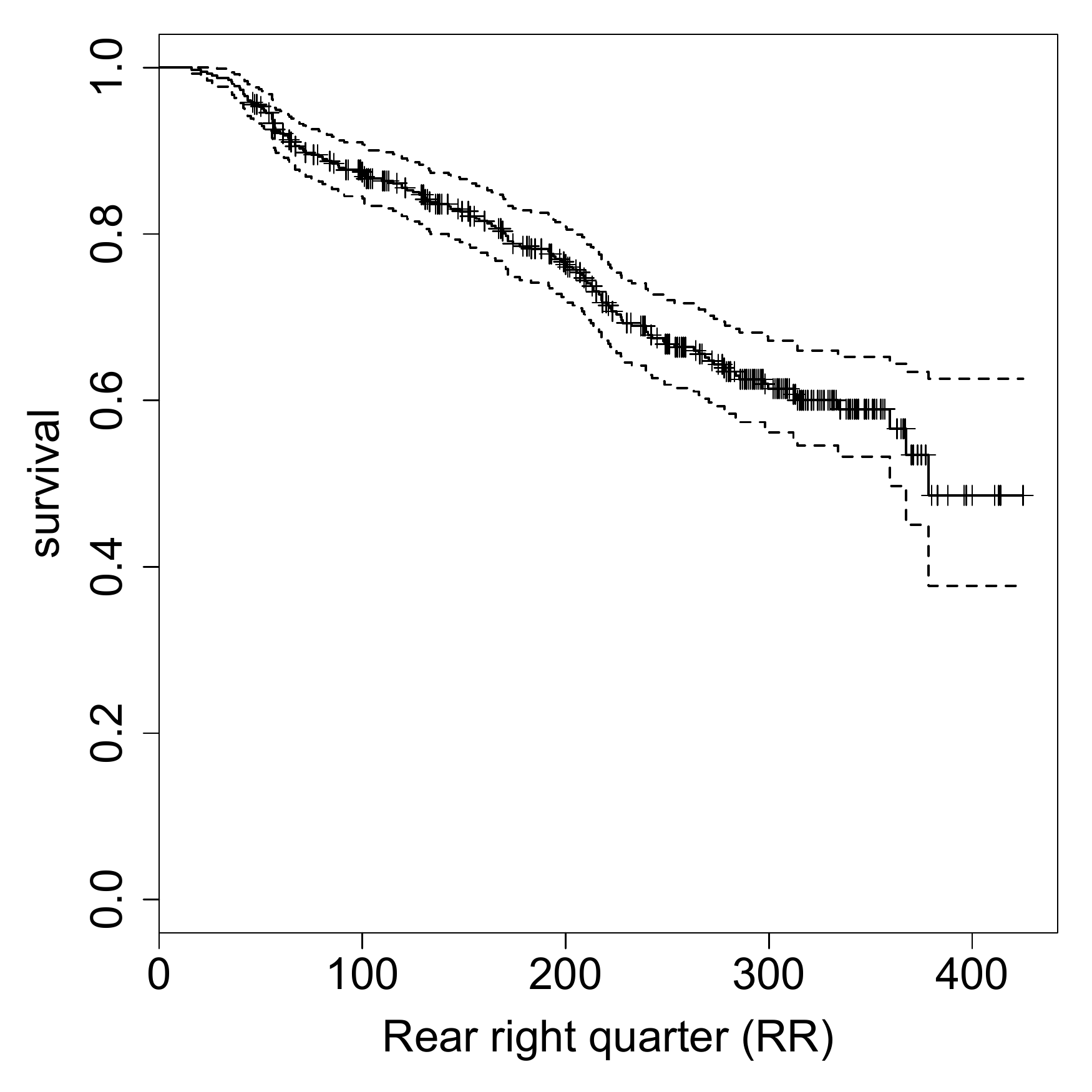}\\
	\end{figure}
	
	\begin{figure}[H]
		\centering
		\caption{Pairs plots of all six udder pairs of the mastitis data based on pseudo-observations generated via Kaplan-Meier estimates of the marginals (see \autoref{fig:KM}). The effect of right censoring is reflected by the empty lower left corner in the pairs plots. Observations shown as $\bullet$ are event times for both udder quarters; $\leftarrow$ is an event time only for the vertical axis; $\downarrow$ is an event time only for the horizontal axis; censored in both components is shown as \hspace{-.3cm}  $\myrelArrowI{}{\myrelArrowI{\myrelArrowI{\myrelArrowII{\leftarrow}{}}{}}{}}\hspace{-.42cm}\downarrow$.}
		\label{fig:ScatterplotsMastitis}
		\includegraphics[width=.4\linewidth]{"Cows_FL-FR"}\includegraphics[width=.4\linewidth]{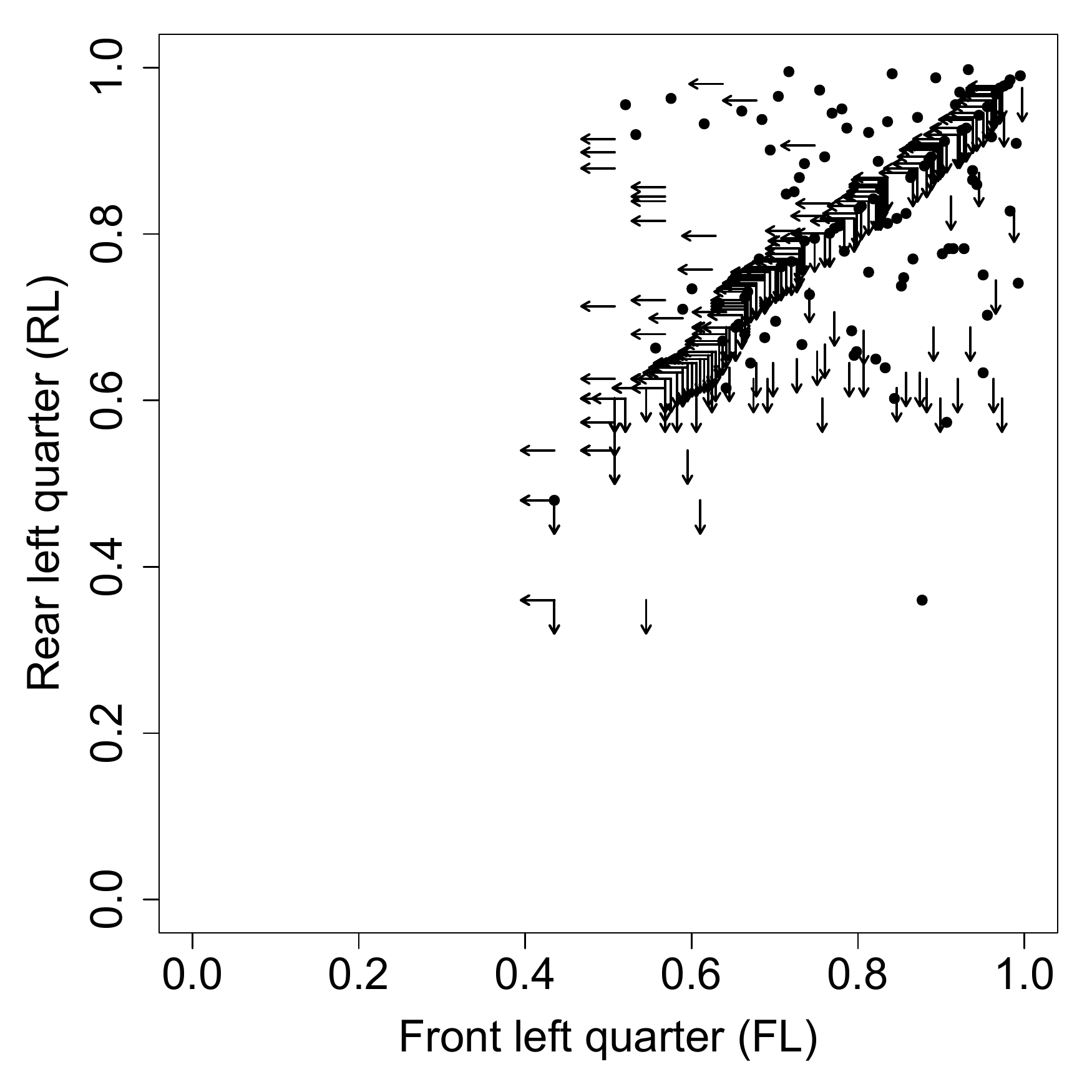}\\
		\includegraphics[width=.4\linewidth]{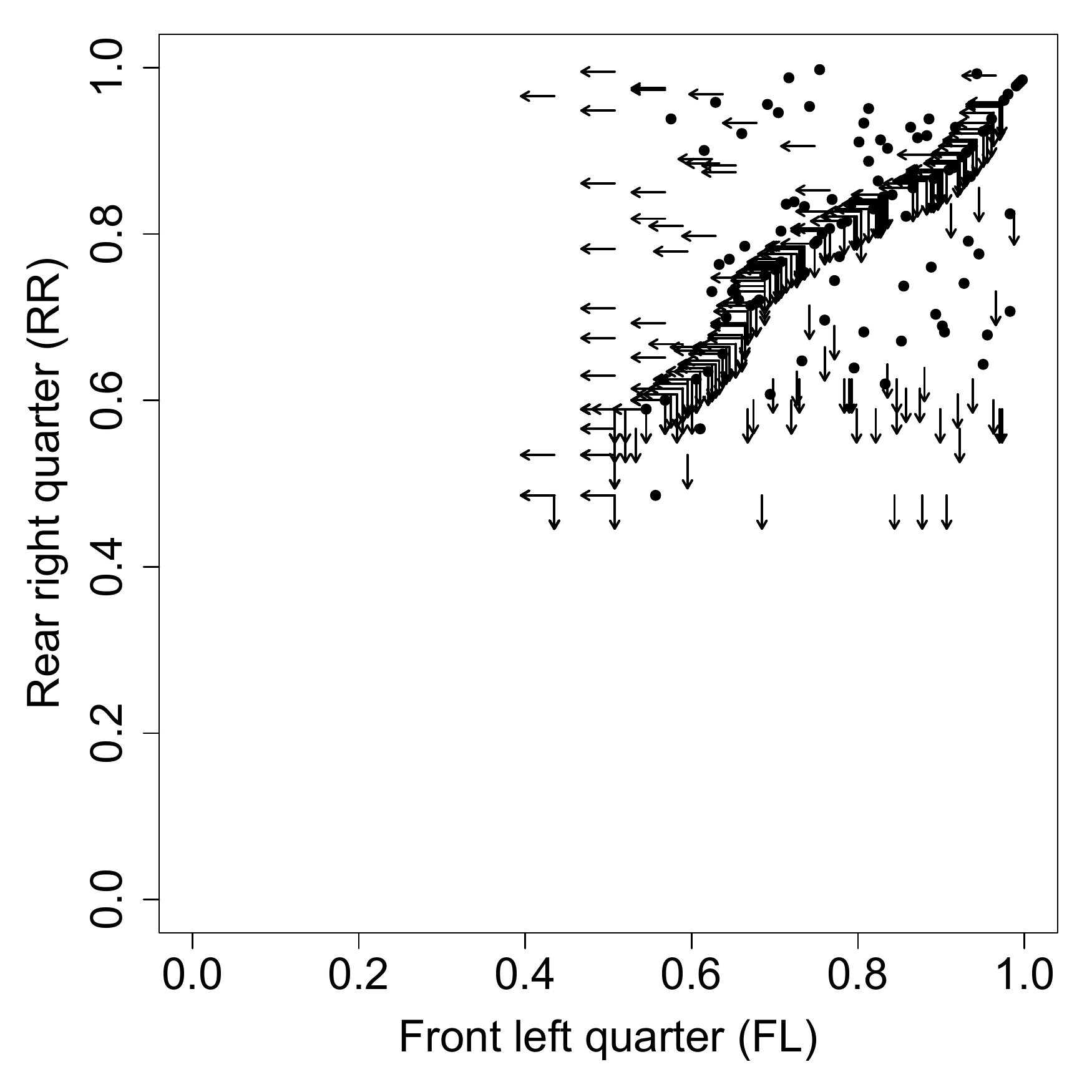}\includegraphics[width=.4\linewidth]{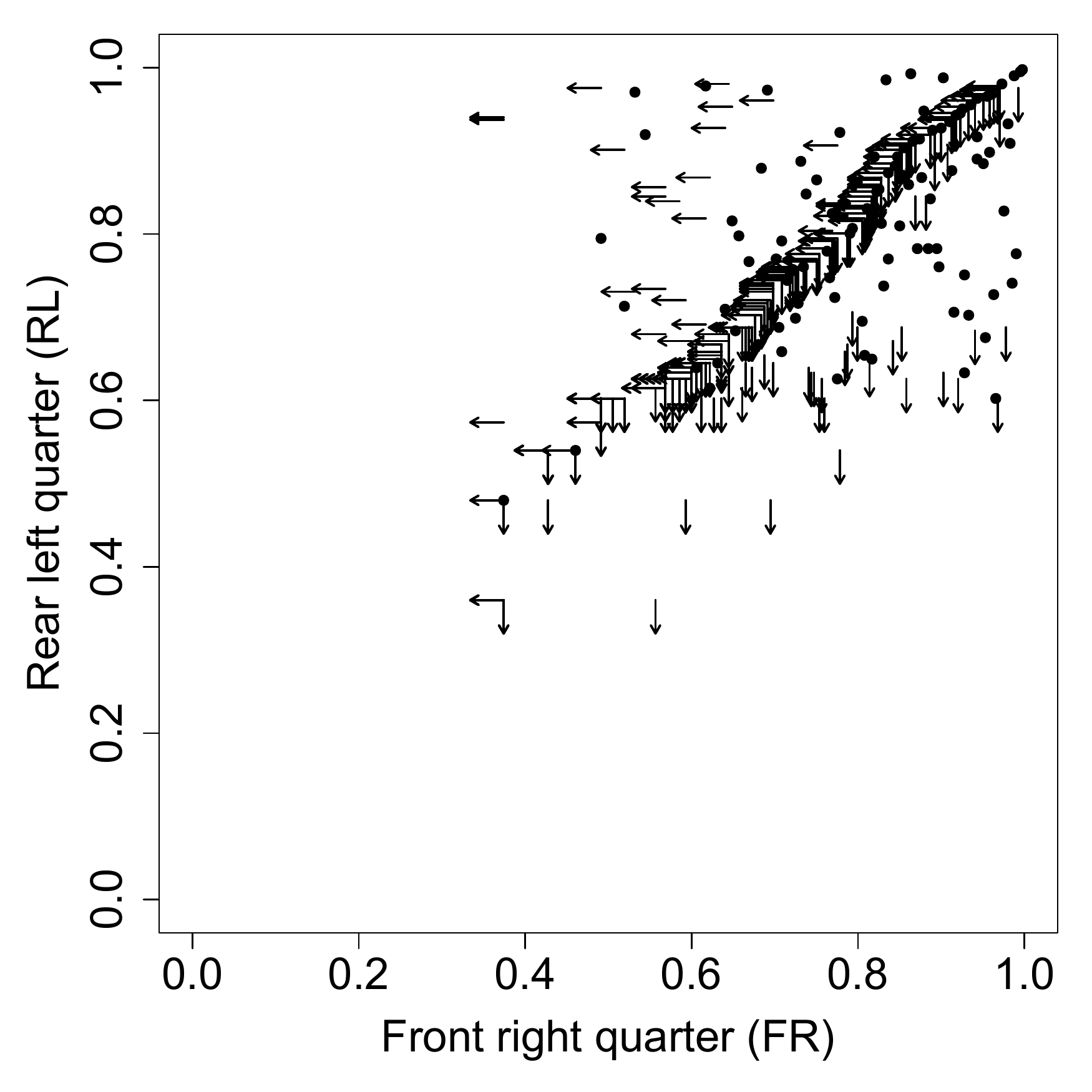}\\
		\includegraphics[width=.4\linewidth]{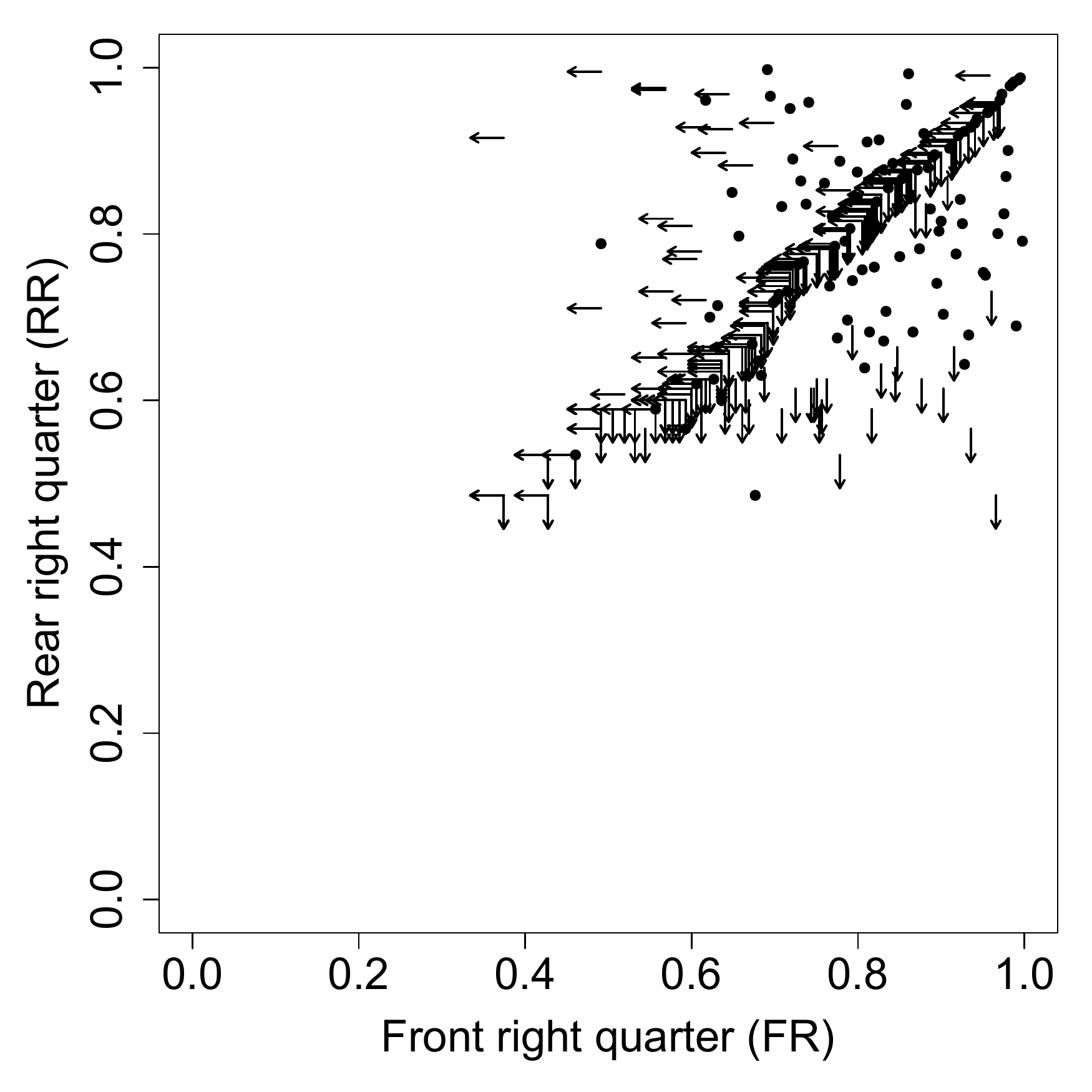}\includegraphics[width=.4\linewidth]{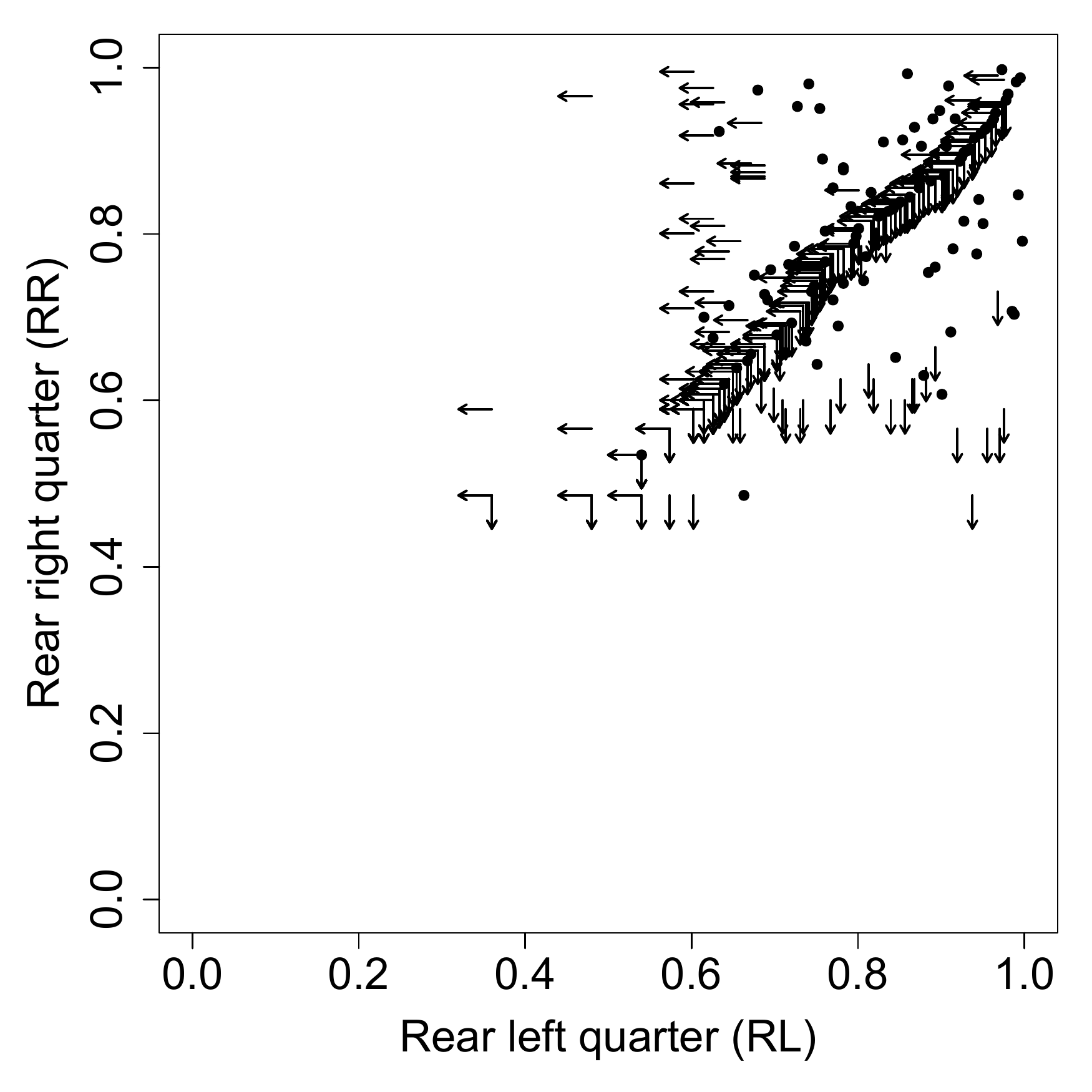}\\
		\vspace{-1cm}
	\end{figure}

	\newpage
	\subsection{Extended simulation study in three dimensions}\label{Sec:ExtSimStudy}
	To give further insights into the finite sample performance of the proposed estimation approach, additional simulation settings are considered. The simulation settings are chosen as described in Section 5.1 of the main text. See Table 2 in the main text for details on the marginal settings and the censoring mechanisms. Further, recall that the same copula family is assumed in $\mathcal{T}_1$ (either Clayton (C) or Gumbel (G)) and a Frank (F) copula is assumed in $\mathcal{T}_2$. All results are based on $200$ replications for sample sizes $200$ and $500$. While in the main text strong dependencies of $\tau_{12} = \tau_{23} = 0.6$ and $\tau_{13;2} = 0.3$ are investigated, we now investigate $\tau_{12} = \tau_{23} = \tau_{13;2} = 0.3$ (moderate dependencies) and $\tau_{12} = \tau_{23} = \tau_{13;2} = 0.1$ (weak dependencies). The results for the extra simulation settings are in line with the analysis given in case of strong dependencies. Note, however, that estimating low dependencies in the presence of heavy censoring is particularly challenging. \autoref{Table:Overview} gives an overview of the presented results.   
	
	\begin{table}[ht]
		\small
		\centering
		\captionof{table}{Overview of considered simulation settings with references to corresponding tables.}
		\label{Table:Overview}
		\hspace*{-.5cm}
		\begin{tabular}{ccccc}
			\midrule
			Kendall's $\tau$ values & Copula family in $\mathcal{T}_1$ & Censoring & Sample size & Table and page\\
			\midrule\midrule
			\multirow{8}{*}{$\tau_{12} = \tau_{23} = \tau_{13;2} = 0.3$} & \multirow{4}{*}{Clayton} & \multirow{2}{*}{$65\%$} & 200 & \multirow{2}{*}{\autoref{Table:CCF_65_200+500_3} (\autopageref{Table:CCF_65_200+500_3})}\\
			& & & 500 & \\
			\cmidrule{3-5}
			& & $25\%$ & 500 & \multirow{2}{*}{\autoref{Table:CCF_0+25_500_3} (\autopageref{Table:CCF_0+25_500_3})}\\
			& & complete data & 500 & \\
			\cmidrule{2-5}
			& \multirow{4}{*}{Gumbel} & \multirow{2}{*}{$65\%$} & 200 & \multirow{2}{*}{\autoref{Table:GGF_65_200+500_3} (\autopageref{Table:GGF_65_200+500_3})}\\
			& & & 500 & \\
			\cmidrule{3-5}
			& & $25\%$ & 500 & \multirow{2}{*}{\autoref{Table:GGF_0+25_500_3} (\autopageref{Table:GGF_0+25_500_3})}\\
			& & complete data & 500 & \\
			\hline\midrule
			\multirow{8}{*}{$\tau_{12} = \tau_{23} = \tau_{13;2} = 0.1$} & \multirow{4}{*}{Clayton} & \multirow{2}{*}{$65\%$} & 200 & \multirow{2}{*}{\autoref{Table:CCF_65_200+500_1} (\autopageref{Table:CCF_65_200+500_1})}\\
			& & & 500 & \\
			\cmidrule{3-5}
			& & $25\%$ & 500 & \multirow{2}{*}{\autoref{Table:CCF_0+25_500_1} (\autopageref{Table:CCF_0+25_500_1})}\\
			& & complete data & 500 & \\
			\cmidrule{2-5}
			& \multirow{4}{*}{Gumbel} & \multirow{2}{*}{$65\%$} & 200 & \multirow{2}{*}{\autoref{Table:GGF_65_200+500_1} (\autopageref{Table:GGF_65_200+500_1})}\\
			& & & 500 & \\
			\cmidrule{3-5}
			& & $25\%$ & 500 & \multirow{2}{*}{\autoref{Table:GGF_0+25_500_1} (\autopageref{Table:GGF_0+25_500_1})}\\
			& & complete data & 500 & \\
			\hline\hline      
		\end{tabular}
	\end{table}

	\begin{landscape}
		\begin{table}[ht]
			\centering
			\captionof{table}{Performance measures for the estimation of the copula parameters and Kendall's $\tau$ values in case of 65\% common right-censored event time data with sample sizes 200 and 500. The copula combination Clayton (C), Clayton (C), Frank (F) with true $\tau_{12}=\tau_{23}=\tau_{13;2}=0.3$ is investigated. Known margins, parametrically estimated margins (MLE) and nonparametrically estimated margins (KME) are considered.}
			\label{Table:CCF_65_200+500_3}
			\begin{tabular}{cccccccccccccc}
				\midrule
				&  &  &  & \multicolumn{5}{c}{Copula parameter} &  \multicolumn{5}{c}{Kendall's $\tau$} \\
				&  &  &  & $\theta$ & $\bar{\theta}$ & $\hat{b}\left(\bar{\theta}\right)$ & $s^2\left(\bar{\theta}\right)$ & $\widehat{mse}\left(\bar{\theta}\right)$ & $\tau$ & $\bar{\tau}$ & $\hat{b}\left(\bar{\tau}\right)$ & $s^2\left(\bar{\tau}\right)$ & $\widehat{mse}\left(\bar{\tau}\right)$ \\
				\hline
				\cmidrule{1-14}
				\multirow{9}{*} {\begin{sideways} $n = 200$, $65\%$ censoring \end{sideways}} & \multirow{3}{*} {\begin{sideways} Known \end{sideways}} & C & $\theta_{12}$ & 0.86 & 0.90 & 0.0410 & 0.1461 & 0.1478 & 0.30 & 0.30 & -0.0015 & 0.0078 & 0.0078 \\ 
				&  & C & $\theta_{23}$ & 0.86 & 0.87 & 0.0170 & 0.0768 & 0.0771 & 0.30 & 0.30 & -0.0023 & 0.0045 & 0.0046 \\ 
				&  & F & $\theta_{13;2}$ & 2.92 & 3.08 & 0.1652 & 0.8452 & 0.8725 & 0.30 & 0.31 & 0.0092 & 0.0063 & 0.0064 \\ 
				\cmidrule{2-14}
				& \multirow{3}{*} {\begin{sideways} MLE \end{sideways}} & C & $\theta_{12}$ & 0.86 & 0.91 & 0.0495 & 0.1525 & 0.1550 & 0.30 & 0.30 & 0.0002 & 0.0080 & 0.0080 \\ 
				&  & C & $\theta_{23}$ & 0.86 & 0.88 & 0.0274 & 0.0858 & 0.0865 & 0.30 & 0.30 & -0.0004 & 0.0050 & 0.0050 \\ 
				&  & F & $\theta_{13;2}$ & 2.92 & 3.09 & 0.1771 & 0.8877 & 0.9190 & 0.30 & 0.31 & 0.0100 & 0.0065 & 0.0066 \\ 
				\cmidrule{2-14}
				& \multirow{3}{*} {\begin{sideways} KME \end{sideways}} & C & $\theta_{12}$ & 0.86 & 0.91 & 0.0506 & 0.1576 & 0.1602 & 0.30 & 0.30 & 0.0001 & 0.0082 & 0.0082 \\ 
				&  & C & $\theta_{23}$ & 0.86 & 0.88 & 0.0201 & 0.0856 & 0.0860 & 0.30 & 0.30 & -0.0023 & 0.0051 & 0.0051 \\ 
				&  & F & $\theta_{13;2}$ & 2.92 & 3.13 & 0.2138 & 0.9328 & 0.9786 & 0.30 & 0.31 & 0.0129 & 0.0068 & 0.0069 \\
				\hline
				\cmidrule{1-14}
				\multirow{9}{*} {\begin{sideways} $n = 500$, $65\%$ censoring \end{sideways}} & \multirow{3}{*} {\begin{sideways} Known \end{sideways}} & C & $\theta_{12}$ & 0.86 & 0.86 & 0.0070 & 0.0580 & 0.0581 & 0.30 & 0.30 & -0.0032 & 0.0034 & 0.0034 \\ 
				&  & C & $\theta_{23}$ & 0.86 & 0.86 & 0.0058 & 0.0288 & 0.0288 & 0.30 & 0.30 & -0.0010 & 0.0017 & 0.0017 \\ 
				&  & F & $\theta_{13;2}$ & 2.92 & 3.02 & 0.1038 & 0.3549 & 0.3657 & 0.30 & 0.31 & 0.0069 & 0.0027 & 0.0027 \\ 
				\cmidrule{2-14}
				& \multirow{3}{*} {\begin{sideways} MLE \end{sideways}} & C & $\theta_{12}$ & 0.86 & 0.87 & 0.0087 & 0.0612 & 0.0612 & 0.30 & 0.30 & -0.0030 & 0.0036 & 0.0036 \\ 
				&  & C & $\theta_{23}$ & 0.86 & 0.86 & 0.0074 & 0.0308 & 0.0309 & 0.30 & 0.30 & -0.0008 & 0.0018 & 0.0018 \\ 
				&  & F & $\theta_{13;2}$ & 2.92 & 3.02 & 0.1065 & 0.3635 & 0.3748 & 0.30 & 0.31 & 0.0071 & 0.0027 & 0.0028 \\ 
				\cmidrule{2-14}
				& \multirow{3}{*} {\begin{sideways} KME \end{sideways}} & C & $\theta_{12}$ & 0.86 & 0.86 & 0.0065 & 0.0614 & 0.0615 & 0.30 & 0.30 & -0.0035 & 0.0036 & 0.0036 \\ 
				&  & C & $\theta_{23}$ & 0.86 & 0.86 & 0.0018 & 0.0304 & 0.0304 & 0.30 & 0.30 & -0.0021 & 0.0018 & 0.0018 \\ 
				&  & F & $\theta_{13;2}$ & 2.92 & 3.03 & 0.1154 & 0.3665 & 0.3798 & 0.30 & 0.31 & 0.0078 & 0.0027 & 0.0028 \\ 
				\hline
				\hline
			\end{tabular}
		\end{table}
	\end{landscape}

	\begin{landscape}
		\begin{table}[ht]
			\centering
			\captionof{table}{Performance measures for the estimation of the copula parameters and Kendall's $\tau$ values in
				case of complete and 25\% common right-censored event time data with sample size 500. The copula combination Clayton (C), Clayton (C), Frank (F) with true $\tau_{12}=\tau_{23}=\tau_{13;2}=0.3$ is investigated. Known margins, parametrically estimated margins (MLE) and nonparametrically estimated margins (ECDF/KME) are considered.}
			\label{Table:CCF_0+25_500_3}
			\begin{tabular}{cccccccccccccc}
				\midrule
				&  &  &  & \multicolumn{5}{c}{Copula parameter} &  \multicolumn{5}{c}{Kendall's $\tau$} \\
				&  &  &  & $\theta$ & $\bar{\theta}$ & $\hat{b}\left(\bar{\theta}\right)$ & $s^2\left(\bar{\theta}\right)$ & $\widehat{mse}\left(\bar{\theta}\right)$ & $\tau$ & $\bar{\tau}$ & $\hat{b}\left(\bar{\tau}\right)$ & $s^2\left(\bar{\tau}\right)$ & $\widehat{mse}\left(\bar{\tau}\right)$ \\
				\hline
				\cmidrule{1-14}
				\multirow{9}{*} {\begin{sideways} $n = 500$, $25\%$ censoring \end{sideways}} & \multirow{3}{*} {\begin{sideways} Known \end{sideways}} & C & $\theta_{12}$ & 0.86 & 0.87 & 0.0170 & 0.0145 & 0.0148 & 0.30 & 0.30 & 0.0029 & 0.0008 & 0.0009 \\ 
				&  & C & $\theta_{23}$ & 0.86 & 0.86 & 0.0014 & 0.0102 & 0.0102 & 0.30 & 0.30 & -0.0005 & 0.0006 & 0.0006 \\ 
				&  & F & $\theta_{13;2}$ & 2.92 & 2.96 & 0.0425 & 0.1144 & 0.1162 & 0.30 & 0.30 & 0.0030 & 0.0009 & 0.0009 \\ 
				\cmidrule{2-14}
				& \multirow{3}{*} {\begin{sideways} MLE \end{sideways}} & C & $\theta_{12}$ & 0.86 & 0.87 & 0.0175 & 0.0185 & 0.0188 & 0.30 & 0.30 & 0.0027 & 0.0011 & 0.0011 \\ 
				&  & C & $\theta_{23}$ & 0.86 & 0.86 & -0.0009 & 0.0136 & 0.0136 & 0.30 & 0.30 & -0.0014 & 0.0008 & 0.0008 \\ 
				&  & F & $\theta_{13;2}$ & 2.92 & 2.96 & 0.0401 & 0.1181 & 0.1197 & 0.30 & 0.30 & 0.0028 & 0.0009 & 0.0009 \\ 
				\cmidrule{2-14}
				& \multirow{3}{*} {\begin{sideways} KME \end{sideways}} & C & $\theta_{12}$ & 0.86 & 0.86 & 0.0037 & 0.0180 & 0.0180 & 0.30 & 0.30 & -0.0006 & 0.0011 & 0.0011 \\ 
				&  & C & $\theta_{23}$ & 0.86 & 0.84 & -0.0141 & 0.0140 & 0.0142 & 0.30 & 0.30 & -0.0047 & 0.0009 & 0.0009 \\ 
				&  & F & $\theta_{13;2}$ & 2.92 & 2.96 & 0.0454 & 0.1220 & 0.1241 & 0.30 & 0.30 & 0.0032 & 0.0009 & 0.0009 \\ 			\hline
				\cmidrule{1-14}
				\multirow{9}{*} {\begin{sideways} $n = 500$, complete data \end{sideways}} & \multirow{3}{*} {\begin{sideways} Known \end{sideways}} & C & $\theta_{12}$ & 0.86 & 0.87 & 0.0136 & 0.0067 & 0.0069 & 0.30 & 0.30 & 0.0027 & 0.0004 & 0.0004 \\ 
				&  & C & $\theta_{23}$ & 0.86 & 0.86 & 0.0059 & 0.0071 & 0.0071 & 0.30 & 0.30 & 0.0008 & 0.0004 & 0.0004 \\ 
				&  & F & $\theta_{13;2}$ & 2.92 & 2.97 & 0.0536 & 0.0847 & 0.0875 & 0.30 & 0.30 & 0.0042 & 0.0007 & 0.0007 \\ 
				\cmidrule{2-14}
				& \multirow{3}{*} {\begin{sideways} MLE \end{sideways}} & C & $\theta_{12}$ & 0.86 & 0.86 & 0.0052 & 0.0108 & 0.0108 & 0.30 & 0.30 & 0.0004 & 0.0007 & 0.0007 \\ 
				&  & C & $\theta_{23}$ & 0.86 & 0.85 & -0.0025 & 0.0108 & 0.0108 & 0.30 & 0.30 & -0.0015 & 0.0007 & 0.0007 \\ 
				&  & F & $\theta_{13;2}$ & 2.92 & 2.96 & 0.0397 & 0.0821 & 0.0837 & 0.30 & 0.30 & 0.0030 & 0.0006 & 0.0006 \\ 
				\cmidrule{2-14}
				& \multirow{3}{*} {\begin{sideways} ECDF \end{sideways}} & C & $\theta_{12}$ & 0.86 & 0.88 & 0.0222 & 0.0112 & 0.0117 & 0.30 & 0.30 & 0.0045 & 0.0007 & 0.0007 \\ 
				&  & C & $\theta_{23}$ & 0.86 & 0.87 & 0.0138 & 0.0115 & 0.0117 & 0.30 & 0.30 & 0.0024 & 0.0007 & 0.0007 \\ 
				&  & F & $\theta_{13;2}$ & 2.92 & 2.96 & 0.0423 & 0.0858 & 0.0876 & 0.30 & 0.30 & 0.0032 & 0.0007 & 0.0007 \\ 			\hline
				\hline
			\end{tabular}
		\end{table}
	\end{landscape}

	\begin{landscape}
		\begin{table}[ht]
			\centering
			\captionof{table}{Performance measures for the estimation of the copula parameters and Kendall's $\tau$ values in case of 65\% common right-censored event time data with sample sizes 200 and 500. The copula combination Gumbel (G), Gumbel (G), Frank (F) with true $\tau_{12}=\tau_{23}=\tau_{13;2}=0.3$ is investigated. Known margins, parametrically estimated margins (MLE) and nonparametrically estimated margins (KME) are considered.}
			\label{Table:GGF_65_200+500_3}
			\begin{tabular}{cccccccccccccc}
				\midrule
				&  &  &  & \multicolumn{5}{c}{Copula parameter} &  \multicolumn{5}{c}{Kendall's $\tau$} \\
				&  &  &  & $\theta$ & $\bar{\theta}$ & $\hat{b}\left(\bar{\theta}\right)$ & $s^2\left(\bar{\theta}\right)$ & $\widehat{mse}\left(\bar{\theta}\right)$ & $\tau$ & $\bar{\tau}$ & $\hat{b}\left(\bar{\tau}\right)$ & $s^2\left(\bar{\tau}\right)$ & $\widehat{mse}\left(\bar{\tau}\right)$ \\
				\hline
				\cmidrule{1-14}
				\multirow{9}{*} {\begin{sideways} $n = 200$, $65\%$ censoring \end{sideways}} & \multirow{3}{*} {\begin{sideways} Known \end{sideways}} & G & $\theta_{12}$ & 1.43 & 1.44 & 0.0086 & 0.0121 & 0.0122 & 0.30 & 0.30 & 0.0001 & 0.0029 & 0.0029 \\ 
				&  & G & $\theta_{23}$ & 1.43 & 1.44 & 0.0084 & 0.0082 & 0.0082 & 0.30 & 0.30 & 0.0013 & 0.0020 & 0.0020 \\ 
				&  & F & $\theta_{13;2}$ & 2.92 & 3.02 & 0.0992 & 0.8832 & 0.8930 & 0.30 & 0.30 & 0.0034 & 0.0067 & 0.0067 \\ 
				\cmidrule{2-14}
				& \multirow{3}{*} {\begin{sideways} MLE \end{sideways}} & G & $\theta_{12}$ & 1.43 & 1.44 & 0.0148 & 0.0138 & 0.0140 & 0.30 & 0.30 & 0.0026 & 0.0032 & 0.0032 \\ 
				&  & G & $\theta_{23}$ & 1.43 & 1.44 & 0.0131 & 0.0106 & 0.0108 & 0.30 & 0.30 & 0.0028 & 0.0025 & 0.0025 \\ 
				&  & F & $\theta_{13;2}$ & 2.92 & 3.04 & 0.1179 & 0.9273 & 0.9411 & 0.30 & 0.30 & 0.0047 & 0.0070 & 0.0070 \\ 
				\cmidrule{2-14}
				& \multirow{3}{*} {\begin{sideways} KME \end{sideways}} & G & $\theta_{12}$ & 1.43 & 1.47 & 0.0464 & 0.0163 & 0.0184 & 0.30 & 0.32 & 0.0170 & 0.0035 & 0.0038 \\ 
				&  & G & $\theta_{23}$ & 1.43 & 1.47 & 0.0365 & 0.0112 & 0.0125 & 0.30 & 0.31 & 0.0139 & 0.0025 & 0.0027 \\ 
				&  & F & $\theta_{13;2}$ & 2.92 & 3.07 & 0.1511 & 0.9618 & 0.9846 & 0.30 & 0.31 & 0.0074 & 0.0072 & 0.0072 \\ 			\hline
				\cmidrule{1-14}
				\multirow{9}{*} {\begin{sideways} $n = 500$, $65\%$ censoring \end{sideways}} & \multirow{3}{*} {\begin{sideways} Known \end{sideways}} & G & $\theta_{12}$ & 1.43 & 1.44 & 0.0080 & 0.0053 & 0.0053 & 0.30 & 0.30 & 0.0022 & 0.0012 & 0.0012 \\ 
				&  & G & $\theta_{23}$ & 1.43 & 1.43 & 0.0030 & 0.0032 & 0.0033 & 0.30 & 0.30 & 0.0003 & 0.0008 & 0.0008 \\ 
				&  & F & $\theta_{13;2}$ & 2.92 & 2.96 & 0.0454 & 0.3610 & 0.3630 & 0.30 & 0.30 & 0.0018 & 0.0027 & 0.0027 \\ 
				\cmidrule{2-14}
				& \multirow{3}{*} {\begin{sideways} MLE \end{sideways}} & G & $\theta_{12}$ & 1.43 & 1.44 & 0.0081 & 0.0064 & 0.0064 & 0.30 & 0.30 & 0.0018 & 0.0015 & 0.0015 \\ 
				&  & G & $\theta_{23}$ & 1.43 & 1.43 & 0.0038 & 0.0040 & 0.0040 & 0.30 & 0.30 & 0.0005 & 0.0010 & 0.0010 \\ 
				&  & F & $\theta_{13;2}$ & 2.92 & 2.96 & 0.0446 & 0.3685 & 0.3705 & 0.30 & 0.30 & 0.0017 & 0.0028 & 0.0028 \\ 
				\cmidrule{2-14}
				& \multirow{3}{*} {\begin{sideways} KME \end{sideways}} & G & $\theta_{12}$ & 1.43 & 1.45 & 0.0210 & 0.0074 & 0.0078 & 0.30 & 0.31 & 0.0077 & 0.0017 & 0.0017 \\ 
				&  & G & $\theta_{23}$ & 1.43 & 1.44 & 0.0127 & 0.0042 & 0.0044 & 0.30 & 0.30 & 0.0048 & 0.0010 & 0.0010 \\ 
				&  & F & $\theta_{13;2}$ & 2.92 & 2.98 & 0.0620 & 0.3768 & 0.3806 & 0.30 & 0.30 & 0.0031 & 0.0028 & 0.0028 \\ \hline
				\hline
			\end{tabular}
		\end{table}
	\end{landscape}

	\begin{landscape}
		\begin{table}[ht]
			\centering
			\captionof{table}{Performance measures for the estimation of the copula parameters and Kendall's $\tau$ values in
				case of complete and 25\% common right-censored event time data with sample size 500. The copula combination Gumbel (G), Gumbel (G), Frank (F) with true $\tau_{12}=\tau_{23}=\tau_{13;2}=0.3$ is investigated. Known margins, parametrically estimated margins (MLE) and nonparametrically estimated margins (ECDF/KME) are considered.}
			\label{Table:GGF_0+25_500_3}
			\begin{tabular}{cccccccccccccc}
				\midrule
				&  &  &  & \multicolumn{5}{c}{Copula parameter} &  \multicolumn{5}{c}{Kendall's $\tau$} \\
				&  &  &  & $\theta$ & $\bar{\theta}$ & $\hat{b}\left(\bar{\theta}\right)$ & $s^2\left(\bar{\theta}\right)$ & $\widehat{mse}\left(\bar{\theta}\right)$ & $\tau$ & $\bar{\tau}$ & $\hat{b}\left(\bar{\tau}\right)$ & $s^2\left(\bar{\tau}\right)$ & $\widehat{mse}\left(\bar{\tau}\right)$ \\
				\hline
				\cmidrule{1-14}
				\multirow{9}{*} {\begin{sideways} $n = 500$, $25\%$ censoring \end{sideways}} & \multirow{3}{*} {\begin{sideways} Known \end{sideways}} & G & $\theta_{12}$ & 1.43 & 1.44 & 0.0071 & 0.0029 & 0.0029 & 0.30 & 0.30 & 0.0025 & 0.0007 & 0.0007 \\ 
				&  & G & $\theta_{23}$ & 1.43 & 1.43 & 0.0022 & 0.0025 & 0.0025 & 0.30 & 0.30 & 0.0002 & 0.0006 & 0.0006 \\ 
				&  & F & $\theta_{13;2}$ & 2.92 & 2.95 & 0.0312 & 0.1125 & 0.1135 & 0.30 & 0.30 & 0.0020 & 0.0009 & 0.0009 \\ 
				\cmidrule{2-14}
				& \multirow{3}{*} {\begin{sideways} MLE \end{sideways}} & G & $\theta_{12}$ & 1.43 & 1.44 & 0.0090 & 0.0037 & 0.0037 & 0.30 & 0.30 & 0.0031 & 0.0009 & 0.0009 \\ 
				&  & G & $\theta_{23}$ & 1.43 & 1.43 & 0.0041 & 0.0030 & 0.0030 & 0.30 & 0.30 & 0.0010 & 0.0007 & 0.0007 \\ 
				&  & F & $\theta_{13;2}$ & 2.92 & 2.94 & 0.0245 & 0.1139 & 0.1145 & 0.30 & 0.30 & 0.0014 & 0.0009 & 0.0009 \\ 
				\cmidrule{2-14}
				& \multirow{3}{*} {\begin{sideways} KME \end{sideways}} & G & $\theta_{12}$ & 1.43 & 1.44 & 0.0136 & 0.0039 & 0.0041 & 0.30 & 0.31 & 0.0053 & 0.0009 & 0.0009 \\ 
				&  & G & $\theta_{23}$ & 1.43 & 1.44 & 0.0077 & 0.0031 & 0.0031 & 0.30 & 0.30 & 0.0027 & 0.0007 & 0.0007 \\ 
				&  & F & $\theta_{13;2}$ & 2.92 & 2.95 & 0.0295 & 0.1168 & 0.1177 & 0.30 & 0.30 & 0.0019 & 0.0009 & 0.0009 \\  			\hline
				\cmidrule{1-14}
				\multirow{9}{*} {\begin{sideways} $n = 500$, complete data \end{sideways}} & \multirow{3}{*} {\begin{sideways} Known \end{sideways}} & G & $\theta_{12}$ & 1.43 & 1.44 & 0.0069 & 0.0023 & 0.0023 & 0.30 & 0.30 & 0.0026 & 0.0005 & 0.0005 \\ 
				&  & G & $\theta_{23}$ & 1.43 & 1.43 & 0.0028 & 0.0023 & 0.0023 & 0.30 & 0.30 & 0.0006 & 0.0006 & 0.0006 \\ 
				&  & F & $\theta_{13;2}$ & 2.92 & 2.96 & 0.0474 & 0.0895 & 0.0917 & 0.30 & 0.30 & 0.0036 & 0.0007 & 0.0007 \\ 
				\cmidrule{2-14}
				& \multirow{3}{*} {\begin{sideways} MLE \end{sideways}} & G & $\theta_{12}$ & 1.43 & 1.43 & 0.0041 & 0.0027 & 0.0027 & 0.30 & 0.30 & 0.0011 & 0.0006 & 0.0006 \\ 
				&  & G & $\theta_{23}$ & 1.43 & 1.43 & 0.0004 & 0.0028 & 0.0028 & 0.30 & 0.30 & -0.0008 & 0.0007 & 0.0007 \\ 
				&  & F & $\theta_{13;2}$ & 2.92 & 2.95 & 0.0305 & 0.0873 & 0.0882 & 0.30 & 0.30 & 0.0021 & 0.0007 & 0.0007 \\ 
				\cmidrule{2-14}
				& \multirow{3}{*} {\begin{sideways} ECDF \end{sideways}} & G & $\theta_{12}$ & 1.43 & 1.44 & 0.0134 & 0.0031 & 0.0033 & 0.30 & 0.31 & 0.0055 & 0.0007 & 0.0007 \\ 
				&  & G & $\theta_{23}$ & 1.43 & 1.44 & 0.0093 & 0.0030 & 0.0031 & 0.30 & 0.30 & 0.0035 & 0.0007 & 0.0007 \\ 
				&  & F & $\theta_{13;2}$ & 2.92 & 2.95 & 0.0337 & 0.0892 & 0.0904 & 0.30 & 0.30 & 0.0024 & 0.0007 & 0.0007 \\ 			\hline
				\hline
			\end{tabular}
		\end{table}
	\end{landscape}

	\begin{landscape}
		\begin{table}[ht]
			\centering
			\captionof{table}{Performance measures for the estimation of the copula parameters and Kendall's $\tau$ values in case of 65\% common right-censored event time data with sample sizes 200 and 500. The copula combination Clayton (C), Clayton (C), Frank (F) with true $\tau_{12}=\tau_{23}=\tau_{13;2}=0.1$ is investigated. Known margins, parametrically estimated margins (MLE) and nonparametrically estimated margins (KME) are considered.}
			\label{Table:CCF_65_200+500_1}
			\begin{tabular}{cccccccccccccc}
				\midrule
				&  &  &  & \multicolumn{5}{c}{Copula parameter} &  \multicolumn{5}{c}{Kendall's $\tau$} \\
				&  &  &  & $\theta$ & $\bar{\theta}$ & $\hat{b}\left(\bar{\theta}\right)$ & $s^2\left(\bar{\theta}\right)$ & $\widehat{mse}\left(\bar{\theta}\right)$ & $\tau$ & $\bar{\tau}$ & $\hat{b}\left(\bar{\tau}\right)$ & $s^2\left(\bar{\tau}\right)$ & $\widehat{mse}\left(\bar{\tau}\right)$ \\
				\hline
				\cmidrule{1-14}
				\multirow{9}{*} {\begin{sideways} $n = 200$, $65\%$ censoring \end{sideways}} & \multirow{3}{*} {\begin{sideways} Known \end{sideways}} & C & $\theta_{12}$ & 0.22 & 0.27 & 0.0487 & 0.0550 & 0.0573 & 0.10 & 0.11 & 0.0106 & 0.0074 & 0.0075 \\ 
				&  & C & $\theta_{23}$ & 0.22 & 0.26 & 0.0341 & 0.0353 & 0.0365 & 0.10 & 0.11 & 0.0077 & 0.0051 & 0.0052 \\ 
				&  & F & $\theta_{13;2}$ & 0.91 & 1.02 & 0.1120 & 0.6702 & 0.6827 & 0.10 & 0.11 & 0.0101 & 0.0075 & 0.0076 \\ 
				\cmidrule{2-14}
				& \multirow{3}{*} {\begin{sideways} MLE \end{sideways}} & C & $\theta_{12}$ & 0.22 & 0.27 & 0.0511 & 0.0565 & 0.0592 & 0.10 & 0.11 & 0.0113 & 0.0075 & 0.0076 \\ 
				&  & C & $\theta_{23}$ & 0.22 & 0.26 & 0.0363 & 0.0364 & 0.0377 & 0.10 & 0.11 & 0.0084 & 0.0052 & 0.0053 \\ 
				&  & F & $\theta_{13;2}$ & 0.91 & 1.02 & 0.1108 & 0.7013 & 0.7136 & 0.10 & 0.11 & 0.0099 & 0.0078 & 0.0079 \\ 
				\cmidrule{2-14}
				& \multirow{3}{*} {\begin{sideways} KME \end{sideways}} & C & $\theta_{12}$ & 0.22 & 0.28 & 0.0564 & 0.0575 & 0.0606 & 0.10 & 0.11 & 0.0132 & 0.0076 & 0.0077 \\ 
				&  & C & $\theta_{23}$ & 0.22 & 0.26 & 0.0386 & 0.0361 & 0.0376 & 0.10 & 0.11 & 0.0094 & 0.0052 & 0.0053 \\ 
				&  & F & $\theta_{13;2}$ & 0.91 & 1.02 & 0.1103 & 0.7159 & 0.7281 & 0.10 & 0.11 & 0.0098 & 0.0080 & 0.0081 \\ 
				\hline
				\cmidrule{1-14}
				\multirow{9}{*} {\begin{sideways} $n = 500$, $65\%$ censoring \end{sideways}} & \multirow{3}{*} {\begin{sideways} Known \end{sideways}} & C & $\theta_{12}$ & 0.22 & 0.23 & 0.0099 & 0.0237 & 0.0238 & 0.10 & 0.10 & -0.0002 & 0.0037 & 0.0037 \\ 
				&  & C & $\theta_{23}$ & 0.22 & 0.23 & 0.0101 & 0.0137 & 0.0138 & 0.10 & 0.10 & 0.0017 & 0.0021 & 0.0021 \\ 
				&  & F & $\theta_{13;2}$ & 0.91 & 0.96 & 0.0507 & 0.2776 & 0.2801 & 0.10 & 0.10 & 0.0047 & 0.0032 & 0.0032 \\ 
				\cmidrule{2-14}
				& \multirow{3}{*} {\begin{sideways} MLE \end{sideways}} & C & $\theta_{12}$ & 0.22 & 0.23 & 0.0107 & 0.0241 & 0.0242 & 0.10 & 0.10 & 0.0001 & 0.0037 & 0.0037 \\ 
				&  & C & $\theta_{23}$ & 0.22 & 0.23 & 0.0094 & 0.0134 & 0.0135 & 0.10 & 0.10 & 0.0014 & 0.0021 & 0.0021 \\ 
				&  & F & $\theta_{13;2}$ & 0.91 & 0.96 & 0.0514 & 0.2776 & 0.2802 & 0.10 & 0.10 & 0.0048 & 0.0032 & 0.0032 \\ 
				\cmidrule{2-14}
				& \multirow{3}{*} {\begin{sideways} KME \end{sideways}} & C & $\theta_{12}$ & 0.22 & 0.24 & 0.0134 & 0.0234 & 0.0236 & 0.10 & 0.10 & 0.0013 & 0.0036 & 0.0036 \\ 
				&  & C & $\theta_{23}$ & 0.22 & 0.23 & 0.0087 & 0.0135 & 0.0136 & 0.10 & 0.10 & 0.0011 & 0.0021 & 0.0021 \\ 
				&  & F & $\theta_{13;2}$ & 0.91 & 0.96 & 0.0552 & 0.2797 & 0.2828 & 0.10 & 0.11 & 0.0052 & 0.0032 & 0.0033 \\ 			\hline
				\hline
			\end{tabular}
		\end{table}
	\end{landscape}

	\begin{landscape}
		\begin{table}[ht]
			\centering
			\captionof{table}{Performance measures for the estimation of the copula parameters and Kendall's $\tau$ values in
				case of complete and 25\% common right-censored event time data with sample size 500. The copula combination Clayton (C), Clayton (C), Frank (F) with true $\tau_{12}=\tau_{23}=\tau_{13;2}=0.1$ is investigated. Known margins, parametrically estimated margins (MLE) and nonparametrically estimated margins (ECDF/KME) are considered.}
			\label{Table:CCF_0+25_500_1}
			\begin{tabular}{cccccccccccccc}
				\midrule
				&  &  &  & \multicolumn{5}{c}{Copula parameter} &  \multicolumn{5}{c}{Kendall's $\tau$} \\
				&  &  &  & $\theta$ & $\bar{\theta}$ & $\hat{b}\left(\bar{\theta}\right)$ & $s^2\left(\bar{\theta}\right)$ & $\widehat{mse}\left(\bar{\theta}\right)$ & $\tau$ & $\bar{\tau}$ & $\hat{b}\left(\bar{\tau}\right)$ & $s^2\left(\bar{\tau}\right)$ & $\widehat{mse}\left(\bar{\tau}\right)$ \\
				\hline
				\cmidrule{1-14}
				\multirow{9}{*} {\begin{sideways} $n = 500$, $25\%$ censoring \end{sideways}} & \multirow{3}{*} {\begin{sideways} Known \end{sideways}} & C & $\theta_{12}$ & 0.22 & 0.23 & 0.0031 & 0.0078 & 0.0078 & 0.10 & 0.10 & -0.0001 & 0.0013 & 0.0013 \\ 
				&  & C & $\theta_{23}$ & 0.22 & 0.22 & 0.0024 & 0.0054 & 0.0054 & 0.10 & 0.10 & -0.0000 & 0.0009 & 0.0009 \\ 
				&  & F & $\theta_{13;2}$ & 0.91 & 0.95 & 0.0395 & 0.0908 & 0.0924 & 0.10 & 0.10 & 0.0040 & 0.0011 & 0.0011 \\ 
				\cmidrule{2-14}
				& \multirow{3}{*} {\begin{sideways} MLE \end{sideways}} & C & $\theta_{12}$ & 0.22 & 0.22 & -0.0002 & 0.0080 & 0.0080 & 0.10 & 0.10 & -0.0015 & 0.0013 & 0.0013 \\ 
				&  & C & $\theta_{23}$ & 0.22 & 0.22 & -0.0009 & 0.0054 & 0.0054 & 0.10 & 0.10 & -0.0013 & 0.0009 & 0.0009 \\ 
				&  & F & $\theta_{13;2}$ & 0.91 & 0.95 & 0.0379 & 0.0887 & 0.0902 & 0.10 & 0.10 & 0.0038 & 0.0010 & 0.0010 \\ 
				\cmidrule{2-14}
				& \multirow{3}{*} {\begin{sideways} KME \end{sideways}} & C & $\theta_{12}$ & 0.22 & 0.22 & 0.0017 & 0.0077 & 0.0077 & 0.10 & 0.10 & -0.0007 & 0.0012 & 0.0012 \\ 
				&  & C & $\theta_{23}$ & 0.22 & 0.22 & 0.0009 & 0.0057 & 0.0057 & 0.10 & 0.10 & -0.0007 & 0.0009 & 0.0009 \\ 
				&  & F & $\theta_{13;2}$ & 0.91 & 0.94 & 0.0355 & 0.0894 & 0.0907 & 0.10 & 0.10 & 0.0036 & 0.0010 & 0.0011 \\ 		
				\hline
				\cmidrule{1-14}
				\multirow{9}{*} {\begin{sideways} $n = 500$, complete data \end{sideways}} & \multirow{3}{*} {\begin{sideways} Known \end{sideways}} & C & $\theta_{12}$ & 0.22 & 0.22 & 0.0024 & 0.0036 & 0.0036 & 0.10 & 0.10 & 0.0003 & 0.0006 & 0.0006 \\ 
				&  & C & $\theta_{23}$ & 0.22 & 0.23 & 0.0031 & 0.0036 & 0.0036 & 0.10 & 0.10 & 0.0006 & 0.0006 & 0.0006 \\ 
				&  & F & $\theta_{13;2}$ & 0.91 & 0.94 & 0.0285 & 0.0675 & 0.0683 & 0.10 & 0.10 & 0.0029 & 0.0008 & 0.0008 \\ 
				\cmidrule{2-14}
				& \multirow{3}{*} {\begin{sideways} MLE \end{sideways}} & C & $\theta_{12}$ & 0.22 & 0.22 & -0.0005 & 0.0040 & 0.0040 & 0.10 & 0.10 & -0.0009 & 0.0007 & 0.0007 \\ 
				&  & C & $\theta_{23}$ & 0.22 & 0.22 & 0.0012 & 0.0038 & 0.0038 & 0.10 & 0.10 & -0.0002 & 0.0006 & 0.0006 \\ 
				&  & F & $\theta_{13;2}$ & 0.91 & 0.93 & 0.0197 & 0.0657 & 0.0660 & 0.10 & 0.10 & 0.0019 & 0.0008 & 0.0008 \\ 
				\cmidrule{2-14}
				& \multirow{3}{*} {\begin{sideways} ECDF \end{sideways}} & C & $\theta_{12}$ & 0.22 & 0.23 & 0.0100 & 0.0043 & 0.0044 & 0.10 & 0.10 & 0.0033 & 0.0007 & 0.0007 \\ 
				&  & C & $\theta_{23}$ & 0.22 & 0.23 & 0.0112 & 0.0042 & 0.0043 & 0.10 & 0.10 & 0.0038 & 0.0007 & 0.0007 \\ 
				&  & F & $\theta_{13;2}$ & 0.91 & 0.93 & 0.0214 & 0.0670 & 0.0675 & 0.10 & 0.10 & 0.0021 & 0.0008 & 0.0008 \\  			\hline
				\hline
			\end{tabular}
		\end{table}
	\end{landscape}

	\begin{landscape}
		\begin{table}[ht]
			\centering
			\captionof{table}{Performance measures for the estimation of the copula parameters and Kendall's $\tau$ values in case of 65\% common right-censored event time data with sample sizes 200 and 500. The copula combination Gumbel (G), Gumbel (G), Frank (F) with true $\tau_{12}=\tau_{23}=\tau_{13;2}=0.1$ is investigated. Known margins, parametrically estimated margins (MLE) and nonparametrically estimated margins (KME) are considered.}
			\label{Table:GGF_65_200+500_1}
			\begin{tabular}{cccccccccccccc}
				\midrule
				&  &  &  & \multicolumn{5}{c}{Copula parameter} &  \multicolumn{5}{c}{Kendall's $\tau$} \\
				&  &  &  & $\theta$ & $\bar{\theta}$ & $\hat{b}\left(\bar{\theta}\right)$ & $s^2\left(\bar{\theta}\right)$ & $\widehat{mse}\left(\bar{\theta}\right)$ & $\tau$ & $\bar{\tau}$ & $\hat{b}\left(\bar{\tau}\right)$ & $s^2\left(\bar{\tau}\right)$ & $\widehat{mse}\left(\bar{\tau}\right)$ \\
				\hline
				\cmidrule{1-14}
				\multirow{9}{*} {\begin{sideways} $n = 200$, $65\%$ censoring \end{sideways}} & \multirow{3}{*} {\begin{sideways} Known \end{sideways}} & G & $\theta_{12}$ & 1.11 & 1.12 & 0.0087 & 0.0049 & 0.0050 & 0.10 & 0.10 & 0.0036 & 0.0030 & 0.0030 \\ 
				&  & G & $\theta_{23}$ & 1.11 & 1.12 & 0.0068 & 0.0036 & 0.0036 & 0.10 & 0.10 & 0.0029 & 0.0023 & 0.0023 \\ 
				&  & F & $\theta_{13;2}$ & 0.91 & 0.97 & 0.0641 & 0.6779 & 0.6821 & 0.10 & 0.10 & 0.0050 & 0.0076 & 0.0076 \\ 
				\cmidrule{2-14}
				& \multirow{3}{*} {\begin{sideways} MLE \end{sideways}} & G & $\theta_{12}$ & 1.11 & 1.12 & 0.0091 & 0.0048 & 0.0049 & 0.10 & 0.10 & 0.0039 & 0.0030 & 0.0030 \\ 
				&  & G & $\theta_{23}$ & 1.11 & 1.12 & 0.0070 & 0.0037 & 0.0038 & 0.10 & 0.10 & 0.0030 & 0.0024 & 0.0024 \\ 
				&  & F & $\theta_{13;2}$ & 0.91 & 0.97 & 0.0649 & 0.6984 & 0.7026 & 0.10 & 0.11 & 0.0050 & 0.0078 & 0.0079 \\ 
				\cmidrule{2-14}
				& \multirow{3}{*} {\begin{sideways} KME \end{sideways}} & G & $\theta_{12}$ & 1.11 & 1.13 & 0.0231 & 0.0058 & 0.0063 & 0.10 & 0.11 & 0.0145 & 0.0034 & 0.0036 \\ 
				&  & G & $\theta_{23}$ & 1.11 & 1.13 & 0.0179 & 0.0041 & 0.0044 & 0.10 & 0.11 & 0.0114 & 0.0025 & 0.0027 \\ 
				&  & F & $\theta_{13;2}$ & 0.91 & 0.98 & 0.0717 & 0.6913 & 0.6964 & 0.10 & 0.11 & 0.0058 & 0.0078 & 0.0078 \\ 
				\hline
				\cmidrule{1-14}
				\multirow{9}{*} {\begin{sideways} $n = 500$, $65\%$ censoring \end{sideways}} & \multirow{3}{*} {\begin{sideways} Known \end{sideways}} & G & $\theta_{12}$ & 1.11 & 1.11 & 0.0032 & 0.0017 & 0.0018 & 0.10 & 0.10 & 0.0013 & 0.0011 & 0.0011 \\ 
				&  & G & $\theta_{23}$ & 1.11 & 1.11 & 0.0004 & 0.0013 & 0.0013 & 0.10 & 0.10 & -0.0006 & 0.0008 & 0.0008 \\ 
				&  & F & $\theta_{13;2}$ & 0.91 & 0.92 & 0.0156 & 0.2675 & 0.2678 & 0.10 & 0.10 & 0.0009 & 0.0031 & 0.0031 \\ 
				\cmidrule{2-14}
				& \multirow{3}{*} {\begin{sideways} MLE \end{sideways}} & G & $\theta_{12}$ & 1.11 & 1.11 & 0.0030 & 0.0018 & 0.0018 & 0.10 & 0.10 & 0.0012 & 0.0012 & 0.0012 \\ 
				&  & G & $\theta_{23}$ & 1.11 & 1.11 & 0.0007 & 0.0013 & 0.0013 & 0.10 & 0.10 & -0.0004 & 0.0009 & 0.0009 \\ 
				&  & F & $\theta_{13;2}$ & 0.91 & 0.92 & 0.0147 & 0.2674 & 0.2676 & 0.10 & 0.10 & 0.0008 & 0.0031 & 0.0031 \\ 
				\cmidrule{2-14}
				& \multirow{3}{*} {\begin{sideways} KME \end{sideways}} & G & $\theta_{12}$ & 1.11 & 1.12 & 0.0097 & 0.0020 & 0.0021 & 0.10 & 0.11 & 0.0063 & 0.0013 & 0.0013 \\ 
				&  & G & $\theta_{23}$ & 1.11 & 1.12 & 0.0051 & 0.0014 & 0.0014 & 0.10 & 0.10 & 0.0031 & 0.0009 & 0.0009 \\ 
				&  & F & $\theta_{13;2}$ & 0.91 & 0.93 & 0.0233 & 0.2635 & 0.2641 & 0.10 & 0.10 & 0.0018 & 0.0030 & 0.0030 \\ 
				\hline
				\hline
			\end{tabular}
		\end{table}
	\end{landscape}

	\begin{landscape}
		\begin{table}[ht]
			\centering
			\captionof{table}{Performance measures for the estimation of the copula parameters and Kendall's $\tau$ values in
				case of complete and 25\% common right-censored event time data with sample size 500. The copula combination Gumbel (G), Gumbel (G), Frank (F) with true $\tau_{12}=\tau_{23}=\tau_{13;2}=0.1$ is investigated. Known margins, parametrically estimated margins (MLE) and nonparametrically estimated margins (ECDF/KME) are considered.}
			\label{Table:GGF_0+25_500_1}
			\begin{tabular}{cccccccccccccc}
				\midrule
				&  &  &  & \multicolumn{5}{c}{Copula parameter} &  \multicolumn{5}{c}{Kendall's $\tau$} \\
				&  &  &  & $\theta$ & $\bar{\theta}$ & $\hat{b}\left(\bar{\theta}\right)$ & $s^2\left(\bar{\theta}\right)$ & $\widehat{mse}\left(\bar{\theta}\right)$ & $\tau$ & $\bar{\tau}$ & $\hat{b}\left(\bar{\tau}\right)$ & $s^2\left(\bar{\tau}\right)$ & $\widehat{mse}\left(\bar{\tau}\right)$ \\
				\hline
				\cmidrule{1-14}
				\multirow{9}{*} {\begin{sideways} $n = 500$, $25\%$ censoring \end{sideways}} & \multirow{3}{*} {\begin{sideways} Known \end{sideways}} & G & $\theta_{12}$ & 1.11 & 1.11 & 0.0031 & 0.0013 & 0.0013 & 0.10 & 0.10 & 0.0015 & 0.0009 & 0.0009 \\ 
				&  & G & $\theta_{23}$ & 1.11 & 1.11 & 0.0008 & 0.0011 & 0.0011 & 0.10 & 0.10 & -0.0002 & 0.0007 & 0.0007 \\ 
				&  & F & $\theta_{13;2}$ & 0.91 & 0.94 & 0.0364 & 0.0898 & 0.0911 & 0.10 & 0.10 & 0.0037 & 0.0010 & 0.0011 \\ 
				\cmidrule{2-14}
				& \multirow{3}{*} {\begin{sideways} MLE \end{sideways}} & G & $\theta_{12}$ & 1.11 & 1.11 & 0.0033 & 0.0014 & 0.0014 & 0.10 & 0.10 & 0.0016 & 0.0009 & 0.0009 \\ 
				&  & G & $\theta_{23}$ & 1.11 & 1.11 & 0.0015 & 0.0012 & 0.0012 & 0.10 & 0.10 & 0.0003 & 0.0008 & 0.0008 \\ 
				&  & F & $\theta_{13;2}$ & 0.91 & 0.94 & 0.0335 & 0.0889 & 0.0900 & 0.10 & 0.10 & 0.0034 & 0.0010 & 0.0010 \\ 
				\cmidrule{2-14}
				& \multirow{3}{*} {\begin{sideways} KME \end{sideways}} & G & $\theta_{12}$ & 1.11 & 1.12 & 0.0072 & 0.0015 & 0.0015 & 0.10 & 0.10 & 0.0048 & 0.0009 & 0.0010 \\ 
				&  & G & $\theta_{23}$ & 1.11 & 1.12 & 0.0044 & 0.0012 & 0.0012 & 0.10 & 0.10 & 0.0027 & 0.0007 & 0.0008 \\ 
				&  & F & $\theta_{13;2}$ & 0.91 & 0.94 & 0.0344 & 0.0885 & 0.0897 & 0.10 & 0.10 & 0.0035 & 0.0010 & 0.0010 \\   			\hline
				\cmidrule{1-14}
				\multirow{9}{*} {\begin{sideways} $n = 500$, complete data \end{sideways}} & \multirow{3}{*} {\begin{sideways} Known \end{sideways}} & G & $\theta_{12}$ & 1.11 & 1.11 & 0.0025 & 0.0011 & 0.0011 & 0.10 & 0.10 & 0.0013 & 0.0007 & 0.0007 \\ 
				&  & G & $\theta_{23}$ & 1.11 & 1.11 & 0.0008 & 0.0010 & 0.0010 & 0.10 & 0.10 & -0.0001 & 0.0007 & 0.0007 \\ 
				&  & F & $\theta_{13;2}$ & 0.91 & 0.93 & 0.0253 & 0.0679 & 0.0685 & 0.10 & 0.10 & 0.0025 & 0.0008 & 0.0008 \\ 
				\cmidrule{2-14}
				& \multirow{3}{*} {\begin{sideways} MLE \end{sideways}} & G & $\theta_{12}$ & 1.11 & 1.11 & 0.0015 & 0.0011 & 0.0011 & 0.10 & 0.10 & 0.0004 & 0.0007 & 0.0007 \\ 
				&  & G & $\theta_{23}$ & 1.11 & 1.11 & 0.0002 & 0.0011 & 0.0011 & 0.10 & 0.10 & -0.0006 & 0.0007 & 0.0007 \\ 
				&  & F & $\theta_{13;2}$ & 0.91 & 0.92 & 0.0168 & 0.0664 & 0.0667 & 0.10 & 0.10 & 0.0016 & 0.0008 & 0.0008 \\ 
				\cmidrule{2-14}
				& \multirow{3}{*} {\begin{sideways} ECDF \end{sideways}} & G & $\theta_{12}$ & 1.11 & 1.12 & 0.0062 & 0.0012 & 0.0012 & 0.10 & 0.10 & 0.0041 & 0.0008 & 0.0008 \\ 
				&  & G & $\theta_{23}$ & 1.11 & 1.12 & 0.0045 & 0.0011 & 0.0012 & 0.10 & 0.10 & 0.0028 & 0.0007 & 0.0007 \\ 
				&  & F & $\theta_{13;2}$ & 0.91 & 0.93 & 0.0187 & 0.0673 & 0.0676 & 0.10 & 0.10 & 0.0018 & 0.0008 & 0.0008 \\  			\hline
				\hline
			\end{tabular}
		\end{table}
	\end{landscape}

	\subsection{Vine copula bootstrapping results for the mastitis data}\label{Sec:VineCopBootRes}
	The following tables show the results of the copula bootstrap, as applied to the four vine copula models that best describe the mastitis data (see Table 9 in the main text). All results are based on 100 replications. Each table contains in lines 1-3 the results for the copula parameters, in lines 4-6 the results for the Kendall's $\tau$ values and in lines 7-9 the results for the lower tail-dependence coefficients $\lambda^\text{L}$ (LTD). Since Frank (F) copulas do not exhibit any tail-dependence, the latter are only reported for Clayton (C) copulas. First, the underlying model is given. Second, we give estimation results in case of $65\%$ censoring (as in the mastitis data). Third, we provide estimation results in case of no censoring. The latter serves as a benchmark to assess the impact of information loss due to censoring. Each time, we list the mean parameter estimate together with the corresponding standard error (in parenthesis).
	
	\begin{enumerate}
		\item[] \textbf{1st best model}
		\begin{itemize}
			\item global likelihood estimation
		\end{itemize}
		\vspace*{-.5cm}
	\end{enumerate}
	\renewcommand{\arraystretch}{1.4}
	\begin{table}[H]
		\centering
		\scriptsize
		\hspace*{-.5cm}	\begin{tabular}{lccccccc}
			\hline
			\midrule
			\multirow{3}{*}{\begin{sideways} Parameter \end{sideways}} & model	&  $F; \ \widehat{\theta}_{13}: 6.56$ & $F; \ \widehat{\theta}_{34}: 6.34$ & $F; \ \widehat{\theta}_{24}: 6.99$ & $F; \ \widehat{\theta}_{13;2}: 1.68$ & $F; \ \widehat{\theta}_{24;3}: 2.79$ & $F; \ \widehat{\theta}_{14;23}: 3.71$ \\ 
			& 	$65\%$ censoring	& 6.727 (0.797) & 6.376 (0.746) & 7.102 (0.770) & 1.792 (0.549) & 2.886 (0.553) & 3.740 (0.650) \\ 
			&	complete data	& 6.641 (0.406) & 6.324 (0.377) & 7.045 (0.444) & 1.712 (0.295)  & 2.794 (0.304) & 3.715 (0.353) \\
			\midrule
			\multirow{3}{*}{\begin{sideways} Kendall's $\tau$ \end{sideways}}&	model	&  $F; \ \widehat{\tau}_{13}: 0.54$ & $F; \ \widehat{\tau}_{34}: 0.53$ & $F; \ \widehat{\tau}_{24}: 0.56$ & $F; \ \widehat{\tau}_{13;2}: 0.18$ & $F; \ \widehat{\tau}_{24;3}: 0.29$ & $F; \ \widehat{\tau}_{14;23}: 0.37$ \\ 
			&	$65\%$ censoring	 & 0.547 (0.036) & 0.531 (0.036) & 0.565 (0.034) & 0.192 (0.055) & 0.295 (0.049) & 0.366 (0.050) \\ 
			&	complete data    & 0.545 (0.019) & 0.530 (0.019) & 0.563 (0.020) & 0.184 (0.030) & 0.288 (0.027) & 0.366 (0.027) \\ 
			\midrule
			\multirow{3}{*}{\begin{sideways} LTD \end{sideways}}&	model	&   &  &  &  &  &  \\ 
			&	$65\%$ censoring	 & -- & -- & -- & -- & -- & -- \\ 
			&	complete data    &  &  &  & &  &  \\
			\hline
			\hline
		\end{tabular}
	\end{table}

	\begin{enumerate}
		\item[] 
		\begin{itemize}
			\item $\mathcal{T}_1$-sequential likelihood estimation
		\end{itemize}
		\vspace*{-.5cm}
	\end{enumerate}
	\begin{table}[H]
		\centering
		\scriptsize
		\hspace*{-.5cm}	\begin{tabular}{lccccccc}
			\hline
			\midrule
			\multirow{3}{*}{\begin{sideways} Parameter \end{sideways}} &	model &  $F; \ \widehat{\theta}_{13}: 6.38$ & $F; \ \widehat{\theta}_{34}: 6.34$ & $F; \ \widehat{\theta}_{24}: 6.77$ & $F; \ \widehat{\theta}_{13;2}: 1.67$ & $F; \ \widehat{\theta}_{24;3}: 2.81$ & $F; \ \widehat{\theta}_{14;23}: 3.72$ \\ 
			&	$65\%$ censoring	& 6.561 (0.808) & 6.367 (0.786) & 6.860 (0.802) & 1.771 (0.565) & 2.917 (0.566) & 3.733 (0.629) \\ 
			&	complete data	& 6.641 (0.406) & 6.324 (0.377) & 7.045 (0.444) & 1.712 (0.295) & 2.794 (0.304) & 3.715 (0.353) \\
			\midrule
			\multirow{3}{*}{\begin{sideways} Kendall's $\tau$ \end{sideways}}&	model &  $F; \ \widehat{\tau}_{13}: 0.53$ & $F; \ \widehat{\tau}_{34}: 0.53$ & $F; \ \widehat{\tau}_{24}: 0.55$ & $F; \ \widehat{\tau}_{13;2}: 0.18$ & $F; \ \widehat{\tau}_{24;3}: 0.29$ & $F; \ \widehat{\tau}_{14;23}: 0.37$ \\ 
			&	$65\%$ censoring	& 0.540 (0.038) & 0.530 (0.038) & 0.553 (0.036) & 0.189 (0.057) & 0.298 (0.050) & 0.365 (0.049) \\ 
			&	complete data	& 0.537 (0.019) & 0.530 (0.019) & 0.553 (0.020) & 0.183 (0.030) & 0.291 (0.027) & 0.366 (0.027) \\ 
			\midrule
			\multirow{3}{*}{\begin{sideways} LTD \end{sideways}}&	model	&   &  &  &  &  &  \\ 
			&	$65\%$ censoring	 & -- & -- & -- & -- & -- & -- \\ 
			&	complete data    &  &  &  & &  &  \\
			\hline
			\hline
		\end{tabular}
	\end{table} 
	
	\begin{enumerate}
		\item[] \textbf{2nd best model}
		\begin{itemize}
			\item global likelihood estimation
		\end{itemize}
		\vspace*{-.5cm}
	\end{enumerate}
	\begin{table}[H]
		\centering
		\scriptsize
		\hspace*{-.5cm}	\begin{tabular}{lccccccc}
			\hline
			\midrule
			\multirow{3}{*}{\begin{sideways} Parameter \end{sideways}} &	model &  $C;  \ \widehat{\theta}_{13}: 3.78 $ & $F; \ \widehat{\theta}_{34}: 6.39$ & $F; \ \widehat{\theta}_{24}: 6.93$ & $F; \ \widehat{\theta}_{13;2}: 1.51$ & $F; \ \widehat{\theta}_{24;3}: 2.78$ & $F; \ \widehat{\theta}_{14;23}: 3.48$ \\ 
			&	$65\%$ censoring	& 3.824 (0.580) & 6.420 (0.754) & 7.052 (0.742) & 1.624 (0.530) & 2.858 (0.510) & 3.491 (0.614) \\ 
			&	complete data	& 3.810 (0.214) & 6.364 (0.384) & 6.972 (0.447) & 1.538 (0.286) & 2.768 (0.298) & 3.516 (0.335) \\
			\midrule
			\multirow{3}{*}{\begin{sideways} Kendall's $\tau$ \end{sideways}}&	model &  $C; \ \widehat{\tau}_{13}: 0.65$ & $F; \ \widehat{\tau}_{34}: 0.53$ & $F; \ \widehat{\tau}_{24}: 0.56$ & $F; \ \widehat{\tau}_{13;2}: 0.16$ & $F; \ \widehat{\tau}_{24;3}: 0.29$ & $F; \ \widehat{\tau}_{14;23}: 0.35$ \\ 
			&	$65\%$ censoring	& 0.653 (0.035) & 0.533 (0.036) & 0.562 (0.033) & 0.175 (0.054) & 0.293 (0.046) & 0.346 (0.049) \\ 
			&	complete data	& 0.655 (0.013) & 0.532 (0.019) & 0.560 (0.020) & 0.167 (0.030) & 0.286 (0.027) & 0.350 (0.027) \\ 
			\midrule
			\multirow{3}{*}{\begin{sideways} LTD \end{sideways}}&	model & $C; \ \widehat{\lambda}^\text{L}_{13}: 0.83$  &  &  & & &  \\ 
			&	$65\%$ censoring	& 0.831 (0.024) & -- & -- & -- & -- & -- \\ 
			&	complete data	& 0.833 (0.008) &  &   &  &  &  \\ 
			\hline
			\hline
		\end{tabular}
	\end{table}

	\begin{enumerate}
		\item[] 
		\begin{itemize}
			\item $\mathcal{T}_1$-sequential likelihood estimation
		\end{itemize}
		\vspace*{-.5cm}
	\end{enumerate}
	\begin{table}[H]
		\centering
		\scriptsize
		\hspace*{-.5cm}	\begin{tabular}{lccccccc}
			\hline
			\midrule
			\multirow{3}{*}{\begin{sideways} Parameter \end{sideways}} &	model &  $C; \ \widehat{\theta}_{13}: 3.60$ & $F; \ \widehat{\theta}_{34}: 6.34$ & $F; \ \widehat{\theta}_{24}: 6.77$ & $F; \ \widehat{\theta}_{13;2}: 1.49$ & $F; \ \widehat{\theta}_{24;3}: 2.81$ & $F; \ \widehat{\theta}_{14;23}: 3.48$ \\ 
			&	$65\%$ censoring	& 3.653 (0.578) & 6.379 (0.785) & 6.863 (0.786) & 1.613 (0.576) & 2.884 (0.534) & 3.493 (0.633) \\ 
			&	complete data	& 3.629 (0.206) & 6.315 (0.382) & 6.806 (0.441) & 1.508 (0.287) & 2.797 (0.297) & 3.523 (0.335) \\
			\midrule
			\multirow{3}{*}{\begin{sideways} Kendall's $\tau$ \end{sideways}}&	model &  $C; \ \widehat{\tau}_{13}: 0.64$ & $F; \ \widehat{\tau}_{34}: 0.53$ & $F; \ \widehat{\tau}_{24}: 0.55$ & $F; \ \widehat{\tau}_{13;2}: 0.16$ & $F; \ \widehat{\tau}_{24;3}: 0.29$ & $F; \ \widehat{\tau}_{14;23}: 0.35$ \\ 
			&	$65\%$ censoring & 0.643 (0.037) & 0.531 (0.038) & 0.554 (0.035) & 0.173 (0.059) & 0.295 (0.048) & 0.346 (0.051) \\ 
			&	complete data	& 0.644 (0.013) & 0.530 (0.019) & 0.553 (0.020) & 0.164 (0.030) & 0.289 (0.027) & 0.350 (0.027) \\ 
			\midrule
			\multirow{3}{*}{\begin{sideways} LTD \end{sideways}}&	model & $C; \ \widehat{\lambda}^\text{L}_{13}: 0.82$  &  &  & & &  \\ 
			&	$65\%$ censoring	& 0.824 (0.026) & -- & -- & -- & -- & -- \\ 
			&	complete data	& 0.826 (0.009) &  &   &  &  &  \\ 
			\hline
			\hline
		\end{tabular}
	\end{table} 
	
	\begin{enumerate}
		\item[] \textbf{3rd best model}
		\begin{itemize}
			\item global likelihood estimation
		\end{itemize}
		\vspace*{-.5cm}
	\end{enumerate}
	\begin{table}[H]
		\centering
		\scriptsize
		\hspace*{-.5cm}	\begin{tabular}{lccccccc}
			\hline
			\midrule
			\multirow{3}{*}{\begin{sideways} Parameter \end{sideways}} &	model &  $F; \ \widehat{\theta}_{13}: 6.51$ & $F; \ \widehat{\theta}_{34}: 6.36$ & $C; \ \widehat{\theta}_{24}: 4.10$ & $F; \ \widehat{\theta}_{13;2}: 1.57$ & $F; \ \widehat{\theta}_{24;3}:2.79 $ & $F; \ \widehat{\theta}_{14;23}: 3.86$ \\ 
			&	$65\%$ censoring	& 6.676 (0.792) & 6.406 (0.719) & 4.134 (0.612) & 1.668 (0.549) & 2.889 (0.550) & 3.887 (0.650) \\ 
			&	complete data	& 6.592 (0.407) & 6.343 (0.376) & 4.138 (0.237) & 1.606 (0.288) & 2.798 (0.302) & 3.869 (0.357) \\
			\midrule
			\multirow{3}{*}{\begin{sideways} Kendall's $\tau$ \end{sideways}}&	model &  $F; \ \widehat{\tau}_{13}: 0.54$ & $F; \ \widehat{\tau}_{34}: 0.53$ & $C; \ \widehat{\tau}_{24}: 0.67$ & $F; \ \widehat{\tau}_{13;2}: 0.17$ & $F; \ \widehat{\tau}_{24;3}: 0.29$ & $F; \ \widehat{\tau}_{14;23}: 0.38$ \\ 
			&	$65\%$ censoring	& 0.545 (0.036) & 0.533 (0.035) & 0.671 (0.033) & 0.179 (0.056) & 0.296 (0.049) & 0.377 (0.050) \\ 
			&	complete data	& 0.543 (0.019) & 0.531 (0.019) & 0.674 (0.013) & 0.174 (0.030) & 0.289 (0.027) & 0.377 (0.027) \\ 
			\midrule
			\multirow{3}{*}{\begin{sideways} LTD \end{sideways}}&	model &   &  & $C; \ \widehat{\lambda}^\text{L}_{24}: 0.84$ & & &  \\ 
			&	$65\%$ censoring	& -- & -- & 0.843 (0.022) & -- & -- & -- \\ 
			&	complete data	&   &  & 0.845 (0.008)  &  &  &  \\ 
			\hline
			\hline
		\end{tabular}
	\end{table} 
	
	\begin{enumerate}
		\item[] 
		\begin{itemize}
			\item $\mathcal{T}_1$-sequential likelihood estimation
		\end{itemize}
		\vspace*{-.5cm}
	\end{enumerate}
	\begin{table}[H]
		\centering
		\scriptsize
		\hspace*{-.5cm}	\begin{tabular}{lccccccc}
			\hline
			\midrule
			\multirow{3}{*}{\begin{sideways} Parameter \end{sideways}} &	model &  $F; \ \widehat{\theta}_{13}: 6.38 $ & $F; \ \widehat{\theta}_{34}: 6.34$ & $C; \ \widehat{\theta}_{24}: 3.90$ & $F; \ \widehat{\theta}_{13;2}: 1.54 $ & $F; \ \widehat{\theta}_{24;3}: 2.76$ & $F; \ \widehat{\theta}_{14;23}: 3.86$ \\ 
			&	$65\%$ censoring	& 6.562 (0.808) & 6.373 (0.791) & 3.907 (0.598) & 1.634 (0.550) & 2.860 (0.553) & 3.862 (0.635) \\ 
			&	complete data	& 6.460 (0.403) & 6.315 (0.375) & 3.937 (0.228) & 1.568 (0.288) & 2.769 (0.301) & 3.868 (0.358) \\
			\midrule
			\multirow{3}{*}{\begin{sideways} Kendall's $\tau$ \end{sideways}}&	model &  $F; \ \widehat{\tau}_{13}: 0.53$ & $F; \ \widehat{\tau}_{34}: 0.53$ & $C; \ \widehat{\tau}_{24}: 0.66$ & $F; \ \widehat{\tau}_{13;2}: 0.17$ & $F; \ \widehat{\tau}_{24;3}: 0.29$ & $F; \ \widehat{\tau}_{14;23}: 0.38$ \\ 
			&	$65\%$ censoring	& 0.540 (0.038) & 0.530 (0.039) & 0.658 (0.034) & 0.175 (0.056) & 0.293 (0.049) & 0.375 (0.048) \\ 
			&	complete data	& 0.537 (0.019) & 0.530 (0.019) & 0.663 (0.013) & 0.170 (0.030) & 0.286 (0.027) & 0.377 (0.027) \\ 
			\midrule
			\multirow{3}{*}{\begin{sideways} LTD \end{sideways}}&	model &   &  & $C; \ \widehat{\lambda}^\text{L}_{24}: 0.84$ & & &  \\ 
			&	$65\%$ censoring	& -- & -- & 0.834 (0.023) & -- & -- & -- \\ 
			&	complete data	&   &  & 0.838 (0.009)  &  &  &  \\ 	
			\hline
			\hline
		\end{tabular}
	\end{table}

	\begin{enumerate}
		\item[] \textbf{4th best model}
		\begin{itemize}
			\item global likelihood estimation
		\end{itemize}
		\vspace*{-.5cm}
	\end{enumerate}
	\begin{table}[H]
		\centering
		\scriptsize
		\hspace*{-.5cm}	\begin{tabular}{lccccccc}
			\hline
			\midrule
			\multirow{3}{*}{\begin{sideways} Parameter \end{sideways}} &	model &  $C; \ \widehat{\theta}_{13}: 3.75$ & $F; \ \widehat{\theta}_{34}: 6.40$ & $C; \ \widehat{\theta}_{24}: 4.04$ & $F; \ \widehat{\theta}_{13;2}: 1.39 $ & $F; \ \widehat{\theta}_{24;3}: 2.72$ & $F; \ \widehat{\theta}_{14;23}: 3.71$ \\ 
			&	$65\%$ censoring	& 3.782 (0.612) & 6.437 (0.742) & 4.090 (0.590) & 1.491 (0.552) & 2.780 (0.509) & 3.747 (0.625) \\ 
			&	complete data	& 3.788 (0.212) & 6.377 (0.380) & 4.070 (0.235) & 1.414 (0.274) & 2.706 (0.292) & 3.744 (0.339) \\
			\midrule
			\multirow{3}{*}{\begin{sideways} Kendall's $\tau$ \end{sideways}}&	model &  $C; \ \widehat{\tau}_{13}: 0.65$ & $F; \ \widehat{\tau}_{34}: 0.53$ & $C; \ \widehat{\tau}_{24}: 0.67$ & $F; \ \widehat{\tau}_{13;2}: 0.15$ & $F; \ \widehat{\tau}_{24;3}: 0.28$ & $F; \ \widehat{\tau}_{14;23}: 0.37$ \\ 
			&	$65\%$ censoring	& 0.650 (0.037) & 0.534 (0.035) & 0.669 (0.032) & 0.161 (0.057) & 0.286 (0.046) & 0.366 (0.048) \\ 
			&	complete data	& 0.654 (0.013) & 0.533 (0.019) & 0.670 (0.013) & 0.154 (0.029) & 0.281 (0.026) & 0.368 (0.026) \\ 
			\midrule
			\multirow{3}{*}{\begin{sideways} LTD \end{sideways}}&	model &  $C; \ \widehat{\lambda}^\text{L}_{13}: 0.83$ &  & $C; \ \widehat{\lambda}^\text{L}_{24}: 0.84$ & & &  \\ 
			&	$65\%$ censoring	& 0.829 (0.025) & -- & 0.841 (0.022) & -- & -- & -- \\ 
			&	complete data	& 0.832 (0.008) &  & 0.843 (0.008) &  &  &  \\ 
			\hline
			\hline
		\end{tabular}
	\end{table} 
	
	\begin{enumerate}
		\item[] 
		\begin{itemize}
			\item $\mathcal{T}_1$-sequential likelihood estimation
		\end{itemize}
		\vspace*{-.5cm}
	\end{enumerate}
	\begin{table}[H]
		\centering
		\scriptsize
		\hspace*{-.5cm}	\begin{tabular}{lccccccc}
			\hline
			\midrule
			\multirow{3}{*}{\begin{sideways} Parameter \end{sideways}} &	model &  $C; \ \widehat{\theta}_{13}: 3.60$ & $F; \ \widehat{\theta}_{34}: 6.34$ & $C; \ \widehat{\theta}_{24}: 3.90$ & $F; \ \widehat{\theta}_{13;2}: 1.36$ & $F; \ \widehat{\theta}_{24;3}: 2.71$ & $F; \ \widehat{\theta}_{14;23}: 3.70$ \\ 
			&	$65\%$ censoring	& 3.653 (0.578) & 6.381 (0.791) & 3.914 (0.588) & 1.464 (0.565) & 2.770 (0.530) & 3.703 (0.644) \\ 
			&	complete data	& 3.630 (0.206) & 6.314 (0.378) & 3.926 (0.229) & 1.389 (0.275) & 2.690 (0.291) & 3.741 (0.339) \\
			\midrule
			\multirow{3}{*}{\begin{sideways} Kendall's $\tau$ \end{sideways}}&	model &  $C; \ \widehat{\tau}_{13}: 0.64$ & $F; \ \widehat{\tau}_{34}: 0.53$ & $C; \ \widehat{\tau}_{24}: 0.66$ & $F; \ \widehat{\tau}_{13;2}: 0.15$ & $F; \ \widehat{\tau}_{24;3}: 0.28$ & $F; \ \widehat{\tau}_{14;23}: 0.37$ \\ 
			&	$65\%$ censoring	& 0.643 (0.037) & 0.531 (0.038) & 0.659 (0.034) & 0.158 (0.058) & 0.285 (0.048) & 0.363 (0.050) \\ 
			&	complete data	& 0.644 (0.013) & 0.530 (0.019) & 0.662 (0.013) & 0.151 (0.029) & 0.279 (0.026) & 0.368 (0.026) \\ 
			\midrule
			\multirow{3}{*}{\begin{sideways} LTD \end{sideways}}&	model &  $C; \ \widehat{\lambda}^\text{L}_{13}: 0.82$ &  & $C; \ \widehat{\lambda}^\text{L}_{24}: 0.84$ & & &  \\ 
			&	$65\%$ censoring	& 0.824 (0.026) & -- & 0.835 (0.023) & -- & -- & -- \\ 
			&	complete data	& 0.826 (0.009) &  & 0.838 (0.009) &  &  &  \\ 
			\hline
			\hline
		\end{tabular}
	\end{table} 


\end{document}